\documentclass[12pt]{article}
\usepackage{cite}
\input{library.ps} 
\newcommand{\Init}{1 1 Scale Init}
\newcommand{\diagram}[1]{\special{" ProcDict begin #1 end}}
\newcommand{\Bullet}[2]{\diagram{\Init [#1 #2] Goto 0 Bullet}}
\newcommand{\VectorBoson}[5]{\diagram{\Init [[#1 #2] [#3 #4]] #5 VectorBoson}}

\newcommand{\Fermion}[5]{\diagram{\Init [[#1 #2] [#3 #4]] #5 Fermion}}

\newcommand{\ScaGho}[5]{\diagram{\Init [[#1 #2] [#3 #4]] #5 ScaGho}}
\newcommand{\VectorBosonArc}[6]{\diagram{\Init [[#1 #2] [#3 #4]] #5 #6 VectorBosonArc}}

\newcommand{\FermionArc}[6]{\diagram{\Init [[#1 #2] [#3 #4]] #5 #6 FermionArc}}
\newcommand{\ScaGhoArc}[6]{\diagram{\Init [[#1 #2] [#3 #4]] #5 #6 ScaGhoArc}}

\newcommand{\nlsig}{\mbox{O$(3)\,\sigma$-model}}
\newcommand {\SU}[1]{\mbox{SU($#1$)}}

\newcommand{\be}{\begin{equation}}
\newcommand{\ee}{\end{equation}}
\newcommand{\bea}{\begin{eqnarray}}
\newcommand{\eea}{\end{eqnarray}}

\newcommand{\refeq}[1]{\mbox{eq.~(\ref{eq:#1})}}
\newcommand{\refeqs}[2]{\mbox{eqs.~(\ref{eq:#1}),~(\ref{eq:#2})}}
\newcommand{\refEq}[1]{\mbox{Eq.~(\ref{eq:#1})}}

\newcommand{\refnn}[1]{\mbox{(\ref{eq:#1})}}

\newcommand{\Journal}[4]{{#1}{#2} (#4) #3}
\newcommand{\NPB}{{Nucl. Phys. B}}
\newcommand{\PLB}{{Phys. Lett. B}}
\newcommand{\PRe}{{Phys. Rep.\ }}
\newcommand{\PRL}{{Phys. Rev. Lett.\ }}
\newcommand{\CMP}{{Comm. Math. Phys.\ }}
\newcommand{\ANP}{{Ann. Phys. (N.Y.)\ }}

\newcommand{\PRD}{{Phys. Rev. D}}
\newcommand{\RMP}{{Rev. Mod. Phys.\ }}

\newcommand{\half}{\raisebox{.15ex}{\scriptsize$\frac{1}{2}$}}
\newcommand{\third}{\raisebox{.15ex}{\scriptsize$\frac{1}{3}$}}
\newcommand{\fourth}{\raisebox{.15ex}{\scriptsize$\frac{1}{4}$}}
\newcommand{\sixth}{\raisebox{.15ex}{\scriptsize$\frac{1}{6}$}}
\newcommand{\eighth}{\raisebox{.15ex}{\scriptsize$\frac{1}{8}$}}
\newcommand{\twelfth}{\raisebox{.15ex}{\scriptsize$\frac{1}{12}$}}

\newcommand{\twentyfourth}{\raisebox{.15ex}{\scriptsize$\frac{1}{24}$}}
\newcommand{\fourtyeighth}{\raisebox{.15ex}{\scriptsize$\frac{1}{48}$}}
\newcommand{\seventysecond}{\raisebox{.15ex}{\scriptsize$\frac{1}{72}$}}
\newcommand{\onefourfourth}{\raisebox{.15ex}{\scriptsize$\frac{1}{144}$}}
\newcommand{\twoeighteighth}{\raisebox{.15ex}{\scriptsize$\frac{1}{288}$}}
\newcommand{\forthreetwoth}{\raisebox{.15ex}{\scriptsize$\frac{1}{432}$}}
\newcommand{\eightsixfourth}{\raisebox{.15ex}{\scriptsize$\frac{1}{864}$}}

\newcommand{\threehalf}{\raisebox{.15ex}{\scriptsize$\frac{3}{2}$}}

\newcommand{\breuk}[2]{\raisebox{.15ex}{\scriptsize$\frac{#1}{#2}$}}
\newcommand{\medbreuk}[2]{\raisebox{.05ex}{\small$\frac{#1}{#2}$}}

\newcommand{\cD}{\mbox{$\cal D$}}
\newcommand{\cO}{\mbox{$\cal O$}}

\newcommand{\Tr}{\mbox{\,Tr\,}}
\newcommand{\RE}{\mbox{Re}}
\newcommand{\LW}{\mbox{L\"{u}scher-Weisz}}

\newcommand{\myvec}[1]{{\bf#1}}
\newcommand{\myvecC}[1]{{\bf#1}}
\newcommand{\myvecD}[1]{{\bf#1}}

\newcommand{\unit}{{\rm\bf1}}
\newcommand{\real}{\relax{\rm I\kern-.18em R}}
\newcommand{\complex}{\,\,\rule{.2mm}{2.8mm}\!\!{\rm C}}
\newcommand{\integer}{{\rm Z}\!\!{\rm Z}}
\newcommand{\natur}{\relax{\rm I\kern-.18em N}}

\newcommand{\action}[1]{\mbox{$S_{\mbox{\scriptsize #1}}$}}
\newcommand{\dollar}[1]{\raisebox{-2.3ex}{$\stackrel{\mbox{\LARGE\$}}{\mbox{\scriptsize $#1$}}$}}
\newcommand{\txtdollar}[1]{\mbox{$\$_{#1}$}}
\newcommand{\tA}{\tilde{A}}
\newcommand{\Pexp}{\mbox{$P$exp}}
\newcommand{\momconsper}[1]{(2\pi)^4\delta(#1)}
\newcommand{\prop}{D}
\newcommand{\fprop}{D^{\rm full}}
\newcommand{\hk}{\hat{k}}
\newcommand{\WWsym}{\raisebox{-1mm}{\large$\rule{0mm}{0mm}^{/\!\!/}$}\hspace{-1.16mm}\rule[3.3mm]{.65mm}{.13mm}\hspace{-1.8mm}\rule[-.1mm]{.65mm}{.13mm}\rule{0mm}{4mm}\,\,}
\newcommand{\WWint}[1]{\int_{\myvec{#1}}}
\newcommand{\kfeqzero}{|_{k_0 = 0}}
\newcommand{\brs}[2]{(#1,#2)}
\newcommand{\bra}[2]{\langle#1,#2\rangle}
\newcommand{\mod} {{\rm mod}}

\newcommand{\basispl}{
   \put(-.5,-.5){\line(1,0){1}}
   \put(.5,-.5){\line(0,1){1}}
   \put(.5,.5){\line(-1,0){1}}
   \put(-.5,.5){\line(0,-1){1}}}

\newcommand{\plaq}{\setlength{\unitlength}{.5cm}\raisebox{-.2cm}{
   \begin{picture}(1.2,1.2)(-.6,-.6)
   \basispl
   \put(-.5,-.5){\circle*{.2}}
   \put(-.5,.5){\circle*{.2}}
   \put(.5,-.5){\circle*{.2}}
   \put(.5,.5){\circle*{.2}}
   \put(.5,0){\vector(0,1){0}}
   \put(-.6,-.6){\makebox(0,0)[tr]{\footnotesize $x$}}
   \put(-.55,0){\makebox(0,0)[r]{\footnotesize $\nu$}}
   \put(0,-.55){\makebox(0,0)[t]{\footnotesize $\mu$}}
   \end{picture}}}

\newcommand{\twooneplaq}{\setlength{\unitlength}{.5cm}
   \raisebox{-.2cm}{
   \begin{picture}(2.2,1.2)(-1.1,-.6)
   \put(-1,-.5){\line(1,0){2}}
   \put(-1,.5){\line(1,0){2}}
   \put(-1,-.5){\line(0,1){1}}
   \put(1,-.5){\line(0,1){1}}
   \multiput(-1,-.5)(1,0){3}{\circle*{.2}}
   \multiput(-1,.5)(1,0){3}{\circle*{.2}}
   \put(-1.1,-.6){\makebox(0,0)[tr]{\footnotesize $x$}}
   \put(-1.05,0){\makebox(0,0)[r]{\footnotesize $\nu$}}
   \put(-.3,-.55){\makebox(0,0)[t]{\footnotesize $\mu$}}
   \put(1,0){\vector(0,1){0}}
   \end{picture}}}

\newcommand{\hookplaq}{\setlength{\unitlength}{.5cm}
   \raisebox{-.3268cm}{
   \begin{picture}(1.7071,1.7071)(-.7071,-.7071)
   \put(0,0){\line(0,1){1}}
   \put(0,1){\line(1,0){1}}
   \put(1,1){\line(0,-1){1}}
   \put(-.7071,-.7071){\line(1,0){1}}
   \put(0,0){\line(-1,-1){.7071}}
   \put(1,0){\line(-1,-1){.7071}}
   \multiput(0,0)(1,0){2}{\circle*{.2}}
   \multiput(0,1)(1,0){2}{\circle*{.2}}
   \multiput(-.7071,-.7071)(1,0){2}{\circle*{.2}}
   \multiput(0,0)(.25,0){4}{\circle*{.03}}
   \put(1,0.5){\vector(0,1){0}}
   \put(-.8,-.8){\makebox(0,0)[tr]{\footnotesize $x$}}
   \put(-.2,.6){\makebox(0,0)[r]{\footnotesize $\nu$}}
   \put(-.2,-.8){\makebox(0,0)[t]{\footnotesize $\mu$}}
   \put(.9,-.4){\makebox(0,0)[t]{\footnotesize $\lambda$}}
   \end{picture}}}

\newcommand{\cornplaq}{\setlength{\unitlength}{.5cm}
   \raisebox{-.3268cm}{
   \begin{picture}(1.7071,1.7071)(-.7071,-.7071)
   \put(-.7071,-.7071){\line(0,1){1}}
   \put(0,1){\line(1,0){1}}
   \put(1,1){\line(0,-1){1}}
   \put(-.7071,-.7071){\line(1,0){1}}
   \put(0,1){\line(-1,-1){.7071}}
   \put(1,0){\line(-1,-1){.7071}}
   \put(-.7071,-.7071){\circle*{.1}}
   \put(-.7071,.2929){\circle*{.2}}
   \multiput(0,0)(1,0){2}{\circle*{.2}}
   \multiput(0,1)(1,0){2}{\circle*{.2}}
   \multiput(-.7071,-.7071)(1,0){2}{\circle*{.2}}
   \multiput(0,0)(.25,0){4}{\circle*{.03}}
   \multiput(0,0)(0,.25){4}{\circle*{.03}}
   \multiput(0,0)(-.1768,-.1768){4}{\circle*{.03}}
   \put(1,0.5){\vector(0,1){0}}
   \put(-.8,-.8){\makebox(0,0)[tr]{\footnotesize $x$}}
   \put(-.8,-.2){\makebox(0,0)[r]{\footnotesize $\nu$}}
   \put(-.2,-.8){\makebox(0,0)[t]{\footnotesize $\mu$}}
   \put(.9,-.4){\makebox(0,0)[t]{\footnotesize $\lambda$}}
   \end{picture}}}

\newcommand{\twoplaq}{\setlength{\unitlength}{1cm}\raisebox{-.5cm}{
   \begin{picture}(1.2,1.2)(-.6,-.6)
   \basispl
   \put(-.5,-.5){\circle*{.1}}
   \put(-.5,.5){\circle*{.1}}
   \put(.5,-.5){\circle*{.1}}
   \put(.5,.5){\circle*{.1}}
   \put(0,-.5){\circle*{.1}}
   \put(0,.5){\circle*{.1}}
   \put(.5,0){\circle*{.1}}
   \put(-.5,0){\circle*{.1}}
   \put(.5,-.2){\vector(0,1){0}}
   \put(-.55,-.55){\makebox(0,0)[tr]{\footnotesize $x$}}
   \put(-.55,-.2){\makebox(0,0)[r]{\footnotesize $\nu$}}
   \put(-.2,-.55){\makebox(0,0)[t]{\footnotesize $\mu$}}
   \end{picture}}}  
 
\newcommand{\twoplaqbare}{\setlength{\unitlength}{1cm}\raisebox{-.5cm}{
   \begin{picture}(1.2,1.2)(-.6,-.6)
   \basispl
   \put(-.5,0){\circle*{.1}}
   \put(.5,-.2){\vector(0,1){0}}
   \end{picture}}}

\newcommand{\twoplaqsplitA}{\setlength{\unitlength}{1cm}\raisebox{-.5cm}{
   \begin{picture}(1.2,1.2)(-.6,-.6)
   \put(-.5,0){\line(0,-1){.5}}
   \put(-.5,-.5){\line(1,0){1}}
   \put(.5,-.5){\line(0,1){.43}}
   \put(.5,0){\oval(1.8,.14)[l]}
   \put(.5,.06){\line(0,1){.43}}
   \put(.5,.5){\line(-1,0){1}}
   \put(-.5,.5){\line(0,-1){1}}
   \put(-.5,0){\circle*{.1}}
   \put(.5,-.2){\vector(0,1){0}}
   \end{picture}}}

\newcommand{\twoplaqup}{\setlength{\unitlength}{1cm}\raisebox{-.5cm}{
   \begin{picture}(1.2,1.2)(-.6,-.6)
   \put(-.5,0){\line(0,-1){.5}}
   \put(-.5,-.5){\line(1,0){1}}
   \put(.5,-.5){\line(0,1){.5}}
   \put(-.5,0){\line(1,0){1}}
   \put(-.5,0){\circle*{.1}}
   \put(.5,-.2){\vector(0,1){0}}
   \end{picture}}}

\newcommand{\twoplaqdown}{\setlength{\unitlength}{1cm}\raisebox{-.5cm}{
   \begin{picture}(1.2,1.2)(-.6,-.6)
   \put(-.5,0){\line(0,1){.5}}
   \put(-.5,.5){\line(1,0){1}}
   \put(.5,.5){\line(0,-1){.5}}
   \put(-.5,0){\line(1,0){1}}
   \put(-.5,0){\circle*{.1}}
   \put(.5,.3){\vector(0,1){0}}
   \end{picture}}}

\newcommand{\gluonline}{
\hspace{-7.5mm}
\raisebox{1mm}{
\setlength{\unitlength}{25pt}
\renewcommand{\Init}{0.330 0.330 Scale Init}
\begin{picture}(0.990,0.330)
\put(0,0){\VectorBoson{1.000}{0.000}{5.000}{0.000}{}}
\end{picture}
\hspace{5mm}
}}

\newcommand{\ghostline}{
\hspace{-7.5mm}
\raisebox{1mm}{
\setlength{\unitlength}{25pt}
\renewcommand{\Init}{0.330 0.330 Scale Init}
\begin{picture}(0.990,0.330)
\put(0,0){\ScaGho{1.000}{0.000}{5.000}{0.000}{+}}
\end{picture}
\hspace{5mm}
}}

\newcommand{\vertex}{
\hspace{-8mm}
\raisebox{.8mm}{
\setlength{\unitlength}{25pt}
\renewcommand{\Init}{0.330 0.330 Scale Init}
\begin{picture}(0.330,0.330)
\put(0,0){\Bullet{1.000}{0.000}}
\end{picture}
\hspace{-1mm}
}}

\newcommand{\stattree}{
\hspace{-8mm}
\raisebox{.8mm}{
\setlength{\unitlength}{25pt}
\renewcommand{\Init}{0.330 0.330 Scale Init}
\begin{picture}(1.980,0.990)
\put(0,0){\Fermion{1.000}{2.000}{1.000}{-2.000}{}}
\put(0,0){\Fermion{1.000}{-2.000}{11.000}{-2.000}{}}
\put(0,0){\Fermion{11.000}{-2.000}{11.000}{2.000}{}}
\put(0,0){\Fermion{11.000}{2.000}{1.000}{2.000}{}}
\put(1.980,-0.660){\vector(1,0){0}}
\put(3.630,0.000){\vector(0,1){0}}
\put(1.980,0.660){\vector(-1,0){0}}
\put(0.330,0.000){\vector(0,-1){0}}
\put(0,0){\VectorBoson{4.000}{2.000}{7.000}{-2.000}{+}}
\end{picture}
\hspace{13mm}
}}

\newcommand{\statonefour}{
\hspace{-8mm}
\raisebox{1mm}{
\setlength{\unitlength}{25pt}
\renewcommand{\Init}{0.330 0.330 Scale Init}
\begin{picture}(1.980,0.990)
\put(0,0){\Fermion{1.000}{2.000}{1.000}{-2.000}{}}
\put(0,0){\Fermion{1.000}{-2.000}{11.000}{-2.000}{}}
\put(0,0){\Fermion{11.000}{-2.000}{11.000}{2.000}{}}
\put(0,0){\Fermion{11.000}{2.000}{1.000}{2.000}{}}
\put(0,0){\VectorBoson{8.000}{2.000}{1.000}{-1.000}{}}
\put(1.980,-0.660){\vector(1,0){0}}
\put(3.630,0.000){\vector(0,1){0}}
\put(1.980,0.660){\vector(-1,0){0}}
\put(0.330,0.000){\vector(0,-1){0}}
\put(0,0){\VectorBosonArc{7.000}{-2.000}{10.000}{-2.000}{183.974}{}}
\end{picture}
\hspace{13mm}
}}

\newcommand{\statonethree}{
\hspace{-8mm}
\raisebox{1mm}{
\setlength{\unitlength}{25pt}
\renewcommand{\Init}{0.330 0.330 Scale Init}
\begin{picture}(1.980,0.990)
\put(0,0){\Fermion{1.000}{2.000}{1.000}{-2.000}{}}
\put(0,0){\Fermion{1.000}{-2.000}{11.000}{-2.000}{}}
\put(0,0){\Fermion{11.000}{-2.000}{11.000}{2.000}{}}
\put(0,0){\Fermion{11.000}{2.000}{1.000}{2.000}{}}
\put(1.980,-0.660){\vector(1,0){0}}
\put(3.630,0.000){\vector(0,1){0}}
\put(1.980,0.660){\vector(-1,0){0}}
\put(0.330,0.000){\vector(0,-1){0}}
\put(0,0){\Bullet{5.000}{0.000}}
\put(0,0){\Bullet{5.000}{0.000}}
\put(0,0){\VectorBoson{5.000}{0.000}{11.000}{-1.000}{}}
\put(0,0){\VectorBoson{5.000}{0.000}{3.000}{-2.000}{}}
\put(0,0){\VectorBoson{5.000}{0.000}{2.000}{2.000}{+}}
\end{picture}
\hspace{13mm}
}}

\newcommand{\statonevacpol}{
\hspace{-8mm}
\raisebox{1mm}{
\setlength{\unitlength}{25pt}
\renewcommand{\Init}{0.330 0.330 Scale Init}
\begin{picture}(1.980,0.990)
\put(0,0){\Fermion{1.000}{2.000}{1.000}{-2.000}{}}
\put(0,0){\Fermion{1.000}{-2.000}{11.000}{-2.000}{}}
\put(0,0){\Fermion{11.000}{-2.000}{11.000}{2.000}{}}
\put(0,0){\Fermion{11.000}{2.000}{1.000}{2.000}{}}
\put(1.980,-0.660){\vector(1,0){0}}
\put(3.630,0.000){\vector(0,1){0}}
\put(1.980,0.660){\vector(-1,0){0}}
\put(0.330,0.000){\vector(0,-1){0}}
\put(0.330,0.000){\vector(0,-1){0}}
\put(0,0){\FermionArc{4.000}{0.000}{6.000}{0.000}{-180.000}{}}
\put(0,0){\FermionArc{6.000}{0.000}{4.000}{0.000}{-180.000}{}}
\put(0,0){\Fermion{5.000}{-1.000}{6.000}{0.000}{}}
\put(0,0){\Fermion{4.000}{0.000}{5.000}{1.000}{}}
\put(0,0){\Fermion{5.680}{0.680}{4.320}{-0.680}{}}
\put(0,0){\Fermion{4.120}{-0.400}{5.400}{0.880}{}}
\put(0,0){\Fermion{4.600}{-0.880}{5.880}{0.400}{}}
\put(0,0){\VectorBoson{2.000}{-2.000}{4.160}{-0.560}{}}
\put(0,0){\VectorBoson{8.000}{2.000}{5.840}{0.600}{}}
\end{picture}
\hspace{13mm}
}}

\newcommand{\vacpoltot}{
\hspace{-7mm}
\raisebox{1mm}{
\setlength{\unitlength}{25pt}
\renewcommand{\Init}{0.330 0.330 Scale Init}
\begin{picture}(1.650,0.990)
\put(0,0){\VectorBoson{1.000}{0.000}{3.000}{0.000}{}}
\put(0,0){\VectorBoson{7.000}{0.000}{9.000}{0.000}{}}
\put(0,0){\FermionArc{3.000}{0.000}{7.000}{0.000}{-180.000}{}}
\put(0,0){\FermionArc{7.000}{0.000}{3.000}{0.000}{-180.000}{}}
\put(0,0){\Fermion{3.000}{0.000}{5.000}{2.000}{}}
\put(0,0){\Fermion{5.000}{-2.000}{7.000}{0.000}{}}
\put(0,0){\Fermion{3.600}{-1.400}{6.400}{1.400}{}}
\put(0,0){\Fermion{4.160}{-1.800}{6.760}{0.800}{}}
\put(0,0){\Fermion{3.200}{-0.800}{5.800}{1.800}{}}
\end{picture}
\hspace{10mm}
}}

\newcommand{\vacpolins}{
\hspace{-7mm}
\raisebox{1mm}{
\setlength{\unitlength}{25pt}
\renewcommand{\Init}{0.330 0.330 Scale Init}
\begin{picture}(1.650,0.990)
\put(0,0){\VectorBoson{1.000}{0.000}{4.320}{0.000}{}}
\put(0,0){\VectorBoson{5.640}{0.000}{9.000}{0.000}{}}
\put(0,0){\FermionArc{4.320}{0.000}{5.640}{0.000}{180.000}{}}
\put(0,0){\FermionArc{4.320}{0.000}{5.640}{0.000}{-180.000}{}}
\put(1.650,0.000){\makebox(0,0){\small$i$}}
\end{picture}
\hspace{10mm}
}}

\newcommand{\vacpolmeas}{
\hspace{-7mm}
\raisebox{1mm}{
\setlength{\unitlength}{25pt}
\renewcommand{\Init}{0.330 0.330 Scale Init}
\begin{picture}(1.650,0.990)
\put(0,0){\VectorBoson{1.000}{0.000}{4.320}{0.000}{}}
\put(0,0){\VectorBoson{5.640}{0.000}{9.000}{0.000}{}}
\put(0,0){\FermionArc{4.320}{0.000}{5.640}{0.000}{180.000}{}}
\put(0,0){\FermionArc{4.320}{0.000}{5.640}{0.000}{-180.000}{}}
\put(0,0){\Fermion{4.520}{-0.480}{5.480}{0.480}{}}
\put(0,0){\Fermion{4.520}{0.480}{5.480}{-0.480}{}}
\end{picture}
\hspace{10mm}
}}

\newcommand{\vacpolghthree}{
\hspace{-7mm}
\raisebox{1mm}{
\setlength{\unitlength}{25pt}
\renewcommand{\Init}{0.330 0.330 Scale Init}
\begin{picture}(1.650,0.990)
\put(0,0){\VectorBoson{1.000}{0.000}{3.000}{0.000}{}}
\put(0,0){\VectorBoson{7.000}{0.000}{9.000}{0.000}{}}
\put(0,0){\ScaGhoArc{3.000}{0.000}{7.000}{0.000}{-180.000}{+}}
\put(0,0){\ScaGhoArc{7.000}{0.000}{3.000}{0.000}{-180.000}{+}}
\put(0,0){\Bullet{3.000}{0.000}}
\put(0,0){\Bullet{7.000}{0.000}}
\end{picture}
\hspace{10mm}
}}

\newcommand{\vacpolghfour}{
\hspace{-6mm}
\raisebox{6.5mm}{
\setlength{\unitlength}{25pt}
\renewcommand{\Init}{0.330 0.330 Scale Init}
\begin{picture}(1.650,0.990)
\put(0,0){\VectorBoson{1.000}{-2.000}{5.200}{-2.000}{+}}
\put(0,0){\VectorBoson{5.200}{-2.000}{9.000}{-2.000}{+}}
\put(0,0){\Bullet{4.960}{-1.720}}
\put(0,0){\ScaGhoArc{4.960}{-1.720}{4.960}{2.280}{-180.000}{+}}
\put(0,0){\ScaGhoArc{4.960}{2.280}{4.960}{-1.720}{-180.000}{+}}
\end{picture}
\hspace{11mm}
}}

\newcommand{\vacpolVthree}{
\hspace{-7mm}
\raisebox{1mm}{
\setlength{\unitlength}{25pt}
\renewcommand{\Init}{0.330 0.330 Scale Init}
\begin{picture}(1.650,0.990)
\put(0,0){\VectorBoson{1.000}{0.000}{3.000}{0.000}{}}
\put(0,0){\VectorBoson{7.000}{0.000}{9.000}{0.000}{}}
\put(0,0){\VectorBosonArc{3.000}{0.000}{7.000}{0.000}{-180.000}{}}
\put(0,0){\Bullet{3.000}{0.000}}
\put(0,0){\Bullet{7.000}{0.000}}
\put(0,0){\VectorBosonArc{3.000}{0.000}{7.000}{0.000}{180.000}{}}
\end{picture}
\hspace{10mm}
}}

\newcommand{\vacpolVfour}{
\hspace{-6mm}
\raisebox{6.5mm}{
\setlength{\unitlength}{25pt}
\renewcommand{\Init}{0.330 0.330 Scale Init}
\begin{picture}(1.650,0.990)
\put(0,0){\VectorBoson{1.000}{-2.000}{5.200}{-2.000}{+}}
\put(0,0){\VectorBoson{5.200}{-2.000}{9.000}{-2.000}{+}}
\put(0,0){\Bullet{4.960}{-1.720}}
\put(0,0){\VectorBosonArc{4.960}{-1.720}{4.960}{2.280}{-180.000}{}}
\put(0,0){\VectorBosonArc{4.960}{2.280}{4.960}{-1.720}{-180.000}{}}
\put(1.610,-0.488){\makebox(0,0)[b]{\scriptsize$V$}}
\end{picture}
\hspace{11mm}
}}

\newcommand{\vacpolW}{
\hspace{-6mm}
\raisebox{6.5mm}{
\setlength{\unitlength}{25pt}
\renewcommand{\Init}{0.330 0.330 Scale Init}
\begin{picture}(1.650,0.990)
\put(0,0){\VectorBoson{1.000}{-2.000}{5.200}{-2.000}{+}}
\put(0,0){\VectorBoson{5.200}{-2.000}{9.000}{-2.000}{+}}
\put(0,0){\Bullet{4.960}{-1.720}}
\put(0,0){\VectorBosonArc{4.960}{-1.720}{4.960}{2.280}{-180.000}{}}
\put(0,0){\VectorBosonArc{4.960}{2.280}{4.960}{-1.720}{-180.000}{}}
\put(1.584,-0.488){\makebox(0,0)[b]{\scriptsize$W$}}
\end{picture}
\hspace{11mm}
}}

\newcommand{\threegltot}{
\hspace{-6mm}
\raisebox{1mm}{
\setlength{\unitlength}{25pt}
\renewcommand{\Init}{0.330 0.330 Scale Init}
\begin{picture}(1.650,1.650)
\put(0,0){\VectorBoson{5.400}{0.000}{8.000}{0.000}{}}
\put(0,0){\FermionArc{1.400}{0.000}{5.400}{0.000}{-180.000}{}}
\put(0,0){\FermionArc{5.400}{0.000}{1.400}{0.000}{-180.000}{}}
\put(0,0){\Fermion{1.400}{0.000}{3.400}{2.000}{}}
\put(0,0){\Fermion{3.400}{-2.000}{5.400}{0.000}{}}
\put(0,0){\Fermion{2.000}{-1.400}{4.800}{1.400}{}}
\put(0,0){\Fermion{2.560}{-1.800}{5.160}{0.800}{}}
\put(0,0){\Fermion{1.600}{-0.800}{4.200}{1.800}{}}
\put(0,0){\VectorBoson{1.000}{4.000}{2.240}{1.640}{+}}
\put(0,0){\VectorBoson{1.000}{-4.000}{2.240}{-1.640}{}}
\end{picture}
\hspace{9mm}}}

\newcommand{\threeglone}{
\hspace{-6mm}
\raisebox{1mm}{
\setlength{\unitlength}{25pt}
\renewcommand{\Init}{0.330 0.330 Scale Init}
\begin{picture}(1.650,1.650)
\put(0,0){\FermionArc{2.840}{0.000}{4.160}{0.000}{180.000}{}}
\put(0,0){\FermionArc{2.840}{0.000}{4.160}{0.000}{-180.000}{}}
\put(1.162,0.000){\makebox(0,0){\small$i$}}
\put(0,0){\VectorBoson{1.000}{-4.000}{3.240}{-0.640}{}}
\put(0,0){\VectorBoson{1.000}{3.960}{3.200}{0.680}{+}}
\put(0,0){\VectorBoson{8.000}{-0.040}{4.160}{0.000}{+}}
\end{picture}
\hspace{9mm}}}

\newcommand{\threegltwoA}{
\hspace{-6mm}
\raisebox{1mm}{
\setlength{\unitlength}{25pt}
\renewcommand{\Init}{0.330 0.330 Scale Init}
\begin{picture}(1.650,1.650)
\put(0,0){\ScaGhoArc{2.000}{0.000}{6.000}{0.000}{-148.939}{+}}
\put(0,0){\ScaGhoArc{6.000}{0.000}{2.000}{0.000}{-148.939}{+}}
\put(0,0){\VectorBoson{6.000}{0.000}{8.000}{0.000}{}}
\put(0,0){\VectorBoson{1.000}{3.960}{2.000}{0.000}{}}
\put(0,0){\VectorBoson{1.000}{-4.000}{2.000}{0.000}{+}}
\put(0,0){\Bullet{2.000}{0.000}}
\put(0,0){\Bullet{6.000}{0.000}}
\end{picture}
\hspace{9mm}}}

\newcommand{\threegltwoB}{
\hspace{-6mm}
\raisebox{1mm}{
\setlength{\unitlength}{25pt}
\renewcommand{\Init}{0.330 0.330 Scale Init}
\begin{picture}(1.650,1.650)
\put(0,0){\ScaGhoArc{4.000}{2.000}{4.000}{-2.000}{-148.939}{+}}
\put(0,0){\ScaGhoArc{4.000}{-2.000}{4.000}{2.000}{-148.939}{+}}
\put(0,0){\VectorBoson{1.000}{4.000}{4.000}{2.000}{+}}
\put(0,0){\VectorBoson{1.000}{-4.000}{4.000}{-2.000}{}}
\put(0,0){\VectorBosonArc{4.000}{2.000}{8.000}{0.000}{109.898}{}}
\put(0,0){\Bullet{4.000}{2.000}}
\put(0,0){\Bullet{4.000}{-2.000}}
\end{picture}
\hspace{9mm}}}

\newcommand{\threegltwoC}{
\hspace{-6mm}
\raisebox{1mm}{
\setlength{\unitlength}{25pt}
\renewcommand{\Init}{0.330 0.330 Scale Init}
\begin{picture}(1.650,1.650)
\put(0,0){\ScaGhoArc{4.000}{2.000}{4.000}{-2.000}{-148.939}{+}}
\put(0,0){\ScaGhoArc{4.000}{-2.000}{4.000}{2.000}{-148.939}{+}}
\put(0,0){\VectorBoson{1.000}{4.000}{4.000}{2.000}{+}}
\put(0,0){\VectorBoson{1.000}{-4.000}{4.000}{-2.000}{}}
\put(0,0){\Bullet{4.000}{-2.000}}
\put(0,0){\VectorBosonArc{4.000}{-2.000}{8.000}{0.000}{-91.908}{+}}
\put(0,0){\Bullet{4.000}{2.000}}
\end{picture}
\hspace{9mm}}}

\newcommand{\threeglthreeA}{
\hspace{-6mm}
\raisebox{1mm}{
\setlength{\unitlength}{25pt}
\renewcommand{\Init}{0.330 0.330 Scale Init}
\begin{picture}(1.650,1.650)
\put(0,0){\VectorBoson{1.000}{-4.000}{2.200}{-2.000}{}}
\put(0,0){\VectorBoson{1.000}{3.960}{2.200}{2.000}{+}}
\put(0,0){\VectorBoson{8.000}{-0.040}{5.680}{0.040}{+}}
\put(0,0){\ScaGho{2.200}{2.000}{2.200}{-2.000}{+}}
\put(0,0){\ScaGho{2.200}{-2.000}{5.680}{0.000}{+}}
\put(0,0){\ScaGho{5.680}{0.000}{2.200}{2.000}{+}}
\put(0,0){\Bullet{2.200}{2.000}}
\put(0,0){\Bullet{2.200}{-2.000}}
\put(0,0){\Bullet{5.680}{0.000}}
\end{picture}
\hspace{9mm}}}

\newcommand{\threeglthreeB}{
\hspace{-6mm}
\raisebox{1mm}{
\setlength{\unitlength}{25pt}
\renewcommand{\Init}{0.330 0.330 Scale Init}
\begin{picture}(1.650,1.650)
\put(0,0){\VectorBoson{1.000}{-4.000}{2.200}{-2.000}{+}}
\put(0,0){\VectorBoson{1.000}{3.960}{2.200}{2.000}{+}}
\put(0,0){\VectorBoson{8.000}{-0.040}{5.680}{0.040}{+}}
\put(0,0){\ScaGho{2.200}{2.000}{2.200}{-2.000}{-}}
\put(0,0){\ScaGho{2.200}{-2.000}{5.680}{0.000}{-}}
\put(0,0){\ScaGho{5.680}{0.000}{2.200}{2.000}{-}}
\put(0,0){\Bullet{2.200}{2.000}}
\put(0,0){\Bullet{2.200}{-2.000}}
\put(0,0){\Bullet{5.680}{0.000}}
\end{picture}
\hspace{9mm}}}

\newcommand{\threeglfourVA}{
\hspace{-6mm}
\raisebox{1mm}{
\setlength{\unitlength}{25pt}
\renewcommand{\Init}{0.330 0.330 Scale Init}
\begin{picture}(1.650,1.650)
\put(0,0){\VectorBoson{6.000}{0.000}{8.000}{0.000}{}}
\put(0,0){\VectorBoson{1.000}{3.960}{2.000}{0.000}{}}
\put(0,0){\VectorBoson{1.000}{-4.000}{2.000}{0.000}{+}}
\put(0,0){\VectorBosonArc{2.000}{0.000}{6.000}{0.000}{148.939}{}}
\put(0,0){\VectorBosonArc{2.000}{0.000}{6.000}{0.000}{-148.939}{+}}
\put(0,0){\Bullet{2.000}{0.000}}
\put(0,0){\Bullet{6.000}{0.000}}
\put(0.766,0.000){\makebox(0,0)[l]{\scriptsize$V$}}
\end{picture}
\hspace{9mm}}}

\newcommand{\threeglfourVB}{
\hspace{-6mm}
\raisebox{1mm}{
\setlength{\unitlength}{25pt}
\renewcommand{\Init}{0.330 0.330 Scale Init}
\begin{picture}(1.650,1.650)
\put(0,0){\VectorBoson{1.000}{4.000}{4.000}{2.000}{+}}
\put(0,0){\VectorBoson{1.000}{-4.000}{4.000}{-2.000}{}}
\put(0,0){\VectorBosonArc{4.000}{2.000}{8.000}{0.000}{109.898}{}}
\put(0,0){\VectorBosonArc{4.000}{2.000}{4.000}{-2.000}{148.939}{}}
\put(0,0){\VectorBosonArc{4.000}{2.000}{4.000}{-2.000}{-148.939}{}}
\put(0,0){\Bullet{4.000}{2.000}}
\put(0,0){\Bullet{4.000}{-2.000}}
\put(1.294,0.409){\makebox(0,0)[t]{\scriptsize$V$}}
\end{picture}
\hspace{9mm}}}

\newcommand{\threeglfourVC}{
\hspace{-6mm}
\raisebox{1mm}{
\setlength{\unitlength}{25pt}
\renewcommand{\Init}{0.330 0.330 Scale Init}
\begin{picture}(1.650,1.650)
\put(0,0){\VectorBoson{1.000}{4.000}{4.000}{2.000}{+}}
\put(0,0){\VectorBoson{1.000}{-4.000}{4.000}{-2.000}{}}
\put(0,0){\VectorBosonArc{4.000}{2.000}{4.000}{-2.000}{148.939}{}}
\put(0,0){\VectorBosonArc{4.000}{2.000}{4.000}{-2.000}{-148.939}{}}
\put(0,0){\VectorBosonArc{4.000}{-2.000}{8.000}{0.000}{-79.195}{}}
\put(1.307,-0.436){\makebox(0,0)[b]{\scriptsize$V$}}
\put(0,0){\Bullet{4.000}{2.000}}
\put(0,0){\Bullet{4.000}{-2.000}}
\end{picture}
\hspace{9mm}}}

\newcommand{\threeglfourWA}{
\hspace{-6mm}
\raisebox{1mm}{
\setlength{\unitlength}{25pt}
\renewcommand{\Init}{0.330 0.330 Scale Init}
\begin{picture}(1.650,1.650)
\put(0,0){\VectorBoson{6.000}{0.000}{8.000}{0.000}{}}
\put(0,0){\VectorBoson{1.000}{3.960}{2.000}{0.000}{}}
\put(0,0){\VectorBoson{1.000}{-4.000}{2.000}{0.000}{+}}
\put(0,0){\VectorBosonArc{2.000}{0.000}{6.000}{0.000}{148.939}{}}
\put(0,0){\VectorBosonArc{2.000}{0.000}{6.000}{0.000}{-148.939}{+}}
\put(0,0){\Bullet{2.000}{0.000}}
\put(0,0){\Bullet{6.000}{0.000}}
\put(0.766,0.000){\makebox(0,0)[l]{\scriptsize$W$}}
\end{picture}
\hspace{9mm}}}

\newcommand{\threeglfourWB}{
\hspace{-6mm}
\raisebox{1mm}{
\setlength{\unitlength}{25pt}
\renewcommand{\Init}{0.330 0.330 Scale Init}
\begin{picture}(1.650,1.650)
\put(0,0){\VectorBoson{1.000}{4.000}{4.000}{2.000}{+}}
\put(0,0){\VectorBoson{1.000}{-4.000}{4.000}{-2.000}{}}
\put(0,0){\VectorBosonArc{4.000}{2.000}{8.000}{0.000}{109.898}{}}
\put(0,0){\VectorBosonArc{4.000}{2.000}{4.000}{-2.000}{148.939}{}}
\put(0,0){\VectorBosonArc{4.000}{2.000}{4.000}{-2.000}{-148.939}{}}
\put(0,0){\Bullet{4.000}{2.000}}
\put(0,0){\Bullet{4.000}{-2.000}}
\put(1.294,0.409){\makebox(0,0)[t]{\scriptsize$W$}}
\end{picture}
\hspace{9mm}}}

\newcommand{\threeglfourWC}{
\hspace{-6mm}
\raisebox{1mm}{
\setlength{\unitlength}{25pt}
\renewcommand{\Init}{0.330 0.330 Scale Init}
\begin{picture}(1.650,1.650)
\put(0,0){\VectorBoson{1.000}{4.000}{4.000}{2.000}{+}}
\put(0,0){\VectorBoson{1.000}{-4.000}{4.000}{-2.000}{}}
\put(0,0){\VectorBosonArc{4.000}{2.000}{4.000}{-2.000}{148.939}{}}
\put(0,0){\VectorBosonArc{4.000}{2.000}{4.000}{-2.000}{-148.939}{}}
\put(0,0){\VectorBosonArc{4.000}{-2.000}{8.000}{0.000}{-79.195}{}}
\put(1.307,-0.436){\makebox(0,0)[b]{\scriptsize$W$}}
\put(0,0){\Bullet{4.000}{2.000}}
\put(0,0){\Bullet{4.000}{-2.000}}
\end{picture}
\hspace{9mm}}}

\newcommand{\threeglfive}{
\hspace{-6mm}
\raisebox{1mm}{
\setlength{\unitlength}{25pt}
\renewcommand{\Init}{0.330 0.330 Scale Init}
\begin{picture}(1.650,1.650)
\put(0,0){\VectorBoson{1.000}{-4.000}{2.200}{-2.000}{}}
\put(0,0){\VectorBoson{1.000}{3.960}{2.200}{2.000}{+}}
\put(0,0){\VectorBoson{8.000}{-0.040}{5.680}{0.040}{+}}
\put(0,0){\VectorBoson{2.200}{2.000}{5.680}{0.040}{}}
\put(0,0){\VectorBoson{5.680}{0.040}{2.200}{-2.000}{}}
\put(0,0){\VectorBoson{2.200}{-2.000}{2.200}{2.000}{}}
\put(0,0){\Bullet{2.200}{2.000}}
\put(0,0){\Bullet{5.680}{0.040}}
\put(0,0){\Bullet{2.200}{-2.000}}
\end{picture}
\hspace{9mm}}}

\newcommand{\threeglsix}{
\hspace{-6mm}
\raisebox{1mm}{
\setlength{\unitlength}{25pt}
\renewcommand{\Init}{0.330 0.330 Scale Init}
\begin{picture}(1.650,1.650)
\put(0,0){\VectorBoson{3.000}{0.000}{1.000}{-4.000}{+}}
\put(0,0){\VectorBosonArc{3.000}{0.000}{1.000}{4.000}{18.296}{}}
\put(0,0){\VectorBosonArc{3.000}{0.000}{8.000}{0.000}{-84.027}{+}}
\put(0,0){\VectorBosonArc{3.000}{0.000}{4.000}{3.000}{50.503}{}}
\put(0,0){\VectorBosonArc{3.000}{0.000}{5.600}{1.320}{-49.914}{}}
\put(0,0){\VectorBosonArc{5.600}{1.320}{4.000}{3.000}{-203.725}{}}
\put(0,0){\Bullet{3.000}{0.000}}
\end{picture}
\hspace{9mm}}}

\newcommand{\myhat}{\setlength{\unitlength}{1.2cm}{
   \begin{picture}(1,.085)(-.48,0)
   \put(-.5,0){\line(6,1){.5}}
   \put(.5,0){\line(-6,1){.5}}
   \end{picture}}}
\newcommand{\myhatA}{\setlength{\unitlength}{1.8cm}{
   \begin{picture}(1,.057)(-.48,0)
   \put(-.5,0){\line(6,1){.5}}
   \put(.5,0){\line(-6,1){.5}}
   \end{picture}}}

\newcommand{\sn}{\!1\!-}

\begin{document}

\vskip-1cm
\hfill INLO-PUB-1/97
\vskip5mm
\begin{center}
{\LARGE{\bf{{Computation of the one-loop Symanzik}}}}\\[1ex]
{\LARGE{\bf{{coefficients for the square action}}}}\\
\vspace{1cm}
{\large Jeroen Snippe} \\
\vspace{1cm}
Instituut-Lorentz for Theoretical Physics,\\
University of Leiden, PO Box 9506,\\
NL-2300 RA Leiden, The Netherlands.\\ 
\end{center}
\vspace*{5mm}{\narrower\narrower{\noindent
\underline{Abstract}: We compute the one-loop coefficients for an alternative Symanzik improved pure gauge \SU{N} lattice action ($N=2$ and $N=3$). For the standard Symanzik improved action we confirm previous results by L\"{u}scher and Weisz.}}
\vspace{1.5cm}

\section{Introduction}
\label{sec-intro}

When studying pure lattice gauge theory it is important to realize that the lattice spacing $a$ is a rather elusive quantity. The action, when formulated in terms of the natural link variables $U_{\mu}(x)$, does not contain any $a$ dependence. The reason is that the continuum action is scale invariant. The introduction of a lattice spacing by parametrizing
\be
U_{\mu}(x) = \exp(a A_{\mu}(x))
\label{eq:Upert}
\ee
and insisting that $A_{\mu}$ is the (dimensionful) continuum vector potential is only natural in a classical context where one can limit oneself to smooth fields $A_{\mu}$. In a quantum mechanical context, however, defined by the path integral (with coupling constant $g_0$)
\be
Z = \int \cD U \exp(-S_{\rm Lat}[U]/g_0^2),
\label{eq:pathintsym}
\ee
typical field configurations are wildly fluctuating at the scale of one lattice spacing, and \refeq{Upert} is merely a convenient way to make contact with other regularizations of the continuum theory. The introduction of $A_{\mu}$ is not at all necessary for the extraction of physical quantities, for example by means of a Monte Carlo simulation.

Nevertheless, the notion of a lattice spacing is sensible in the quantum mechanical context too. The reason is that \refeq{pathintsym} dynamically produces a length scale, namely the correlation length $\xi$. More generally, it produces masses of particles, $m_i$. Of course due to scale invariance only dimensionless ratios appear, $\bar{m}_i\equiv a m_i$ ($\bar{m}_i$ being the mass in lattice units). This means that the continuum limit is defined by $\bar{m}_i\rightarrow0$. Moreover, the proximity to the continuum can be assessed by considering ratios $\bar{m}_j/\bar{m}_i$ as a function of $a$. For small $a$ (i.e.\ small $\bar{m}_i$) one thus expands
\be
\frac{\bar{m}_j}{\bar{m}_i} = \frac{m_j}{m_i} + {\rm Rest}_{ij}, \ \ \ {\rm Rest}_{ij} = c^{(1)}_{ij} a^2 + c^{(2)}_{ij} a^4 + \cdots,
\label{eq:gencontexp}
\ee
where the coefficients $c^{(n)}_{ij}$ are independent\footnote{For convenience we neglect logarithmic corrections for the moment.} of $a$. Note that in the presence of external physical mass scales, for example as set by a finite volume $L^3$, one can of course also consider ratios with $2\pi/L$.

For the conventional Wilson action~\cite{wilsonaction} all coefficients $c^{(n)}_{ij}$ generically are non-zero. Improvement means reducing the lattice artefacts (`Rest') for all values of $i$ and $j$ that are associated with physical masses. By Wilson's renormalization group argument~\cite{wilson} it should be possible to find a lattice action that produces ${\rm Rest} = 0$. In a certain approximation, the implementation of this approach has been put on a practical level by Hasenfratz and Niedermayer et\ al.~\cite{hasenfratz}. In this paper we follow the less ambitious perturbative approach due to Symanzik~\cite{symanzik}.

Symanzik improvement amounts to expanding `Rest' with respect to $a$, and systematically removing the $a^2,a^4,\cdots$ terms. Usually only the removal of the $a^2$ term is pursued, and we will do so in this paper. From now on we mean by Symanzik improvement this limited reduction only. For pure gauge theories, Symanzik improvement is believed to be attainable~\cite{weiszI,luscherLW2} by adding to the standard Wilson action a finite number of extra Wilson loops, for example\footnote{Taking the liberty to deviate from the notation in ref.~\cite{luscherLW2} we assign $c_4$ to the $2\times2$ Wilson loop.}:
\bea
S(\{c_i(g_0^2)\})\!\!\!\!&=&\!\!\!\!\!\sum_x\!\Tr\!\left\{\rule{0mm}{5mm}\!c_0(g_0^2)\!\!\sum_{\mu\not=\nu}\!\left\langle\sn\,\plaq\,\right\rangle 
\!+\!2c_1(g_0^2)\!\!\sum_{\mu\not=\nu}\!\left\langle\sn\,\twooneplaq\, \right\rangle\!+\!\medbreuk{4}{3}c_2(g_0^2)\!\!\!\!\!\sum_{\mu\not=\nu\not=\lambda}\!\!\!\left\langle\sn\cornplaq \,\right\rangle\right.\nonumber\\
&&\!\!\!\!\!\left.+4c_3(g_0^2)\!\!\!\!\sum_{\mu\not=\nu\not=\lambda}\!\!\left\langle\sn\!\!\hookplaq
\,\right\rangle+c_4(g_0^2)\!\sum_{\mu\not=\nu}\left\langle\sn\twoplaq\, \right\rangle\right\}.\label{eq:squareaction}
\eea

Due to asymptotic freedom~\cite{politzer}, the continuum limit $a\rightarrow0$ is achieved by approaching the critical point, $g_0\rightarrow0$. Therefore it is sensible to calculate $c_i(g_0^2)$ perturbatively as a function of $g_0^2$. To this end one expands
\be
c_i(g_0^2) = \sum_{m=0}^{\infty} c_i^{(m)} (g_0^2)^m.
\label{eq:expandcoeff}
\ee
Symanzik improvement to $n$ loops amounts to fixing $c_i^{(0)}\cdots c_i^{(n)}$ such that physical quantities show only $\cO(a^4,a^2 g_0^{2(n+1)})$ deviations from their continuum values. At tree level ($n=0$) the analysis is rather easy, and many alternative Symanzik-improved actions have been introduced. 
However, for a long time the availability of a one-loop Symanzik-improved action was limited to the \LW~action~\cite{luscherLW1}. In ref.~\cite{squarelett} we have presented the one-loop coefficients for a second action, namely the square action that was introduced in ref.~\cite{oldnew}. The purpose of the present paper is to give the details of our computation. 

In present-day Monte Carlo simulations $g_0^2$ is typically of order $1$, and this value seems to be too large for Symanzik improvement to be effective. For this reason a phenomenological extension of Symanzik improvement, called tadpole improvement, was introduced~\cite{lepage}. In this approach a mean-field parameter~\cite{parisi} $u_0\approx\langle U_{\mu}(x)\rangle$ is included in a tree-level or one-loop Symanzik-improved action by dividing each link variable by $u_0$. This has been argued~\cite{lepage,alford} to capture most of the higher-order corrections in $g_0$, thus approximating an all-loop Symanzik-improved action. In practice, Monte Carlo data for tadpole-improved actions often, but not always, show dramatic improvement. As a side remark, let us mention that the goal of an {\em exact}\, all-loop Symanzik action might be attainable to sufficient precision by using Monte Carlo techniques. In ref.~\cite{luscherQCD} this has been studied for the fermionic sector of QCD, where the leading lattice artefacts to be cancelled are of the order $a$ instead of $a^2$.

We believe that our introduction of an alternative one-loop Symanzik-improved action is useful for several reasons. It allows one to make the following analyses of improvement:
\begin{itemize}
\item Estimate remaining lattice artefacts by comparing the effectiveness of two different Symanzik-improved actions.
\item Study the universality and consistency of tadpole improvement.
\end{itemize}
(At tree-level such analyses were performed in, e.g., ref.~\cite{alford}).
Moreover, a one-loop calculation has some spin-offs:
\begin{itemize}
\item The Lambda parameter ratio $\Lambda_{\mbox{\scriptsize improved}}/\Lambda_{\mbox{\scriptsize Wilson}}$, which is needed for doing simulations with both actions at the same physical scale.
\item The tadpole parameter $u_0$ to one-loop order. \end{itemize}

The outline of this paper is as follows. In section~\ref{sec-onshell} we discuss {\em on-shell}\, improvement~\cite{luscherLW2}, which implies that only two improvement conditions need to be satisfied in order to cancel the $\cO(a^2)$ errors in all spectral quantities. In section~\ref{sec-lattpert} we set up lattice perturbation theory. All propagators and vertices needed are listed in the appendices~\ref{app-struct}--\ref{app-vert}. In section~\ref{sec-statquark} we calculate the static quark potential to one-loop order. This allows us to extract the Lambda and tadpole parameters, and also one improvement condition. In section~\ref{sec-twisted} we discuss perturbation theory in a twisted finite volume, which subsequently is used to obtain two improvement conditions and, again, the value of the Lambda parameter. As an extra check on the universality of on-shell improvement we compute one additional spectral quantity. We conclude with  section~\ref{sec-summary}, where the main results are listed in a compact way. For the sake of readability we start in section~\ref{sec-backgroundtwist} with discussing the physical setting of the twisted finite volume computation. We will do so from a somewhat more conventional perspective than was given by L\"{u}scher and Weisz in refs.~\cite{luscherLW1,luscherLW3}.

Our calculation follows closely the well-documented work of L\"{u}scher, Weisz and Wohlert \cite{weiszI, luscherLW2, luscherLW1, luscherLW3, weiszII}. Therefore we only go into full detail at points where we take a different approach, and at some points where the earlier treatments can be complemented. Special attention will be paid to the many checks we have performed to convince ourselves of the correctness of the final results.

\vspace{-1mm}
\section{Twisted boundary conditions and electric flux}
\label{sec-backgroundtwist}

In this section we review twisted boundary conditions in \SU{N}\ pure gauge theories that were introduced by 't Hooft~\cite{thooftTW1}, and apply the analysis to the setting of section~\ref{sec-twisted}. For convenience we restrict ourselves to the continuum formulation. However, twist is also well defined on the lattice~\cite{gro15}, and the generalization of the results below to the lattice formulation~\cite{luscherLW3} used in section~\ref{sec-twisted} is straightforward.

Twisted boundary conditions are defined as follows (no summation convention):
\be
A_{\mu}(x+L_{\nu}\hat{\nu})=A^{\Omega_{\nu}}_{\mu}(x),
\label{eq:deftwistmatr}
\ee
where $\hat{\nu}$ is the unit vector in the positive $x_{\nu}$ direction. $L_{\nu}$ denotes the size of the (rectangular) space-time in the $\nu$ direction, and $A^{\Omega_{\nu}}_{\mu}$ is a gauge transform of $A_{\mu}$:
\be
A_{\mu}^{\Omega}(x)\equiv\Omega(x)(A_{\mu}(x)+\partial_{\mu})\Omega^{-1}(x),\ \ \ \Omega(x)\in\SU{N}.
\label{eq:defgaugetransf}
\ee
A structural requirement is: 
\be
\Omega_{\mu}(x+L_{\nu}\hat{\nu})\Omega_{\nu}(x) = \Omega_{\nu}(x+L_{\mu}\hat{\mu})\Omega_{\mu}(x) z^{n_{\mu\nu}},\ \ \ (z\equiv e^{2\pi i/N}),
\label{eq:cocycle}
\ee
for each $\mu,\nu$ plane. This is a consistency relation, necessary because there are two ways to relate $x$ to $x+L_{\mu}\hat{\mu}+L_{\nu}\hat{\nu}$. $n_{\mu\nu}=-n_{\nu\mu}$ are integers, only defined modulo $N$ (i.e.\ $n_{\mu\nu}\in\integer_N$). This guarantees that $Z_{\mu\nu}\equiv\exp(2\pi i n_{\mu\nu}/N)$ is a center element of \SU{N}. In {\em pure} gauge theories it is not required that $Z_{\mu\nu}=1$, because $A_{\mu}(x)$ is invariant under gauge transformations with a center element. If $n_{0i}=0\,(\mod N)$, it is possible to render the twist matrices $\Omega_i$ $x$-independent by a gauge transformation. Below we will assume that this has been done.

How pure gauge theory is quantized in the presence of twist is described in refs.~\cite{thooftTW1,baalTHESIS}. The outcome (neglecting instanton effects~\cite{tho2}) is that physical Hilbert space  associated with a twist $n_{ij}$ splits up in $N^3$ sectors, which are eigenspaces of gauge transformations compatible with the boundary conditions. The Hamiltonian can be diagonalized separately in each of these sectors (which are usually labeled by a vector $\myvecD{e}\in\integer_N^3$). In this context an important operator is the Poyakov line
\be
P_j(\myvecC{x}^{(j)})\equiv \Tr\left[\Pexp\left(\int_0^{L_j}d x_j A_j(\myvecC{x})\right)\Omega_j\right]
\label{eq:defPoltwist}
\ee
($\myvecC{x}^{(1)}\equiv(x_2,x_3)$, and cyclic). Under a gauge transformation compatible with the boundary conditions it picks up a phase factor ($\in\integer_N$), and from this it can be shown that it maps the $\myvecD{e}^{\rm th}$ electric sector into the $(\myvecD{e}+\hat{j})^{\rm th}$ sector.

Twisted boundary conditions thus allow for $N^6$ quantum sectors: $N^3$ due to (physically) periodic boundary conditions in space, labeled by $n_{ij}$, and within each of these: $N^3$ sectors labeled by $\myvecD{e}$. These sectors have a beautiful interpretation, due to 't Hooft~\cite{thooftTW1}: $\myvecD{e}$ corresponds to electric flux, $\myvecD{m}$ to magnetic flux (with $n_{ij}=-\sum_k\varepsilon_{ijk} m_k$). For shortness we will not repeat 't Hooft's arguments (see also ref.~\cite{baalTHESIS}) why this is a natural identification. 

We would now like to apply the above results to the specific geometry used in section~\ref{sec-twisted}. To this end we restrict ourselves to a magnetic flux $\myvecD{m}=(0,0,-1)$. We also choose $L_1=L_2\equiv L$ and send $L_3\rightarrow\infty$, thus obtaining a `twisted tube' (this terminology is due to L\"{u}scher and Weisz). It can be proven that the vacuum valley of the twisted tube consists of only one point, $A_{\mu}=0$ (up to a gauge transformation), which is why it admits a straightforward perturbative analysis, contrary to the purely periodic case $\myvecD{m}=\myvecD{0}$~\cite{luscherTORUS}.

What we want to do in the remainder of this section, is to classify the lowest-lying excitations above the quantum vacuum $|0\rangle$. To this end we use a path integral approach. We can get a handle on the spectrum by considering the {\em gauge invariant} correlation function of two Polyakov lines at a distance $t$ in (Euclidean) time,
\be
\langle P^{\dagger}_1(x_2,x_3;t)P_1(x_2,x_3;0)\rangle.
\label{eq:corPol}
\ee
Since $|0\rangle$ is a state with zero electric flux, this correlation function is associated with the creation of a (gauge invariant, i.e.\ physical) state at time~$0$, with electric flux $\myvecD{e}=(1,0,0)$, and its annihilation at time~$t$. Therefore for large values of $t$ it will be dominated by the exponential behavior $\exp(-M t)$, where $M$ is the energy of the lowest-lying Hamiltonian eigenstate that has non-zero overlap with $\hat{P}_1(x_2,x_3)|0\rangle$.

In order to compute \refnn{corPol} perturbatively, the allowed fields in the path integral must be parametrized. To this end one defines the $\myvec{x}$-independent matrices
\be
\Gamma[\myvecD{n}] = \Omega_1^{-n_2} \Omega_2^{n_1} z^{\half(n_1+n_2)(n_1+n_2-1)},\ \ \ \myvecD{n}\in\integer_N^3.
\label{eq:defGammak}
\ee
The choice of the phase factor is a matter of convenience. The important property of $\Gamma[\myvecD{n}]$ is that it satisfies $\Omega_{\nu}\Gamma[\myvecD{n}]\Omega_{\nu}^{-1}=z^{n_{\nu}}$ ($\nu=1,2$). From the second paper in ref.~\cite{gro15} we know that
\be
\frac{1}{N}\Tr\left(\Gamma[\myvecD{n}]\Gamma^{\dagger}[\myvecD{n}']\right) = \delta_{\myvecD{n},\myvecD{n}' (\mod N)}.
\ee
This allows us to Fourier transform $A_{\mu}(x)$ in the following way:
\be
A_{\mu}(x) = \frac{g_0}{(2\pi)^2 L^2 N} \sum_{k_1,k_2} \int_{\real^2}d k_0\, dk_3\,\,e^{i k x} \tilde{A}_{\mu}(k) \Gamma_{k}.
\label{eq:Ftrconttwist}
\ee
Here $k x\equiv\sum_{\mu=0}^3 k_{\mu}x_{\mu}$ and $\sum_{k_1,k_2}$ runs over
\be
k_{\nu}=m n_{\nu},\ \ \ m\equiv\frac{2\pi}{NL},\ \ \ n_{\nu}\in\integer,\ \ \ (\nu=1,2).
\label{eq:momquant}
\ee
Also we adopted the notation of ref.~\cite{luscherLW3}:
\be
\Gamma_{k}\equiv\Gamma[(n_1,n_2,0)],\ \ \ (k_{\nu} = m n_{\nu},\ \ \ \nu=1,2).
\label{eq:Gamntok}
\ee
In contrast to the purely periodic case, $\tilde{A}_{\mu}(k)\in\complex$ carries no \SU{N}\ index. Instead, the momenta in the twisted directions are quantized in units of $m$ rather than $2\pi/L=N m $. The total number of degrees of freedom is preserved, because $\Tr A_{\mu}(x)=0$ implies
\be
\tilde{A}_{\mu}(k)=0\ \ \ {\rm for\ } n_{1,2}=0\,(\mod N).
\label{eq:zeromomabs}
\ee

Using this formalism one quickly derives (with $t=x_0$)
\bea
P_1(x_2,x_3;t) &=& \frac{g_0}{(2\pi)^2 L} \sum_{n\in\integer}\int_{\real^2}d k_0\, dk_3\,\, e^{ikx} \tilde{A}_1(k)\,\, + \cO(A_1^2),
\label{eq:expPol}\\
&&k=(k_0,0,(2\pi/L)n + m, k_3).
\label{eq:momPol}
\eea
Note that $k_2$ is now quantized in units of $N m$, but carries an additional momentum $m$. Physically, this is the Poynting vector $\myvecD{e}\times\myvecD{m}$ due to the electric flux $\myvecD{e}=(1,0,0)$ created by the Polyakov line~\cite{poynting}.

For the moment we neglect the $\cO(A_1^2)$ corrections in \refeq{expPol}. We then obtain
\be
\langle P^{\dagger}_1(x_2,x_3;t)P_1(x_2,x_3;0)\rangle = \frac{g_0^2 N}{8\pi^2} \sum_{n\in\integer}\int_{\real^2}d k_0\, dk_3\,\, e^{ik_0 t}\prop_{11}^{\rm full}(k),
\label{eq:corPolexp}
\ee
where $\prop_{11}^{\rm full}(k)$ is the dressed propagator, defined through the 2-point function:
\be
\langle\tilde{A}^{*}_{\mu}(k)\tilde{A}_{\nu}(k')\rangle \!=\! \medbreuk{1}{2}((2\pi)^2L^2 N)\delta_{n'_1,n_1}\delta_{n'_2,n_2}\delta(k'_0-k_0)\delta(k'_3-k_3) \chi_k \prop_{\mu\nu}^{\rm full}(k).
\label{eq:defDfullintro}
\ee
The factor $\chi_k\equiv(1-\delta_{n_1,0(\mod N)}\delta_{n_2,0(\mod N)})$ is implied by \refeq{zeromomabs}: modes with zero momentum (mod$(Nm)$) in the twisted directions are absent on the twisted tube. In \refeq{defDfullintro}, $k_{\nu}=m n_{\nu}$ and $k'_{\nu}=m n'_{\nu}$ ($\nu=1,2$). In \refeq{corPolexp}, the momentum $k$ is restricted to \refeq{momPol}.

Now it is well known that 
\be
\lim_{t\rightarrow\infty}\int_{\real}d k_0 e^{ik_0 t}\prop_{11}^{\rm full}(k)\sim e^{-E(\myvecD{k})t},
\ee 
where $k=(iE(\myvecD{k});\myvecD{k})$ is the pole of $\prop_{11}^{\rm full}(k)$. To lowest order in perturbation theory this gives the usual formula $E(\myvecD{k})=|\myvecD{k}|\equiv\sqrt{k_1^2+k_2^2+k_3^2}$. Here it is essential that the Polyakov line, due to its gauge invariance, couples only to physical (i.e.\ transversal) polarizations. Indeed, from \refeq{expPol} we see that $P_1(x_2,x_3)$ couples to the polarization $\varepsilon_{\mu}=\delta_{\mu,1}$, which is transversal due to \refeq{momPol}.

Inserting this result in \refeq{corPolexp} we see that the leading exponential decay comes from $\myvecD{k}=(0,m,0)$. Thus the lightest particle with electric flux $(1,0,0)$ has a mass $m$ ($+\cO(g_0^2)$). This is what L\"{u}scher and Weisz called an A~meson. From \refeqs{momquant}{zeromomabs} we see that it is the lightest particle on the twisted tube. 

There are more particles with the same mass. To lowest order in perturbation theory these are associated with the other poles of the bare propagator for $|\myvecD{k}|=m$, i.e.\ at $\myvecD{k}=(m,0,0)$; $(0,-m,0)$; $(-m,0,0)$. Of course they are created by Polyakov lines with electric fluxes $\myvecD{e}=(0,-1,0)$; $(-1,0,0)$; $(0,1,0)$. It can be proven that the total of four particles now found have equal masses to all orders in $g_0$. This is simply a consequence of symmetry considerations. Namely, the Hamiltonian has two CP symmetries, one corresponding to (1) $x_1\rightarrow-x_1$, and the other corresponding to (2) $x_1\leftrightarrow x_2$. We will not prove this here (for the proof see ref.~\cite{luscherLW3}). Let us only mention that (for $N\geq3$) the charge conjugation C (defined by $A_{\mu}(x) \rightarrow W A_{\mu}^{*}(x) W^{-1}$, where $W\in\SU{N}$ can be freely chosen) is not a symmetry, because it violates the boundary conditions (it changes $\myvecD{m}$ into $-\myvecD{m}$). The behavior of the electric fluxes under the two CP~transformations (referred to as ${\rm CP}^{(1,2)}$) is given in table~\ref{tab:CP}.  
\begin{table}
\begin{center}
\begin{tabular}{|c|c|c|}
\hline
$(e_1,e_2)$     & ${\rm CP}^{(1)}(e_1,e_2)$ & ${\rm CP}^{(2)}(e_1,e_2)$\\
\hline
$(1,0)$     & $(1,0)$ & $(0,-1)$\\
\hline
$(0,1)$     & $(0,-1)$ & $(-1,0)$\\
\hline
$(-1,0)$     & $(-1,0)$ & $(0,1)$\\
\hline
$(0,-1)$     & $(0,1)$ & $(1,0)$\\
\hline
\end{tabular}
\caption{Properties of the electric fluxes created by the Polyakov lines $P_1$, $P_2$, $P_1^{\dagger}$ and $P_2^{\dagger}$, under two CP~transformations; ${\rm P}^{(1)}(x_1,x_2)=(-x_1,x_2)$ and ${\rm P}^{(2)}(x_1,x_2)=(x_2,x_1)$.}
\label{tab:CP}
\end{center}
\end{table}

One may wonder about the $\cO(A_1^2)$ corrections in \refeq{momPol}. In principle they bring about $\cO(g_0^2)$ corrections to the above analysis, because they give rise to 3-point (or in general $n$-point, $n\geq3$) functions in the expansion of~\refnn{corPol}. The leading Feynman diagram of this type is given by (in coordinate space)
\be
\setlength{\unitlength}{25pt}
\renewcommand{\Init}{0.350 0.350 Scale Init}
\raisebox{0cm}{
\begin{picture}(2.800,2.4)(1.2,-1.05)
\put(0,0){\Fermion{3.000}{2.000}{3.000}{-2.000}{}}
\put(0,0){\Fermion{13.000}{-2.000}{13.000}{2.000}{}}
\put(4.550,0.000){\vector(0,1){0}}
\put(1.050,0.000){\vector(0,-1){0}}
\put(0,0){\Bullet{5.640}{-0.080}}
\put(0,0){\Bullet{5.640}{-0.080}}
\put(0,0){\VectorBoson{5.640}{-0.080}{13.000}{0.760}{}}
\put(0,0){\VectorBoson{5.640}{-0.080}{3.000}{-1.440}{}}
\put(0,0){\VectorBoson{5.640}{-0.080}{3.000}{0.880}{}}
\put(0,0){\Fermion{3.000}{3.000}{5.640}{3.000}{}}
\put(0,0){\Fermion{3.000}{-3.000}{13.000}{-3.000}{}}
\put(1.050,-1.050){\vector(-1,0){0}}
\put(4.550,-1.050){\vector(1,0){0}}
\put(1.050,1.050){\vector(-1,0){0}}
\put(1.974,1.050){\vector(1,0){0}}
\put(1.512,1.148){\makebox(0,0)[b]{$t_1$}}
\put(2.786,-.950){\makebox(0,0)[b]{$t$}}
\put(5.550,-1.050){.}
\end{picture}}
\ee
For $t\rightarrow\infty$, we must have $t_1/t\rightarrow0$. The reason is that from the above analysis we know that each propagator of length $t'$ suppresses the correlation function by a factor of approximately $\exp(-m t')$. Thus only diagrams in which a single (dressed) propagator runs from the one Polyakov line to the other contribute to the leading behavior for $t\rightarrow\infty$: $\exp(-(m+\cO(g_0^2))t)$. Diagrams like the one above do not contribute to the $\cO(g_0^2)$ corrections to the mass $m$. Instead, their contributions should be associated with the renormalization of the composite operator $P_1(x_2,x_3)$.

There is a further set of four particles which at tree-level are degenerate with the four particles found above. These also carry electric flux $|\myvec{e}|=1$, but they have a different polarization, $\varepsilon_{\mu}=\delta_{\mu,3}$, at momenta $\myvecD{k}=(0,m,0)$; $(m,0,0)$; $(0,-m,0)$; $(-m,0,0)$. We will refer to this set as ${\rm A}^{-}$~mesons, because under appropriate CP transformations they have eigenvalues opposite to their counterparts found above. Those we will call ${\rm A}^{+}$~mesons. Due to the ${\rm CP}^{(1,2)}$~symmetries, all ${\rm A}^{-}$~mesons are exactly degenerate. However, there is no symmetry relating the ${\rm A}^{-}$~mesons to the ${\rm A}^{+}$~mesons, so beyond tree-level they must be expected to have different masses. The ${\rm A}^{-}$~mesons can be created by gauge-invariant operators, but these are of a more complicated nature than the Polyakov lines creating the ${\rm A}^{+}$~mesons.

No more particles with masses $m+\cO(g_0^2)$ exist, because pure gauge theory admits only two physical polarizations. From the bare propagator one sees that the next particles\footnote{It is natural to consider particles with momenta $(0,m,k_3)$, $k_3\not=0$, as A~mesons in motion.} encountered in the spectrum have twisted momenta $|k_1|=|k_2|=m$, and thus their energies equal $\sqrt{2}m+\cO(g_0^2)$. These were called B~mesons by L\"{u}scher and Weisz. In our language they carry electric flux $|e_1|=|e_2|=1$. 
Still higher up are particles with masses $2m+\cO(g_0^2)$ or more. For $g_0\not=0$ (or rather for the renormalized coupling $g_R(L)\not=0$) many of these particles should be expected to be unstable, because the kinematics allow them to decay into A~or B~mesons (only the particles with tree-level masses $2m$ might be safe, depending on the $\cO(g_0^2)$ corrections).

\section{On-shell improvement}
\label{sec-onshell}

It is important to specify precisely what quantities are to be improved. In Symanzik's work~\cite{symanzik} the aim was to improve correlation functions. However, correlation functions depend on the elementary field operators. As a result it is difficult to find a formulation in terms of which this kind of improvement can be implemented. For the \nlsig, Symanzik succeeded by performing a well-chosen field redefinition, but for gauge theories the problem becomes more intractable and to our knowledge no one has solved it up to date.

One can avoid this complication by restricting oneself to `on-shell improvement' as introduced by L\"{u}scher and Weisz ~\cite{luscherLW2}. Instead of improving correlation functions one now pursues only the improvement of spectral quantities. Unlike correlation functions these are necessarily invariant under field redefinitions. Examples of spectral quantities are masses of stable particles and the static quark potential.

In order to determine a minimal set of improvement conditions it is sufficient~\cite{luscherLW2} to perform a small-$a$ expansion of the classical action. The most general action we consider is given by~\refeq{squareaction}, the expansion of which reads (see e.g.\ refs.~\cite{weiszII,overimpr}):
\bea
S(\{c_i(g_0^2)\})
&=&-\frac{a^4}{2}(c_0+8c_1+8c_2+16c_3+16c_4)(g_0^2)\sum_{x,\mu,\nu}
\Tr\!\left(\rule{0mm}{4mm}F_{\mu\nu}(x)\right)^{\!2}\nonumber\\
&&+\frac{a^6}{12}(c_0+20c_1-4c_2+4c_3+64c_4)(g_0^2)\sum_{x,\mu,\nu}
\Tr\!\left(\rule{0mm}{4mm} D_{\mu}F_{\mu\nu}(x)\right)^{\!2}\nonumber\\
&&+\frac{a^6}{3}(c_2+3c_3)(g_0^2)\sum_{x,\mu,
\nu,\lambda}\Tr\!\left(\rule{0mm}{4mm} D_{\mu}F_{\mu\lambda}(x) D_{\nu}F_{\nu\lambda}(x)\right)\nonumber\\
&&+\frac{a^6}{3}c_2(g_0^2)\sum_{x,\mu,\nu,\lambda}\Tr\!\left(\rule{0mm}{4mm} D_\mu F_{\nu\lambda}(x)\right)^{\!2}+\cO(a^8).
\label{eq:expandac}
\eea
In this equation, $F_{\mu\nu} = \partial_{\mu}\bar{A}_{\nu} - \partial_{\nu}\bar{A}_{\mu} + [\bar{A}_{\mu},\bar{A}_{\nu}]$ and $D_{\mu} = \partial_{\mu} + {\rm ad} \bar{A}_{\mu}$, where $\bar{A}_{\mu}$ is related to $U_{\mu}$ by a path ordered exponentiation (see \refeq{Upath}). A crucial observation is that $c_4(g_0^2)$ appears only in the combinations
\be
\tilde{c}_0(g_0^2) \equiv c_0(g_0^2) - 16 c_4(g_0^2), \ \ \  \tilde{c}_1(g_0^2) \equiv c_1(g_0^2) + 4 c_4(g_0^2).
\label{eq:tildecoeff}
\ee
This implies that with the substitutions $c_0(g_0^2)\rightarrow \tilde{c}_0(g_0^2),\ c_1(g_0^2) \rightarrow \tilde{c}_1(g_0^2)$, the entire analysis of ref.~\cite{luscherLW2} can be copied literally. The only subtlety is that positivity of the lattice action must be rechecked for $c_4(g_0^2)\not=0$. Postponing this proof to appendix~\ref{app-positive}, we take over the following conclusions from ref.~\cite{luscherLW2}:
\begin{itemize}
\item The normalization condition
\be
\tilde{c}_0(g_0^2)+8\tilde{c}_1(g_0^2)+8c_2(g_0^2)+16c_3(g_0^2) = 1
\label{eq:contcond}
\ee
can be imposed. This ensures that \refeq{expandac} has a natural continuum limit and, more importantly, determines $\tilde{c}_0(g_0^2)$ as a function of the other coefficients. Since in the path integral the action is weighted by a factor of $1/g_0^2$, \refeq{contcond} is always attainable by a redefinition of the bare coupling constant $g_0$.
\item After imposing the normalization condition, on-shell improvement is expected to be satisfied for all spectral quantities once the coefficients $\tilde{c}_1(g_0^2)$, $c_2(g_0^2)$ and $c_3(g_0^2)$ are tuned correctly. Furthermore, $c_3(g_0^2)$ can be chosen 0 because of the freedom of field redefinitions. Hence in fact only $\tilde{c}_1(g_0^2)$ and $c_2(g_0^2)$ are to be determined, i.e.\ we need only two improvement conditions.
\item The tree-level conditions for on-shell improvement are
\bea
\tilde{c}_1 &=& -\frac{1}{12},\nonumber\\
c_2 &=& 0.
\label{eq:onshelltree}
\eea
\end{itemize}

Moreover, the fact that $c_0(g_0^2)$, $c_1(g_0^2)$ and $c_4(g_0^2)$ enter the analysis only through the combinations $\tilde{c}_0(g_0^2)$ and $\tilde{c}_1(g_0^2)$, implies that one of them, e.g.\ $c_4(g_0^2)$, is a free function. Having noted this, one should yet keep in mind that the one-loop coefficients\footnote{We use the notation $c_i = c_i^{(0)}\!,\ c'_i = c_i^{(1)}$.} $\tilde{c}'_0$, $\tilde{c}'_1$, and $c'_2$ do depend on the tree-level choice of $c_4$. In particular, the \LW\ and square\nolinebreak\ actions will have different one-loop coefficients. Nonetheless $c'_4$ is a free parameter for both actions.

In the remainder of this paper we limit ourselves to actions satisfying \refeq{contcond} and $c_3(g_0^2)=0$, though some intermediate results are valid more generally. In particular we focus on the Wilson action~\cite{wilsonaction} (abbreviation W; $c_0=1$, $c_1=c_2=c_3=c_4=0$) as well as the \LW~\cite{luscherLW2} (LW; $c_0=\breuk{5}{3}$, $c_1=-\twelfth$, $c_2=c_3=c_4=0$) and square~\cite{oldnew} (sq; $c_0=\breuk{16}{9}$, $c_1=-\breuk{1}{9}$, $c_2=c_3=0$, $c_4=\breuk{1}{144}$) Symanzik actions. For the last two actions we determine the coefficients $\tilde{c}'_1$ and $c'_2$ in sections~\ref{sec-statquark} and~\ref{sec-twisted}.

\section{Lattice perturbation theory}
\label{sec-lattpert}

This will be a very short description of lattice perturbation theory. For a thorough setup of the general formalism see refs.~\cite{rothe,luscherLW3}.

The elementary field in pure lattice gauge theories is the link variable $U_{\mu}(x)$. In a perturbative approach it is convenient to use the parametrization \refnn{Upert}. Since $U_{\mu}(x)\in\SU{N}$, $A_{\mu}(x)$ lies in the Lie algebra of \SU{N}:
\be
A_{\mu}(x) = g_0 \sum_a A_{\mu}^a(x)T^a,
\label{eq:defineAa}
\ee
where $A_{\mu}^a(x)\in\real$. We limit ourselves to the fundamental representation, with conventions
\be
\Tr\left(T^a T^b\right) = -\half \delta_{a b},\ \ \ \left[T^a,T^b\right] = f_{a b c} T^c.
\label{eq:Lieconv}
\ee
Note that \refeq{Upert} contains an ordinary exponential, not a path ordered exponential. Therefore \refeq{expandac} receives corrections. In appendix~\ref{app-vert} this provides us with a non-trivial check on the vertices.

Using \refeq{Upert} it is straightforward to expand the action (\ref{eq:squareaction}) to a given order in $g_0$. The only complication is the many terms involved, but this can be handled by a symbolic computer language such as FORM~\cite{vermaseren} or Mathematica~\cite{wolfram}.

Another issue is gauge fixing. We have adopted the gauge choice of ref.~\cite{weiszII}, i.e.\ covariant gauge fixing. The advantage of this choice over the Coulomb-like choice in ref.~\cite{luscherLW3} is that the gauge condition,
\be
\sum_{\mu}\partial^L_{\mu}A_{\mu}(x) \stackrel{\mbox{\scriptsize covariant g.f.}}{=} 0,
\label{eq:covgauge}
\ee
is independent of the lattice action chosen, thus eliminating a potential source of mistakes when changing from the LW action to the square action. Here $\partial^L$ is the left lattice derivative, $\partial^L_{\mu}f(x)\equiv (f(x)-f(x-a\hat{\mu}))/a$.
We implement \refeq{covgauge} by adding to the action
\be
\action{gf} = -\frac{1}{\alpha} a^4\sum_{x}\Tr\left(\sum_{\mu} \partial^L_{\mu}A_{\mu}(x)\right)^2,
\label{eq:Sgf}
\ee
where $\alpha$ is the well-known gauge parameter that should drop out of physical quantities.

Of course the gauge fixing procedure induces ghosts in the usual way. Hence a ghost term \action{ghost} must be added to the action. One more term to be added is the Haar measure term \action{measure} associated with the parametrization \refnn{Upert}, \refnn{defineAa}. The total action thus reads
\be
\frac{1}{g_0^2}\action{total} = \frac{1}{g_0^2}(S(\{c_i(g_0^2)\}) + \action{gf}) + \action{measure} + \action{ghost}. 
\label{eq:Stotal}
\ee
For the calculations in the following sections we need the expansion of $\action{total}/g_0^2$ up to third order in $g_0$. It is given in the appendices~\ref{app-struct}--\ref{app-vert}. Using this expansion, and the Euclidean path integral
\be
Z = \int\exp\left(-\frac{1}{g_0^2} \action{total}\right),
\label{eq:pathint}
\ee
perturbation theory is set up like in the continuum, with the exception that coordinates are now discrete or, equivalently, momenta run over one Brillouin zone only. Correspondingly, momentum-conserving delta functions $\delta(k_1+\cdots+k_n)$ are $2\pi/a$ periodic. In the remaining sections we use lattice units. The $a$~dependence can be reobtained from a dimensional analysis.
 
\section{Static quark potential}
\label{sec-statquark}
\subsection{Generalities}
\label{subsec-statquark-gen}

This section is devoted to the calculation of the static quark potential $V(L)$ to one-loop order. Its definition reads
\be
V(L) = \lim_{T\rightarrow\infty} -\frac{1}{T}\ln\left( \frac{1}{N}\left\langle \RE\Tr U(L\times T)\right\rangle\right).
\label{eq:statquarkdef}
\ee
Here $U(L\times T)$ denotes a rectangular Wilson loop with extension $T$ in the time direction ($\hat{0}$) and $L$ in some spatial direction, say $\hat{1}$. The expectation value $\langle\cdots\rangle$ denotes averaging with respect to the path integral in infinite volume. Since $V(L)$ captures the exponential decay of a gauge invariant correlation function, it can in principle be expressed as an ($L$~dependent) function of the spectrum. Therefore its small-$a$ behavior, or in lattice units large-$L$ behavior, is suitable for a (partial) determination of the on-shell improvement coefficients.

To avoid confusion, note that we will not pursue the computation of $V(L)$ in the confining regime, i.e.\ at large {\em physical} distances $L\gg\xi$ (where $\xi$ is the correlation length). That would not be feasible within the framework of ordinary perturbation theory. However, it is the basic postulate of Symanzik improvement that all physical quantities can be improved simultaneously, so that considering short-distance quantities only is sufficient for a complete determination of the improvement coefficients.

The potential $V(L)$ not only depends on $L$, but also on the coupling $g_0$. Roughly following the notation of refs.~\cite{weiszII,weiszI}, but limiting ourselves to the fundamental representation of the Wilson loop, we expand
\be
-\ln\left( \frac{1}{N}\left\langle \RE\Tr U(L\times T)\right\rangle\right) = \sum_{n=1}^{\infty} \frac{g_0^{2n}}{(2n)!} w_n(L, T).
\label {eq:defwn}
\ee
We restrict ourselves to the tree-level and one-loop contributions ($n=1,2$). In terms of Feynman diagrams they read
\bea
w_1\!\!\! &=&\!\!\! -\ \ \stattree,
\label{eq:w1Feyn}\\
&&\nonumber\\
w_2\!\!\! &=&\!\!\!\! \left(\! 3w_1^2 -\ \ \statonefour\right) - 4\ \ \statonethree 
- 12\ \ \statonevacpol.\label{eq:w2Feyn}\\
&&\nonumber
\eea
The endpoints of the gluon lines (\gluonline) represent source terms\footnote{These are proportional to $g_0$, cf.\ \protect\refeqs{Upert}{defineAa}.} due to the Wilson loop, and hence are to be summed over all lattice sites on the loop. On the other hand, vertices (\vertex) should be summed over all sites in space-time. The striped graph represents the one-loop vacuum polarization:
\vspace{-5mm}
\bea
\vacpoltot &=& \vacpolins +\, \vacpolmeas + \vacpolghfour \nonumber\\
&& +\,\, \vacpolghthree +\, \vacpolVthree \nonumber\\
&&\nonumber\\
&& +\, \vacpolVfour + \vacpolW.
\label{eq:vacpoltot}
\eea
The first diagram on the right hand side represents $c'_i$ insertions (cf.\ \refeq{insertions}), the second one stands for the measure term insertion, \refeq{expSmeas}. The other symbols have the usual meaning, with ghost propagator \ghostline. As in ref.~\cite{weiszII} we have split the four-vertex in two parts, corresponding to the second and third lines of \refeq{expS4}.

The analytic expressions corresponding to the graphs in eqs.~(\ref{eq:w1Feyn})--(\ref{eq:vacpoltot}) can be found in ref.~\cite{weiszII}, both for finite $T$ and for $T\rightarrow\infty$, and in terms of arbitrary coefficients\footnote{In refs.~\protect\cite{weiszI,weiszII}, $c_2\leftrightarrow c_3$ with respect to the convention of refs.~\protect\cite{luscherLW1,luscherLW2,luscherLW3} and this paper.} $c_i$. Below we copy those expressions whenever needed.

At tree level the following result was found~\cite{weiszI,weiszII}:
\be
\lim_{T\rightarrow\infty} \frac{1}{T} w_1(L, T) = 4 C_{f} \WWint{k} \sin^2(\half k_1 L) D_{00}(k)\kfeqzero,
\label{eq:stattreeWW}
\ee
where
\be
\WWint{k} \equiv \prod_{j=1}^3\left(\int_{-\pi}^{\pi}\! \frac{dk_{j}}{2\pi}\right),
\label{eq:WWint}
\ee
$C_f$ is the Casimir in the fundamental representation,
\be
C_f = \frac{1}{2}\left(N - \frac{1}{N}\right),
\label{eq:casimir}
\ee
and $D_{\mu\nu}(k)$ is the gauge field propagator, see \refeq{expS2} and appendix~\ref{app-prop}. Since $\mu=\nu=0$ and $k_0=0$, the form~(\ref{eq:simpleprop}) of the propagator is applicable:
\bea
\prop^{-1}_{00}(k) &\stackrel{k_0=0}{=}& \hat{\myvec{k}}^2 - (\tilde{c}_1-c_2) \hat{\myvec{k}}^{(4)} - c_2 (\hat{\myvec{k}}^2)^2\nonumber\\
&=& \myvec{k}^2 - (\tilde{c}_1-c_2+\twelfth) \myvec{k}^{(4)} - c_2 (\myvec{k}^2)^2 + \cO(k^6).
\label{eq:Dinv00k0eq0}
\eea
The symbol $\hat{\myvec{k}}^{(4)}$ and similarly $\myvec{k}^{(4)}$ are defined in \refeq{hatkn}, however now summing over spatial indices only. By doing a contour integration one finds~\cite{luscherLW2} (up to an irrelevant $L$-independent term)
\be
\lim_{T\rightarrow\infty} \frac{1}{T} w_1(L, T) = - \frac{C_{f}}{2\pi L}\left[ 1 + 3\left(\frac{\tilde{c}_1-c_2+\twelfth}{L^2}\right) + \cO(L^{-4}) \right],
\label{eq:stattreeL}
\ee
which obviously is improved for $\tilde{c}_1-c_2 = -\twelfth$.

Now let us proceed to the one-loop level. In ref.~\cite{weiszII} it is shown that
\be
\lim_{T\rightarrow\infty} \frac{1}{T} w_2(L, T) = 48 C_{f} \WWint{k} \sin^2(\half k_1 L) \tilde{w}_2(\myvec{k}) \left(D_{00}(k)\kfeqzero\right)^2,
\label{eq:statonelWW}
\ee
where the $L$-independent function $\tilde{w}_2(\myvec{k})$ has the expansion\footnote{We adopt the convention that $\cO(a^n)$ may also stand for $a^n \ln a$ terms. Note that in lattice units, $a$ is absorbed in $k$.}
\bea
\tilde{w}_2(\myvec{k}) &=& \myvec{k}^2\left\{\tilde{a}_1 -\beta_0 \ln\myvec{k}^2 + (\tilde{a}_{2} + \tilde{b}_{2} \ln\myvec{k}^2) \frac{\myvec{k}^{(4)}}{\myvec{k}^2}  + (\tilde{a}_{3} + \tilde{b}_{3} \ln\myvec{k}^2) \myvec{k}^2 + \cO(k^4) \right\}.\label{eq:wttstruct}
\eea
The coefficients $\tilde{a}_n$, $\tilde{b}_n$ depend on the tree-level values $\tilde{c}_1$ and $c_2$. In particular, $\tilde{b}_2$ and $\tilde{b}_3$ are zero\footnote{This is to be expected from Symanzik's work on scalar theories~\protect\cite{symanzik}. For the present situation Weisz and Wohlert~\protect\cite{weiszII} have shown it explicitly for $c_4=0$. Extension of their result to other tree-level improved actions is not difficult, as we will see in subsection~\protect\ref{subsec-statquark-part}.} if \refeq{onshelltree} holds, which from now on we will assume. On the other hand,
\be
\beta_0 = \frac{11N}{3(4\pi)^2}
\label{eq:betazero}
\ee
takes precisely the universal value required for renormalization (see \refeq{rengroup} below). In order to express the improvement condition in a clean way we perform the integration in \refeq{statonelWW}. It can be shown that $\tilde{a}_3$ does not contribute to the $L$~dependence of $w_2$, and the result reads (up to an irrelevant constant)
\be
\lim_{T\rightarrow\infty} \frac{1}{T} w_2(L, T) = \frac{6 C_f}{\pi L}\left[\beta_0 \ln\left(\frac{a^2}{L^2}\right) - \tilde{a}_1 + \beta_0 c - 3\tilde{a}_2 \frac{a^2}{L^2} + \cO\!\left(\frac{a^4}{L^4}\right) \right],
\label{eq:statoneLbare}
\ee
where $c$ is a numerical constant ($c=-2\gamma = -1.1544313298\cdots$). For clarity we reinstated the $a$~dependence. Note that for tree-level non-improved lattice actions \refeq{statoneLbare} receives additional $\cO(a^2/L^2)$ corrections.

The total one-loop static quark potential has a finite regularization-independent limit for $a\rightarrow0$,
\be
V(L) = -\frac{C_f}{4\pi L} g_R^2 \left\{ 1+g_R^2 \left[ 3 \tilde{a}_2 \frac{a^2}{L^2} + \cO\!\left(\frac{a^4}{L^4}\right) \right] + \cO(g_R^4)\right\},
\label{eq:Voneloopren}
\ee
in terms of the renormalized, $L$-dependent coupling
\be
g_R^2 = g_0^2 \left\{1 - \left[\beta_0 \ln\left(\frac{a^2}{L^2}\right) + \beta_0 c - \tilde{a}_1 \right] g_0^2 + \cO(g_0^4) \right\}.
\label{eq:gRg0}
\ee
Since $V(L)$ and $g_R$ are $a$-independent physical objects, we deduce from \refeq{Voneloopren} the one-loop improvement condition
\be
\tilde{a}_2 = 0.
\label{eq:imprcond1}
\ee
The calculation of the $c'_i$~dependent coefficient $\tilde{a}_2$ is described in the next subsection.

Also Lambda parameter ratios can be extracted\cite{celmaster,hasenfratz2,weisz3,weiszII}.
As $g_R$ must be independent of both $a$ and the regularization prescription, it follows from \refeq{gRg0} that $g_0$ depends on $a$ and $\tilde{c_1}$,\ $c_2$. This is expressed by the well-known renormalization group flow
\be
-\frac{1}{\beta_0 g_0^2} \stackrel{a\rightarrow0}{=} \ln(a^2\Lambda^2)+ \cO\!\left(\ln\left[-\ln(a^2\Lambda^2)\right]\right),
\label{eq:rengroup}
\ee
where the Lambda parameter $\Lambda$ depends on the lattice action:
\be
\frac{\Lambda^{*}}{\Lambda} = e^{\frac{1}{2\beta_0}\left(\tilde{a}_1(\{c_i\}) - \tilde{a}_1(\{c_i^{*}\})\right)}.
\label{eq:Lamratio}
\ee
In this equation $\{c_i\}$ and $\{c^{*}_i\}$ represent any two choices of tree-level coefficients that satisfy the normalization condition \refnn{contcond}, and $\Lambda$, $\Lambda^{*}$ are the corresponding Lambda parameters.

As a final remark,  also Wilson loops for finite values of $L$ and $T$ are objects of interest. They allow for a gauge invariant definition of the mean field or tadpole parameter $u_0$~\cite{lepage}:
\be
u_0 \equiv \left( \frac{1}{N}\left\langle \RE\Tr U(1\times 1)\right\rangle\right)^{\fourth}.
\label{eq:u0def}
\ee
More generally one might define for $L,T\in\natur$
\be
u_0(L,T) \equiv \left( \frac{1}{N}\left\langle \RE\Tr U(L\times T)\right\rangle\right)^{{\scriptstyle \frac{1}{2(L+T)}}},
\label{eq:u0altdef}
\ee
but the choice $L=T=1$ is probably optimal. The reason is that if $L$ or $T$ becomes larger (i.e.\ closer to the correlation length), $u_0(L,T)$ may start picking up physical signals instead of the lattice artefacts that tadpole improvement aims to correct for. Nevertheless we will also compute $w_1(1,2)$ and $w_1(2,2)$ because for the LW and square actions they are related to the vacuum polarization.

\subsection{Particulars}
\label{subsec-statquark-part}

We continue with the details of the one-loop calculation. It should be kept in mind that we will only compare the Wilson, \LW\ and square actions, and therefore we set to zero all tree-level coefficients except $c_0$, $c_1$ and $c_4$, and all one-loop coefficients except $c'_0$, $c'_1$, $c'_2$ and $c'_4$.

As an appetizer let us start with finite Wilson loops. From ref.~\cite{weiszII} we read
\be
w_1(L,T) = C_f \,\,\WWsym(L,T),
\label{eq:w1extractCf}
\ee
with
\be
\WWsym(L,T) = \int_k \left( \frac{\sin\half k_1 L}{\sin\half k_1} \right)^{\!\!2}
\left( \frac{\sin\half k_0 T}{\sin\half k_0} \right)^{\!\!2} \prop_{1010}(k).
\label{eq:w1genLT}
\ee
$D_{\mu\nu\lambda\rho}$ is the propagator of $-i(\partial_{\sigma}A_{\tau}-\partial_{\tau}A_{\sigma})$,
\be
D_{\mu\nu\lambda\rho} = \left\{[\hat{k}_{\mu}\hat{k}_{\lambda}\prop_{\nu\rho} - (\mu\leftrightarrow\nu)] - [\lambda\leftrightarrow\rho]\right\},
\label{eq:propf}
\ee
and $\int_k$ is the four-dimensional analogue of \refeq{WWint}. Upon insertion of the expressions given in appendix~\ref{app-prop}, the evaluation of $\WWsym(L,T)$ for finite values of $L$ and $T$ can be handled to a high accuracy by standard integration routines (we used NAG\cite{NAG} for all numerics in this subsection). The results for $L,T\in\{1,2\}$ are listed in \mbox{table~\ref{tab:finiteloop}}. The entries $(L,T)=(1,1)$ and $(L,T)=(1,2)$ for the LW~action agree with the results of Weisz and Wohlert. They found $\WWsym(1,1) = 0.366262$ and $\WWsym(1,2) = 0.662624$.
\begin{table}
\begin{center}
\begin{tabular}{|c|r@{.}ll||r@{.}ll|}
\hline
&\multicolumn{3}{c||}{\LW}&\multicolumn{3}{c|}{square}\\
\hline
$\WWsym(1,1)$     & \multicolumn{3}{c||}{0.366262680(2)} & \multicolumn{3}{c|}{0.3587838551(1)}\\
\hline
$\WWsym(1,2)$     & \multicolumn{3}{c||}{0.662626785(2)} & \multicolumn{3}{c|}{0.6542934512(1)}\\
\hline
$\WWsym(2,2)$     & \multicolumn{3}{c||}{1.098143594(2)} & \multicolumn{3}{c|}{1.0887235337(1)}\\
\hline
\end{tabular}
\caption{Several numerical values of eq.~(\protect\ref{eq:w1genLT}) for the \LW\ and square actions. Note that $\protect\WWsym(L,T)=\protect\WWsym(T,L)$.}
\label{tab:finiteloop}
\end{center}
\end{table}

The rest of this subsection is devoted to $\tilde{w}_2(\myvec{k})$, defined in \refeq{statonelWW}.  Eqs. (\ref{eq:w2Feyn}) and (\ref{eq:vacpoltot}) carry over in
\bea
\tilde{w}_2(\myvec{k}) &=& U^{(4)}(k) + U^{(3)}(k) + \pi_{00}(k),
\label{eq:w2notation}
\\
\pi &=& \pi' + \pi^{\rm meas} + \pi^{\rm gh1} + \pi^{\rm gh2} + \pi^{V_3} + \pi^{V_4} + \pi^{W},
\label{eq:pitotal}
\eea
where $k = (0,\myvec{k})$. The corresponding analytic expressions are copied below from ref.~\cite{weiszII}. Propagators and vertices can be looked up in the appendices. For shortness we use the notation $\prop = \prop_{00}$ and adopt the summation convention. 
\bea
U^{(4)}(k) &=& N \prop^{-1}(k)\int_{k',k''} (2\pi)^4\delta(k + k' + k'') \left\{ \rule{0mm}{8mm} \left( \medbreuk{1}{6} - \medbreuk{1}{8} \prop^{-1}(k)\prop(k'')\right) \right. \nonumber\\
&& \left. - \left(1-\prop^{-1}(k)\prop(k'')\right) \frac{\cos\half k'_0}{\widehat{(2k'_0)}} \frac{\partial}{\partial k'_0} \right\} \prop(k'),
\label{eq:U4}\\
U^{(3)}(k) &=& i N \prop^{-1}(k)\int_{k',k''} (2\pi)^4\delta(k+k'+k'') \nonumber\\
&& \times\frac{\cos\half k'_0}{\widehat{(2k'_0)}} \prop_{\mu0}(k')\prop_{\nu0}(k'') V_{0\mu\nu}(k,k',k''),
\label{eq:U3}\\
\pi'_{\mu\nu}(k) &=& -\left(\hat{k}_{\lambda}\delta_{\mu\nu} - \hat{k}_{\mu}\delta_{\lambda\nu}\right) q'_{\mu\lambda}(k)\hat{k}_{\lambda},
\label{eq:piprime}\\
\pi^{\rm meas}_{\mu\nu}(k) &=& -\twelfth N\delta_{\mu\nu},
\label{eq:pimeas}\\
\pi^{\rm gh1}_{\mu\nu}(k) &=& -\twentyfourth N\delta_{\mu\nu},
\label{eq:pigh1}\\
\pi^{\rm gh2}_{\mu\nu}(k) &=& \fourth N \int_{k',k''} (2\pi)^4\delta(k+k'+k'') \frac{\hat{k}_{\mu}\hat{k}_{\nu}-\stackrel{\myhat}{(k'-k'')}_\mu\stackrel{\myhat}{(k'-k'')}_\nu}{\hat{k}^{'2} \hat{k}^{''2}},
\label{eq:pigh2}\\
\pi^{V_3}_{\mu\nu}(k) &=& -\half N \int_{k',k''} (2\pi)^4\delta(k+k'+k'') \prop_{\lambda\lambda'}(k') \prop_{\rho\rho'}(k'')\nonumber\\
&& \times V_{\mu\lambda\rho}(k,k',k'') V_{\nu\lambda'\rho'}(k,k',k''),
\label{eq:piV3}\\
\pi^{V_4}_{\mu\nu}(k) &=& \third N \int_{k'} \prop_{\lambda\rho}(k') \left[ V_{\lambda\rho\mu\nu}(k',-k',k,-k) -  V_{\lambda\mu\rho\nu}(k',k,-k',-k) \right], \nonumber\\
&& \label{eq:piV4}\\
\pi^{W}_{\mu\nu}(k) &=& \half d\!  \left(\hat{k}_{\lambda}\delta_{\mu\nu}\! - \hat{k}_{\mu}\delta_{\lambda\nu}\right)\!\hat{k}_{\lambda} \sum_i c_i\! \int_{k'}\! K_{\lambda\mu}^{(i)}(k',-k',k,-k) \prop_{\mu\lambda\mu\lambda}(k').
\label{eq:piW}
\eea
In \refeq{piprime} $q_{\mu\nu}'$ denotes $q_{\mu\nu}$, \refeq{qgeneral}, with, following the prescription of appendix~\ref{app-struct}, tree-level coefficients $c_i$ replaced by their one-loop counterparts $c'_i$. Note that because \refeq{contcond} is satisfied at tree level (and $c_3(g_0^2) = 0$),
\be
c'_0 = -8c'_1 - 8 c'_2 - 16 c'_4.
\label{eq:c0primegone}
\ee
The tensors $K_{\mu\nu}^{(i)}$ in \refeq{piW} are defined in appendix~\ref{app-vert}. The unusual factor
\be
d \equiv \twelfth(6C_f -  N ) = \left(\sixth N - \fourth \mbox{\small$\frac{1}{N}$} \right)
\label{eq:defdi}
\ee
arises from a contraction~\cite{weiszII} of \refeq{Glebsch4sym}.

It is worthwhile to analyze how the above expressions lead to \refeq{wttstruct}. In the first place, they are completely finite for non-zero $\myvec{k}$ because UV divergences are regularized by the lattice (integrations run from $-\pi$ to $\pi$) and potential IR divergences are regularized by the external momentum $\myvec{k}$. For $\myvec{k}=\myvec{0}$ some of the integrals are logarithmically IR divergent, and this carries over into the $\ln \myvec{k}^2$ corrections\footnote{In non-lattice units this gives rise to the familiar $\ln a\,$ \underline{UV} divergences.} to the polynomial behavior in \refeq{wttstruct} as a function of $k$ (we come back to this point below). As mentioned in appendix~\ref{app-struct}, even (odd) vertices are parity even (odd). This also holds for the two-vertex, i.e.\ the propagator. From this the invariance of eqs.~(\ref{eq:U4})--(\ref{eq:piW}) under $k\rightarrow-k$ can be verified explicitly. Hence $w_2(\myvec{k})$ contains no odd powers in $k$. Also cubic invariance can be checked, and therefore up to order $k^4$ only the combinations $\myvec{k}^2$ and $\myvec{k}^{(4)}$ appear. Finally, there is no constant term because $\pi_{\mu\nu}(0) = 0$ holds~\cite{weiszII} due to gauge invariance while the expressions for $U^{(3,4)}$ clearly are of order $k^2$ due to an overall factor of $\prop^{-1}(k)$.

For use below note that $\pi_{\mu\nu}(0) = 0$ allows us to subtract from each term on the right hand side of \refeq{pitotal} its value for $k=0$.

We now embark on the computation of the various contributions to $\tilde{w}_2(\myvec{k})$. Note that all of them are proportional to $N$, except for $\pi'_{00}$ and $\pi_{00}^W$. We therefore follow Weisz and Wohlert and define
\be
z(\myvec{k}) = \pi'_{00}(k)\kfeqzero; x(\myvec{k}) = \pi_{00}^W(k)\kfeqzero; y(\myvec{k}) = \mbox{\small$\frac{1}{N}$}(\tilde{w}_2(\myvec{k}) - x(\myvec{k}) - z(\myvec{k})).
\label{eq:defxyz}
\ee
$x(\myvec{k})$ and $z(\myvec{k})$ are easily computable,
\bea
x(\myvec{k}) &=& d \left\{\left[c_0 \,\WWsym(1,1) + 8 c_1 \,\WWsym(1,2) + 16 c_4 \,\WWsym(2,2)\right]\hat{\myvec{k}}^2  \right. \nonumber\\
&& \left. - \left[c_1 \,\WWsym(1,2) + 4 c_4 \,\WWsym(2,2)\right]\hat{\myvec{k}}^{(4)} \right\},
\label{eq:xexact}\\
z(\myvec{k}) &=& (\tilde{c}'_1 - c'_2)\hat{\myvec{k}}^{(4)} + c'_2\left(\hat{\myvec{k}}^2\right)^{\!2}.
\label{eq:zexact}
\eea

The elaborate part of the computation is posed by $y(\myvec{k})$. We can save ourselves a fair amount of work by pursuing only the following two objectives:
\begin{enumerate}
\item The evaluation of $\Lambda_{\rm sq(uare)}/\Lambda_{\rm W(ilson)}$. Due to \refeq{Lamratio} this requires computing $\tilde{a}_1^{\rm sq} - \tilde{a}_1^{\rm W}$, where $\tilde{a}_1$ is defined in \refeq{wttstruct}.
\item The evaluation of the improvement condition \refnn{imprcond1}, i.e.\ of the coefficient $\tilde{a}_2$ defined in \refeq{wttstruct}, for the square action. It is sufficient to compute $\tilde{a}_2^{\rm sq} - \tilde{a}_2^{\rm LW}$, because the value of $\tilde{a}_2^{\rm LW}$ was computed by Weisz and Wohlert~\cite{weiszII}, and later verified by L\"{u}scher and Weisz~\cite{luscherLW3}.
\end{enumerate}
We thus need only consider $y^*(\myvec{k})-y^{**}(\myvec{k})$, where `$*$' and `$**$' denote different lattice actions. This implies that we can drop $\pi_{00}^{\rm meas}$, $\pi_{00}^{\rm gh1}$ and $\pi_{00}^{\rm gh2}$. Much more importantly, it gives us the opportunity to circumvent the logarithmic terms in \refeq{wttstruct} due to the following argument.

The most straightforward way to perform the $k\rightarrow0$ expansion of the integrals would be to expand the integrands. Unfortunately this method is too naive for terms that contain both propagators $\prop(k')$ and $\prop(k'')$, because due to momentum conservation the expansion of $\prop(k'')$ gives for each factor $k^2$ a factor $(k'^2 + \cO(k'^4))^{-1}$, thus possibly introducing IR divergences. As an example consider $\pi_{\mu\nu}^{V_3}(k)$ (\refeq{piV3}). For $k=0$ its integrand behaves as $1/k'^2$ near $k'=0$, so the integral is convergent. However, if we expand the integrand with respect to $k^2$, we introduce at $\cO(k^2)$ a logarithmic IR divergence. This would be fatal were it not that due to its IR nature the divergence is completely determined by the small-$k'$ behavior of the integrand. In particular it is the same for all lattice actions satisfying \refeq{contcond}. This implies that subtracting the integrands of two such actions produces an extra factor of $(k'^2 + \cO(k'^4))^{+1}$, and the $\cO(k^2)$ term is finite.

For other contributions to $y(\myvec{k})$ one can give similar arguments. Therefore the naive method does yield the correct $\myvec{k}^2$-coefficients in the expansion of $y^*(\myvec{k})-y^{**}(\myvec{k})$. Furthermore, if `$*$' and `$**$' both denote tree-level improved actions, then the integrands are the same up to {\em sub\hspace{.3mm}}leading order, and the naive method also produces the correct $k^4$-coefficients.

As a side remark we mention that if one is interested in $y(\myvec{k})$ itself, instead of merely its ${c_i}$~dependence, one has to tackle the expansion with respect to $k$ in a more sophisticated way. For example, one might subtract a suitable integrand that has the same IR behavior, but the integral of which is computable by continuum methods. This leads to the $\ln \myvec{k}^2$ factors appearing in \refeq{wttstruct}. From the above discussion we then conclude that the coefficient $\beta_0$ is universal within the class of all lattice actions satisfying the condition (\ref{eq:contcond}), and the coefficients $\tilde{b}_{2,3}$ are universal within the subclass of tree-level improved actions. Using Weisz and Wohlert's work~\cite{weiszII} on the \LW\ action it then follows that the latter values are $\tilde{b}_{2} = 0 $ and  $\tilde{b}_{3} = 0$.

Let us denote by $N a_i$ the contribution of $N y(\myvec{k})$ to $\tilde{a}_i$. From \refeq{Dinv00k0eq0} and the above discussion we conclude that for the computation of $a_1^{\rm sq} - a_1^{\rm W}$ it is sufficient to consider $(\pi^{V_3}_{00}(k)-\pi^{V_3}_{00}(0))/N$, $(\pi^{V_4}_{00}(k)-\pi^{V_4}_{00}(0))/N$, and
\bea
\tilde{U}^{(3)}(k) &\equiv& i \myvec{k}^2\int_{k'} \frac{\cos\half k'_0}{\widehat{(2k'_0)}} \prop_{\mu0}(k')\prop_{\nu0}(k') V_{0\mu\nu}(0,k',-k'),\label{eq:U3tilde}\\
\tilde{U}^{(4)}(k) &\equiv& \myvec{k}^2 \int_{k'}\left( \frac{1}{6} - \frac{\cos\half k'_0}{\widehat{(2k'_0)}} \frac{\partial}{\partial k'_0} \right) \prop(k').
\label{eq:U4tilde}
\eea
Similarly, $a_2^{\rm sq} - a_2^{\rm LW}$ only receives contributions from $(\pi^{V_3}_{00}(k)-\pi^{V_3}_{00}(0))/N$ and $(\pi^{V_4}_{00}(k)-\pi^{V_4}_{00}(0))/N$ ($U^{(3,4)}(k)$ only produce $k^4$ terms of the type $(\myvec{k}^2)^2$).

The integrands of $\tilde{U}^{(3,4)}$ are already expanded with respect to $k$. The expansion of $\pi^{V_4}_{00}(k)$ can be performed rather easily, because the single internal propagator is independent of the external momentum $k$. We used Mathematica\cite{wolfram} to obtain
\bea
&& \frac{1}{N}\left(\pi^{V_4}_{00}(k) - \pi^{V_4}_{00}(0)\right) \nonumber\\
&&  \hspace{1cm}=\frac{\hat{\myvec{k}}^2}{12} \int_{k'} \left\{ 
(\tilde{c}_0 + 8 \tilde{c}_1) \left[ \left( 7 - {\breuk{5}{2}} {{{\hat{k}_1^{'2}}}} \right) D(k') +   {\hat{k}'_0} {\hat{k}'_1}  D_{01} (k') \right] 
\right.\nonumber\\&&\hspace{2cm}
+ \tilde{c}_1 \left[ \left( 84 - 27 {{{\hat{k}_0^{'2}}}} - 60 {{{\hat{k}_1^{'2}}}} + 
   10 {{{\hat{k}_1^{'4}}}} + {\breuk{23}{2}} {{{\hat{k}_0^{'2}}}} {{{\hat{k}_1^{'2}}}} 
     \right) D(k') 
\right.\nonumber\\&&\hspace{2cm}\left.
+ 2 \left( 18 - 7 {{{\hat{k}_1^{'2}}}} \right) {\hat{k}'_0} {\hat{k}'_1}  D_{01} (k') \right]
\nonumber\\&&\hspace{2cm}
+ c_4 \left[ 2 \left( -60 + 23 {{{\hat{k}_0^{'2}}}} \right)  
   \left( 3 - {{{\hat{k}_1^{'2}}}} \right)  {{{\hat{k}_1^{'2}}}} D(k') 
\right.\nonumber\\&&\hspace{2cm}\left.\left.
+  8 \left( -12 + 5 {{{\hat{k}_0^{'2}}}} \right)  
   \left( -3 + {{{\hat{k}_1^{'2}}}} \right) {\hat{k}'_0} {\hat{k}'_1}  D_{01} (k') \right]
\right\} \nonumber\\
&&\hspace{1cm}\hspace{.6ex} - \frac{\hat{\myvec{k}}^{(4)}}{12} \int_{k'} \left\{ 
\tilde{c}_1 \left[ \left( 25 - 6 {{{\hat{k}_0^{'2}}}} - 10 {{{\hat{k}_1^{'2}}}} + 
   2 {{{\hat{k}_0^{'2}}}} {{{\hat{k}_1^{'2}}}} + {\half} {{{\hat{k}_1^{'4}}}} \right) D(k') 
\right.\right.\nonumber\\&&\hspace{2cm}\left.
+  \left( 4 - {{{\hat{k}_1^{'2}}}} \right) {\hat{k}'_0} {\hat{k}'_1}  D_{01} (k') \right]
\nonumber\\&&\hspace{2cm}
+ c_4 \left[ \left( 12 - 3 {{{\hat{k}_0^{'2}}}} - 120 {{{\hat{k}_1^{'2}}}} + 
   38 {{{\hat{k}_0^{'2}}}} {{{\hat{k}_1^{'2}}}} + 38 {{{\hat{k}_1^{'4}}}} -  {\breuk{23}{2}}
   {{{\hat{k}_0^{'2}}}} {{{\hat{k}_1^{'4}}}} \right) D(k') 
\right.\nonumber\\&&\hspace{2cm}\left.\left.
+ 2 \left( 12 - 5 {{{\hat{k}_0^{'2}}}} \right)  
   \left( 4 - {{{\hat{k}_1^{'2}}}} \right)  {\hat{k}'_0} {\hat{k}'_1}D_{01} (k') \right]
 \right\}.
\label{eq:piV4nice}
\eea
For $c_4 = 0$ this agrees with ref.\cite{weiszII}.

The last ingredient needed is the expansion of $(\pi^{V_3}_{00}(k)-\pi^{V_3}_{00}(0))/N$. In principle the analytic expansion of the integrand is straightforward, but to our taste too tedious due to the $k$-dependence of the propagator(s). We therefore decided to evaluate $\Delta\pi_{00}^{V_3,i}(k)\equiv (\pi_{00}^{V_3,{\rm sq}}(k) - \pi_{00}^{V_3,{\rm sq}}(0) - \pi_{00}^{V_3,i}(k) + \pi_{00}^{V_3,i}(0))/N$ ($i={\rm W,LW}$) numerically for a number of values of $ \myvec{k}$. As explained above the result should be of the form
\be
\Delta\pi_{00}^{V_3,i}(k) = \Delta a_1^{V_3,i} \myvec{k}^2 + \Delta a_2^{V_3,i} \myvec{k}^{(4)} + \Delta a_3^{V_3,i}\! \left(\myvec{k}^2\right)^{\!2} + \cO(k^6).
\label{eq:DV3struct}
\ee
(For $i={\rm W}$ there are also logarithmic terms at $\cO(k^4)$). We determined $\Delta a_1^{V_3,{\rm W}}$ by choosing $ \myvec{k} = (\varepsilon,0,0)$ and fitting to $\varepsilon^2$. For the determination of $\Delta a_2^{V_3,{\rm LW}}$ we subtracted results for $ \myvec{k} = (\varepsilon,0,0)$ and $ \myvec{k} = (\varepsilon/\sqrt{2},\varepsilon/\sqrt{2},0)$, thus eliminating the $ \myvec{k}^2$ and $( \myvec{k}^2)^2$ terms. The results were fitted to $\half\varepsilon^4$. These methods introduce numerical instabilities of the order of $\delta/\varepsilon^2$ and $\delta/\varepsilon^4$ respectively, where $\delta$ is the computer precision ($\delta\approx10^{-14}$). One would like to choose $\varepsilon$ not too big, but on the other hand subleading terms (which are of relative order $\varepsilon^2$) must be kept sufficiently small. By chosing $\varepsilon$ in the ranges 0.0001 to 0.001 and 0.01 to 0.1 respectively, we were able to extract $\Delta a_1^{V_3}$ and $\Delta a_2^{V_3}$ with roughly the same precision as we expect we could have attained by an analytic expansion and subsequent numerical integration.

Our final result in the $y(\myvec{k})$ sector is
\bea
a_1^{\rm sq} - a_1^{\rm W} &=& -0.031810197(2),\nonumber\\
a_2^{\rm sq} - a_2^{\rm LW} &=& \phantom{-}0.000087063(4).
\label{eq:yresultssq}
\eea
As a check of our analysis and programs we also computed
\be
a_1^{\rm LW} - a_1^{\rm W} = -0.031361443(2),
\label{eq:yresultLW}
\ee
which agrees with the value found by Weisz and Wohlert~\cite{weiszII}, -0.03136145(1).
When we include the results for $x(\myvec{k})$ and $z(\myvec{k})$, using $\hat{\myvec{k}}^2 = \myvec{k}^2 - \twelfth\myvec{k}^{(4)} + \cO(k^6)$ and table~\ref{tab:finiteloop} (together with $\WWsym(1,1)_{{\rm Wilson}} = \half$), we obtain for the coefficients of $\tilde{w}_2(\myvec{k})$
\bea
\tilde{a}_1^{i} - \tilde{a}_1^{\rm W} &=& \left\{\begin{array}{ll}
\!\!\!-\!\left(0.086580342(3) N\! - 0.082828348(1) \frac{1}{N}\right) & \!\!\!\!{\rm for\ }i = {\rm LW}\\
\!\!\!-\!\left(0.085608020(2) N\! - 0.08069673318(6) \frac{1}{N}\right) & \!\!\!\!{\rm for\ }i = {\rm sq}, \end{array} \right.
\label{eq:a1tresults}\\
\tilde{a}_2^{\rm sq} - \tilde{a}_2^{\rm LW} &=& \left[\tilde{c}'_1 - c_2'\right]_{\rm sq} -  \left[\tilde{c}'_1 - c_2'\right]_{\rm LW} \nonumber\\
&& - \left(0.002158351(4) N - 0.0033681212(1)\frac{1}{N} \right).
\label{eq:a2tresult}
\eea
It follows from \refeq{Lamratio} that
\bea
\frac{\Lambda_{\rm LW}}{\Lambda_{\rm W}} &=& \left\{\begin{array}{ll}
4.1308935(3) & {\rm for\ }N=2\\
5.2921038(3) & {\rm for\ } N=3, \end{array} \right.
\nonumber\\
\frac{\Lambda_{\rm sq}}{\Lambda_{\rm W}} &=& \left\{\begin{array}{ll}
4.0919901(2) & {\rm for\ }N=2\\
5.2089503(2) & {\rm for\ } N=3. \end{array} \right.
\label{eq:Lamresults}
\eea
The values for the LW~action agree with those in ref.~\cite{weiszII}: 4.13089(1) for $N=2$ and 5.29210(1) for $N=3$. Moreover, the Lambda ratios for the square action completely agree with the results in ref.~\cite{PvBLamsq}. These were obtained in a totally different way, namely by using a background field approach.  

From \refeqs{imprcond1}{a2tresult} we read off a one-loop improvement condition. We can substitute the result of L\"{u}scher and Weisz~\cite{luscherLW3} for the LW action,
\be
\left[\tilde{c}'_1 - c_2'\right]_{\rm LW} = \left\{\begin{array}{ll}
-0.01100879(1) & {\rm for\ }N=2\\
-0.02080086(2) & {\rm for\ } N=3, \end{array} \right.
\label{eq:at1LWresult}
\ee
which is consistent with the value extracted by Weisz and Wohlert~\cite{weiszII} from the static quark potential, but more accurate. In this way we obtain
\be
\left[\tilde{c}'_1 - c_2'\right]_{\rm sq} = \left\{\begin{array}{ll}
-0.00837615(2) & {\rm for\ }N=2\\
-0.01544851(3) & {\rm for\ } N=3. \end{array} \right.
\label{eq:at1sqresult}
\ee

We conclude with the observation that the last diagram in \refeq{vacpoltot} gives quantitatively far the most important contribution to these values:
\bea
\frac{-\left.\tilde{a}_2^{\rm LW}\right|_{\pi^W}}{\left[\tilde{c}'_1 - c_2'\right]_{\rm LW}} &=& \left\{\begin{array}{ll}
0.78 & {\rm for\ }N=2\\
0.82 & {\rm for\ } N=3, \end{array} \right.\nonumber\\
\frac{-\left.\tilde{a_2}^{\rm sq}\right|_{\pi^W}}{\left[\tilde{c}'_1 - c_2'\right]_{\rm sq}} &=& \left\{\begin{array}{ll}
0.69 & {\rm for\ }N=2\\
0.75 & {\rm for\ } N=3. \end{array} \right.
\label{eq:Wcontrlarge}
\eea
The other diagrams give much smaller contributions, so that the large ratios are not due to coincidental cancellations. This fits well in the tadpole/mean field picture because the $W$-vertex is not present in the continuum theory, and hence is at least partly responsible for the deviation from 1 of the tadpole parameter $u_0$.

\section{Spectroscopy in a twisted finite volume}
\label{sec-twisted}
\subsection{Introduction and formalism}
\label{subsec-twisted-form}
In this subsection we give a rather short overview of the formalism of pure lattice gauge theory in a partly twisted space-time. For details we refer to refs.~\cite{luscherLW1,luscherLW3}, and also to section~\ref{sec-backgroundtwist}. As mentioned in section~\ref{sec-backgroundtwist} we limit ourselves to infinite volume in the $x_0$, $x_3$ directions, and twist in the $x_1$, $x_2$ directions (with twist quantum $-1$):
\be
A_{\mu}(x + L \hat{\nu}) = \Omega_{\nu}A_{\mu}(x) \Omega_{\nu}^{-1},\ \ \ (\nu=1,2),
\label{eq:twistbc}
\ee
where $\Omega_{\nu}\in\SU{N}$ are $A_{\mu}$ and $x$ independent matrices satisfying
\be
\Omega_1\Omega_2 = z\Omega_2\Omega_1,\ \ \ z\equiv e^{2\pi i/N}.
\label{eq:twistalg}
\ee
\refEq{Ftrconttwist} generalizes to
\be
A_{\mu}(x) = g_0 \dollar{k} e^{i k (x+\half a \hat{\mu})} \tA_{\mu}(k)\Gamma_k, \label{eq:Ftrtwisted}
\ee
where the integration symbol `$\$$' stands for
\be
\dollar{k} \equiv \frac{1}{L^2 N} \sum_{k_{\perp}} \int_{-\pi}^{\pi}\!\! \frac{dk_0}{2\pi} \int_{-\pi}^{\pi}\!\! \frac{dk_3}{2\pi},
\label{eq:defdollartwisted}
\ee
and $\Gamma_k\in\SU{N}$ satisfies
\be
\Omega_{\nu}\Gamma_k \Omega_{\nu}^{-1} = e^{i k_{\nu} L} \Gamma_k,\ \ \ (\nu=1,2).
\label{eq:defpropGam}
\ee
The momentum components $k_{\perp}\equiv(k_1,k_2)$ in \refeq{defdollartwisted} are discretized as in \refeq{momquant}, but on the lattice $\sum_{k_{\perp}}$ runs over, say, $n_{\nu} = 0, 1,\cdots,NL-1$. The solution to \refeq{defpropGam} is given by \refeqs{defGammak}{Gamntok}.

In terms of
\bea
\chi_k &\equiv& \left\{\begin{array}{ll} 
0 & {\rm if\ } n_{\nu} = 0\,(\mod N),\ \ \ (\nu=1,2) \\
1 & {\rm otherwise},
\end{array} \right.
\label{eq:defchi}\\
\brs{k}{k'} &\equiv& n_1 n'_1 + n_2 n'_2 + (n_1 + n_2)(n'_1 + n'_2),\nonumber\\
\bra{k}{k'} &\equiv& n_1 n'_2 - n'_1 n_2,
\label{eq:defbrackets}
\eea
the matrices $\Gamma_{k}$ satisfy~\cite{luscherLW3}
\be
\left\{\!\!\!\! \begin{array}{l}
\begin{array}{ll}
\Gamma_{k'} = \Gamma_{k} & {\rm if\ } \chi_{k'-k} = 0\\
\Gamma_{k} = \unit & {\rm if\ } \chi_k = 0\\
\!\Tr \Gamma_k = 0 & {\rm if\ } \chi_k = 1
\end{array}\\
\ \,\Gamma_k^{\dagger} = z^{-\half\brs{k}{k}} \Gamma_{-k}\\
\ \,\Gamma_{k}\Gamma_{k'} = \Gamma_{k+k'} z^{\half\bra{k}{k'}-\half\brs{k}{k'}}.
\end{array}
\right.
\label{eq:Gammaprop}
\ee
From these properties, \refeq{Ftrtwisted}, and the fact that $A_{\mu}(x)$ is in the Lie algebra of \SU{N}\ it follows that
\bea
\tilde{A}_{\mu}(k)^{*} &=& -z^{\half\brs{k}{k}} \tilde{A}_{\mu}(-k),
\label{eq:realprop}\\
\tilde{A}_{\mu}(k) &=& 0 \ \ \ {\rm if\ } \chi_k = 0.
\label{eq:momholes}
\eea

For later use it is convenient to define a delta function associated with \refeq{defdollartwisted}:
\be
\delta(k) \equiv N L^2 \delta^{\perp}_{k,0}\,\, 2\pi \delta^1(k_0)\,\, 2\pi \delta^1(k_3),
\label{eq:defdeltwisted}
\ee
where $\delta^1$ is the ordinary 1-dimensional ($2\pi$-periodic) delta function, and $\delta^{\perp}_{k,k'}$ is the $N L$-periodic 2-dimensional Kronecker delta,
\be
\delta^{\perp}_{k,k'} \equiv \left\{\begin{array}{ll} 
1 & {\rm if\ } n_{\nu}-n'_{\nu} = 0\,(\mod (NL)),\ \ \ (\nu=1,2) \\
0 & {\rm otherwise}.
\end{array} \right.
\label{eq:defdeltaorth}
\ee

\refEq{momholes} is an extremely nice property of the twisted tube, at least in perturbation theory. It implies that $\tilde{A}_{\mu}(0)$ is not a degree of freedom of the twisted gauge field, and therefore is not to be summed over in loop `integrals'. In this way the twisted tube escapes any IR problems that lure in infinite or finite periodic volumes.

A strongly related effect can be seen very clearly at tree level. As follows from the Feynman rules listed in appendix~\ref{app-twisted}, the bare propagator equals 
\be
\langle\tilde{A}_{\mu}(k)\tilde{A}_{\nu}(k')\rangle_{g_0=0} = \delta(k+k') \left(-\half z^{-\half\brs{k}{k}}\right) \chi_k \prop_{\mu\nu}(k).
\label{eq:bareproptwist}
\ee
(We slightly disagree with ref.~\cite{luscherLW3}). Its mass-shell equation is determined by the ordinary Feynman propagator $\prop_{\mu\nu}(k)$ and thus reads
\be
k_0 = \pm i E_0(\myvec{k}),\ \ \  E_0(\myvec{k}) = \sqrt{k_{\perp}^2 + k_3^2}\ (1+\cO(k^2)).
\label{eq:k0contlim}
\ee
However, since $k_{\perp}$ is discretized and $k_{\perp}=0$ is not permissible, a mass gap emerges, of width $m$ (in the continuum limit\footnote{Exact lattice formulas can be found in appendix~\ref{app-prop}.}). As long as $L$ is chosen so small (in physical units) that asymptotic freedom ensures $g_R^2(L) \ll 1$, the tree-level value of the mass gap receives only small quantum mechanical corrections, which are computable in perturbation theory.

In refs~\cite{luscherLW1,luscherLW3}, L\"{u}scher and Weisz present a Kaluza-Klein picture of the twisted tube by viewing it as a two-dimensional theory; the compact dimensions are considered internal space. In this interpretation $n_1$ and $n_2$ are quantum numbers. The particles defined by $n_{\perp}=(1,0)$ and $n_{\perp}=(1,1)$ (or cubic transformations thereof) are called `A' and `B' mesons. Their masses are $m_{\rm A} = m$ and $m_{\rm B} = \sqrt{2}m$, to lowest order in  the lattice spacing and coupling constant. The mesons possess a `spin' quantum number taking values $\pm$, corresponding to the two physical polarizations of the underlying four-dimensional theory of massless gauge bosons. For an understanding of these particles in terms of the electric fluxes introduced by 't~Hooft, see section~\ref{sec-backgroundtwist}.

The twisted tube brings about many well-defined spectral quantities that can be used to extract the Symanzik coefficients. The simplest quantities are the A and B masses. In subsection~\ref{subsec-twisted-mass} we compute to one-loop order the mass of the A meson with positive spin. We will see that this gives the same improvement condition as was found from the static quark potential. The reason is that in both cases the coefficients $c'_i$ enter via a propagator insertion. We nevertheless decided to undertake the calculation in order to become familiar with perturbation theory on the twisted tube, and also to check \refeq{at1sqresult}.

The Symanzik coefficients $\tilde{c}'_1$ and $c'_2$ can be obtained separately by also computing the one-loop correction to the three-gluon vertex. For suitable external momenta this is a spectral quantity, because it is related to the elastic scattering amplitude of two A mesons~\cite{luscherLW1} (which in turn is related to the energy eigenvalue of a two-particle state). The calculation is described in subsection~\ref{subsec-twisted-coupling}.

\subsection[Mass of the ${\rm A}^{+}$ meson]{Mass of the \boldmath${\rm A}^{+}$\unboldmath\ meson}
\label{subsec-twisted-mass}

The energy spectrum of (improved) lattice field theories is best defined through the transfer matrix~\cite{luscherTRANSF}. The spectrum dictates the exponential decay of the two-point function $\langle A_{\mu}(x)A_{\mu}(y)\rangle$ as a function of the time separation $x_0-y_0$. Equivalently, the spectrum can be read off from the pole structure of $\fprop_{\mu\nu}(k)$, the definition of which in the twisted tube geometry we already encountered in \refeq{defDfullintro} (cf.\ \refeq{bareproptwist}): 
\be
\langle\tilde{A}_{\mu}(k)\tilde{A}_{\nu}(k')\rangle = \delta(k+k') \left(-\half z^{-\half\brs{k}{k}}\right) \chi_k \fprop_{\mu\nu}(k).
\label{eq:fullproptwist}
\ee

We concentrate on the A~meson with positive spin. From the discussion in section~\ref{sec-backgroundtwist} we know that to all orders in $g_0$ it can be represented by the polarization $\varepsilon_{\mu}=\delta_{\mu,1}$ at a momentum $k=(k_0,\myvec{k})$; $\myvec{k}=(0,m,k_3)$. Let us denote the corresponding eigenvalue of $\fprop_{\mu\nu}(k)$ by $d_{\rm full}(k)$ ($=\fprop_{11}(k)$). The energy $E(\myvec{k})$ and the associated wave function renormalization $Z(\myvec{k})$ belonging to this ${\rm A}^{+}$~meson are defined through
\be
d_{\rm full}(k) = \frac{Z(\myvec{k})}{k_0^2 + E^2(\myvec{k})} + ({\rm regular\ in\ }k_0),
\label{eq:dfullaroundpole}
\ee
where the expansion is valid for $k_0$ close to $E(\myvec{k})$. We will assume that $E(\myvec{k})$ is physical in the sense that $E(\myvec{k})\sim a^0$ instead of $\sim a^{-1}$, in non-lattice units. This can be verified from our results below. It implies~\cite{luscherTRANSF} $E(\myvec{k})\in\real$. For future use we note
\be
Z^{-1}(\myvec{k}) = \left.\frac{1}{2 k_0} \frac{\partial}{\partial k_0} d_{\rm full}^{-1}(k)\right|_{k_0 = \pm i E(\myvec{k})}.
\label{eq:Zfullpract}
\ee

In terms of the vacuum polarization $\pi_{\mu\nu}$, defined through
\be
\fprop_{\mu\nu}(k) = \prop_{\mu\nu}(k) + g_0^2 \prop_{\mu\rho}(k) \pi_{\rho\sigma}(k) \prop_{\sigma\nu}(k) + \cO(g_0^2),
\label{eq:fproppert}
\ee
we find
\be
d^{-1}_{\rm full}(k) = d^{-1}_{(0)}(k) + g_0^2 d^{-1}_{(1)}(k) + \cO(g_0^4),\ \ \ d^{-1}_{(1)}(k) = -\pi_{11}(k).
\label{eq:pertexpdeps}
\ee
For the one-loop energy and wave-function renormalization, defined through
\bea
E(\myvec{k}) &=& E_0(\myvec{k}) + g_0^2 E_1(\myvec{k}) + \cO(g_0^4),\nonumber\\
Z(\myvec{k}) &=& Z_0(\myvec{k}) + g_0^2 Z_1(\myvec{k}) + \cO(g_0^4),
\label{eq:pertexpEZ}
\eea
the following formulas are valid:
\bea
E_1(\myvec{k}) \!\!&=&\!\! \frac{1}{2E_0(\myvec{k})} Z_0(\myvec{k}) \left. d^{-1}_{(1)}(k)\right|_{k_0 = i E_0(\myvec{k})}, 
\label{eq:Eoneloopgen}\\
Z_1(\myvec{k}) \!\!&=&\!\! -Z_0^2(\myvec{k}) \!\left\{ \frac{1}{2k_0} \frac{\partial}{\partial k_0} d^{-1}_{(1)}(k) - 2E_0(\myvec{k}) E_1(\myvec{k})  \left( \frac{1}{2k_0} \frac{\partial}{\partial k_0} \right)^{\!2}\! d^{-1}_{(0)}(k) \right\}_{\!k_0 = i E_0(\myvec{k})}.\label{eq:Zoneloopgen}
\eea

For the mass calculation we put $k_3=0$, since $m_{{\rm A}^{+}} \equiv E(\myvec{k}=(0,m,0))$. At tree level, \refeq{simpleprop} and hence \refeq{ZEsmalla} are applicable, so that~\cite{luscherLW1}
\be
m^{(0)}_{{\rm A}^{+}} = m - (\tilde{c_1}-c_2 + \twelfth) m^2 + \cO(m^4).
\label{eq:mA+tree}
\ee
Note that this is improved for the choice \refnn{onshelltree} of coefficients, as it should be. Below we assume \refeq{onshelltree} is satisfied.

At one-loop level we write
\be
m_{{\rm A}^{+}} = m^{(0)}_{{\rm A}^{+}} + g_0^2 m^{(1)}_{{\rm A}^{+}} + \cO(g_0^4),
\label{eq:mA+pert}
\ee
so that from the above analysis it follows that
\be
m^{(1)}_{{\rm A}^{+}} = \left. -Z_0(\myvec{k}) \frac{\pi_{11}(k)}{2m^{(0)}_{{\rm A}^{+}}}\right|_{k=(i m^{(0)}_{{\rm A}^{+}},0,m,0)}.
\label{eq:mA+one}
\ee
Note that this quantity only depends on $m$ and (implicitly) $N$, or equivalently $L$ and $N$. From \refeq{ZEsmalla} we find
\be
\left. Z_0(\myvec{k})\right|_{\myvec{k}=(0,m,0)} = 1+\cO(m^4).
\label{eq:ZAsmalla}
\ee
For the extraction of the one-loop coefficients we can neglect the $\cO(m^4)$ correction term.

The vacuum polarization $\pi_{\mu\nu}(k)$ has of course the same expansion as in section~\ref{sec-statquark}, see eqs.~\refnn{vacpoltot} and \refnn{pitotal}, but the explicit forms \refnn{piprime}--\refnn{piW} change a little on the twisted tube (as follows from appendix~\ref{app-twisted}):
\bea
\pi'_{\mu\nu}(k) &=& -\left(\hat{k}_{\lambda}\delta_{\mu\nu} - \hat{k}_{\mu}\delta_{\lambda\nu}\right) q'_{\mu\lambda}(k)\hat{k}_{\lambda},
\label{eq:piprimetwist}\\
\pi^{\rm meas}_{\mu\nu}(k) &=& -\twelfth N\delta_{\mu\nu},
\label{eq:pimeastwist}\\
\pi^{\rm gh1}_{\mu\nu}(k) &=& \twelfth \delta_{\mu\nu} \dollar{k'}\, g_{-}^2(k,k') \frac{\hat{k}^{'2}_{\mu}}{\hat{k}^{'2}},
\label{eq:pigh1twist}\\
\pi^{\rm gh2}_{\mu\nu}(k) &=& -\eighth\! \dollar{k',k''} \delta(k+k'+k'') g_{-}^2(k,k')  \frac{\hat{k}_{\mu}\hat{k}_{\nu}-\stackrel{\myhat}{(k'-k'')}_\mu\stackrel{\myhat}{(k'-k'')}_\nu}{\hat{k}^{'2} \hat{k}^{''2}},
\label{eq:pigh2twist}\\
\pi^{V_3}_{\mu\nu}(k) &=& \fourth \dollar{k',k''}\, \delta(k+k'+k'') g_{-}^2(k,k') \prop_{\lambda\lambda'}(k') \prop_{\rho\rho'}(k'')\nonumber\\
&& \times V_{\mu\lambda\rho}(k,k',k'') V_{\nu\lambda'\rho'}(k,k',k''),
\label{eq:piV3twist}\\
\pi^{V_4}_{\mu\nu}(k) &=& -\sixth \dollar{k'}\, g_{-}^2(k,k') \prop_{\lambda\rho}(k') \nonumber\\
&& \times \left[ V_{\lambda\rho\mu\nu}(k',-k',k,-k) -  V_{\lambda\mu\rho\nu}(k',k,-k',-k) \right], \label{eq:piV4twist}\\
\pi^{W}_{\mu\nu}(k) &=& \eighth\!  \left(\hat{k}_{\lambda}\delta_{\mu\nu} - \hat{k}_{\mu}\delta_{\lambda\nu}\right)\!\hat{k}_{\lambda} \sum_i c_i\, \dollar{k'} \left(\chi_{k'}+\sixth g_{-}^2(k,k')\right)\nonumber\\
&&\times K_{\lambda\mu}^{(i)}(k',-k',k,-k) \prop_{\mu\lambda\mu\lambda}(k').
\label{eq:piWtwist}
\eea
In these equations
\be
g_{-}(k,k') \equiv 2i\sin \left(\frac{\pi}{N}\bra{k}{k'}\right),
\label{eq:defg-}
\ee
which equals $0$ if $\chi_{k'} = 0$ or $\chi_{k+k'} = 0$. A reasonable check of the above formulas is that they equal their infinite-volume counterparts in leading order for $m\rightarrow0$ (i.e.\ $L/a\rightarrow\infty$ in non-lattice units), which is what one would expect physically. The technical reason is that in this limit the factor $g_{-}^2(k,k')$ is oscillating infinitely faster than the rest of the integrands, so that it can be replaced by its average value: $-2$. Due to the smoothness of the resulting integrand, $\txtdollar{k'}$ can be replaced by $\mbox{\footnotesize $N$}\int_{k'}$ and $\txtdollar{k'}\mbox{\footnotesize$\chi_{k'}$}$ by $\mbox{\footnotesize$(N-N^{-1})$}\int_{k'}$. 

In ref.~\cite{luscherLW3} it is shown that near the continuum limit $m^{(1)}_{{\rm A}^{+}}$ is of the form
\be
\frac{m^{(1)}_{{\rm A}^{+}}}{m} = \frac{\tilde{m}^{(1)}_{{\rm A}^{+}}}{m} - \left( \tilde{c}'_1 - c'_2\right) m^2 + \cO(m^4);\ \ \ \frac{\tilde{m}^{(1)}_{{\rm A}^{+}}}{m} = a_0 + a_1 m^2 + \cO(m^4),
\label{eq:mAsmallm}
\ee
where we separated the contribution of $\pi'_{11}$ (containing all dependence on $c'_i$). The coefficients\footnote{In spite of the same notation, these are different from the coefficients in subsection~\ref{subsec-statquark-part}.} $a_i$ are determined by the other contributions. It is clear that the improvement condition reads
\be
\tilde{c}'_1 - c'_2 = a_1.
\label{eq:imprcondmA+one}
\ee
In \refeq{mAsmallm} there are no quadratically ($1/m^2$) or logarithmically ($m^0\ln m$) divergent terms because $m^{(0)}_{{\rm A}^{+}}$ is independent of $g_0$, while multiplicative renormalization of $g_0$ alone is sufficient to cancel all divergences in the continuum limit. Also the $m^2 \ln m$ term is absent due to tree-level improvement, cf.\ the discussion below \refeq{wttstruct}. Note however that individual diagrams can give $1/m^2$, $\ln m$ or $m^2 \ln m$ contributions to $\tilde{m}^{(1)}_{{\rm A}^{+}}/m$. For example, $\pi^{\rm meas}$ contributes $N/(24 m^2)$. In the Coulomb gauge, odd powers of $m$ can even appear~\cite{luscherLW3}, but not so in the covariant gauge. The reason is that the Coulomb propagator contains $1/\myvec{k}^2$ divergences for $\myvec{k}\rightarrow0$, while the covariant propagator only has $1/k^2$ poles. In any case, $\tilde{m}^{(1)}_{{\rm A}^{+}}/m$ being of the form \refnn{mAsmallm} is a very good global check against computational errors.

Following ref.~\cite{luscherLW3}, we decided {\em not}\ to perform the small-$m$ expansion of eqs. \refnn{pigh1twist}--\refnn{piWtwist} analytically. Instead, we computed the sum of Feynman diagrams numerically for a number of values of $L$, and fitted the results to the expected form \refnn{mAsmallm}. This is much alike to what we did in the previous section to compute the coefficients appearing in \refeq{DV3struct}. However, a difference of practical importance is that the CPU time needed for the evaluation of $\tilde{m}^{(1)}_{{\rm A}^{+}}$ increases for decreasing $m$ (see below), while for $\Delta \pi_{00}^{V_3}(k)$ it is independent of the fit variable $k$. Another difference is that $\tilde{m}^{(1)}_{{\rm A}^{+}}$ depends on $N$. Due to the twist it is difficult to unravel this $N$ dependence analytically, and like in ref.~\cite{luscherLW3} we did separate computations for $N=2$ and $N=3$.

We did not use entirely the same approach as in ref.~\cite{luscherLW3}. The most important differences are:
\begin{enumerate}
\item We used covariant instead of Coulomb gauge fixing.
\item L\"{u}scher and Weisz fully automized the generation of vertex subprograms. That is, they wrote meta-programs that take a Wilson loop as input, and give vertex subprograms as output. We wrote the subprograms by hand, using the vertices that we had already used before for the computation of the static quark potential. As mentioned in section~\ref{sec-lattpert}, the generation of the vertices themselves was automized by means of Mathematica~\cite{wolfram}. As a precaution against programming errors, we always performed a number of numerical tests of our subprograms against the corresponding Mathematica representations.

A further technical difference is that L\"{u}scher and Weisz used PL/I~\cite{fike} for the generation of vertices.
\item L\"{u}scher and Weisz followed a rather unconventional way to extract the coefficients $a_0$ and $a_1$. For example, to find $a_0$ they constructed from the data for, say, $L=6,8,\cdots30$ a new data series for $L=8,10,\cdots28$ that is improved in the sense that the $m^2$ term in \refeq{mAsmallm} is cancelled, leaving only $\cO(m^4)$ deviations. This procedure was then iterated. We preferred doing a least-squares fit, which for our data appeared to give somewhat more stable results.
\end{enumerate}

Once the vertex (and propagator) subprograms are ready, the programming of eqs. \refnn{pigh1twist}--\refnn{piWtwist} is easy. The main point of interest is the integration routine, since it illustrates once more the merit of the twisted tube. One should appreciate that the integration routine is required to be extraordinary accurate, because many digits are lost when extracting the coefficients, in particular $a_1$. We required a relative accuracy of $10^{-13}$ or better. Whether or not an integration routine is capable of attaining such a high accuracy in a reasonable amount of time depends on the the number of integration variables and the smoothness of the integrand. Now for the twisted tube the `integration' symbol $\$$, \refeq{defdollartwisted}, involves only two integration variables, $k_0$ and $k_3$. The components $k_1$ and $k_2$ are to be summed over, costing a rather cheap\footnote{Nevertheless this factor is, together with stability considerations, the reason why very small values of $m$ cannot be reached.} factor of $(N^2-1)L^2$. More importantly, the integrand is periodic (over one Brillouin zone) and analytic on the domain of integration of $k_0$ and $k_3$ (due to the mass gap, poles are shifted into the complex plane). Such a situation is ideal for constructing efficient integration routines, especially if one performs a change of variables~\cite{luscherLW3} that shifts the pole further away from the domain of integration. As a result, we typically needed only $50^2$ points to approximate $\int_{k_1,k_2}$ within the required accuracy (L\"{u}scher and Weisz quote 32 rather than 50, apparently due to the different pole structure in the Coulomb gauge, or a more efficient change of variables).
  
Concerning the analyticity of the integrand in $k_0$ and $k_3$, we stress that the implementation of momentum conservation for diagrams with two internal propagators is to be chosen wisely. The reason is as follows. Due to tree-level on-shellness, the external component $k_0$ of the momentum $k=(k_0,0,m,0)$ is imaginary, and approximately equal to $i\, m$. Hence if one took $k'_0,k'_3\in\real$ as the integration variables, the $k''$ propagator would be singular for $\{k'_0\approx k'_3\approx0,\  k'_1=k'_2=-m\}$ since in that case $k''\approx(-im,m,0,0)$ and $k^{''2}\approx0$. It is better to shift $k'_0=\tilde{k}_0-\half k_0$ (hence $k''_0=-\tilde{k}_0-\half k_0$), where $\tilde{k}_0$ (together with $k'_3$) is the $\real$-valued integration variable.

Our results for individual lattice sizes are summarized in table~\ref{tab:massdata}. Note that the cancellation of quadratic divergences decreases the accuracy by a factor of $1/m^2$ (i.e.\ $1/(am)^2$ in non-lattice units).
\begin{table}
\begin{center}
{
\small
\begin{tabular}{|r|c|c||c|c|}
\hline
&\multicolumn{2}{c||}{\LW}&\multicolumn{2}{c|}{square}\\
\hline
\multicolumn{1}{|c|}{L}&N=2&N=3&N=2&N=3\\
\hline
4& -0.021583156919197& -0.040661797276715& -0.020085367798214& -0.038849497567066\\
\hline
6& -0.019599952046337& -0.040131915718096& -0.018908108548525& -0.039452155788701\\
\hline
8& -0.018461969616743& -0.039274516206705& -0.018063236142647& -0.038897070150935\\
\hline
10& -0.017889223811184& -0.038810590019211& -0.017632105742522& -0.038571321147693\\
\hline
12& -0.017569982136115& -0.038547675035293& -0.017390811874792& -0.038382470120619\\
\hline
14& -0.017375041457146& -0.038385977285131& -0.017243147194999& -0.038265025623147\\
\hline
16& -0.017247595811504& -0.038279865552201& -0.017146489209537& -0.038187471819923\\
\hline
18& -0.017159817570292& -0.038206617036467& -0.017079864357098& -0.038133728057528\\
\hline
20& -0.017096835258186& -0.038153983851014& -0.017032035100709& -0.038095009442688\\
\hline
22& -0.017050132787137& -0.038114916592240& -0.016996555876247& -0.038066217556170\\
\hline
24& -0.017014554076415& -0.038085133354648& -0.016969519951262& -0.038044238228378\\
\hline
26& -0.016986831369867& -0.038061914071922& {\rm ---} & {\rm ---} \\
\hline
\end{tabular}
}
\caption{Raw data for $\tilde{m}^{(1)}_{{\rm A}^{+}}/m$. For each entry the absolute accuracy is estimated to be $(\frac{NL}{2\pi})^2\cdot10^{-14}$.}
\label{tab:massdata}
\end{center}
\end{table}
As mentioned above, we used least-squares fits to determine the coefficients $a_0$ and $a_1$. We checked that the data is consistent with the absence of odd terms in $m$, and also with the absence of terms $\ln m$ and $m^2 \ln m$. We then fitted the data to the form $a_0 + a_1 m^2 + a_2 m^4 + b_2 m^4 \ln m + a_3 m^6 + b_3 m^6 \ln m$. Our fits strongly suggest that $b_2 = 0$, but we have been unable to find a rigorous analytic proof for this.

We found the most accurate results by using a minimal set of points for the fit, $L=16\cdots26$ for the \LW\ action, and $L=14\cdots24$ for the square action. The disadvantage of this approach is that the error estimate of the coefficients is necessarily ad hoc. As error estimate of $a_i$ ($i=0,1$) we typically used $a_i-\bar{a}_i$, where $\bar{a}_i$ was obtained by dropping either $a_3$ or $b_3$ as a fit parameter. The justification for this procedure is that that the coefficients $a_3$ and $b_3$ themselves can barely be determined from our data. As an a posteriori justification, the results below show the error estimate to be realistic whenever a comparison can be made to other results.

Using this method we obtain
\bea
a_0^{\rm LW} &=& \left\{\begin{array}{ll}
-0.0168265791(7) & {\rm for\ }N=2\\
-0.0379274963(15) & {\rm for\ } N=3, \end{array} \right.
\label{eq:a0LW}\\
a_1^{\rm LW} &=& \left\{\begin{array}{ll}
-0.01100890(15) & {\rm for\ }N=2\\
-0.0208015(5) & {\rm for\ } N=3, \end{array} \right.
\label{eq:a1LW}\\
a_0^{\rm sq} &=& \left\{\begin{array}{ll}
-0.0168265790(15) & {\rm for\ }N=2\\
-0.037927497(2) & {\rm for\ } N=3, \end{array} \right.
\label{eq:a0sq}\\
a_1^{\rm sq} &=& \left\{\begin{array}{ll}
-0.0083763(3) & {\rm for\ }N=2\\
-0.0154489(7) & {\rm for\ } N=3. \end{array} \right.
\label{eq:a1sq}
\eea
Note that the continuum coefficient $a_0$ is independent of the action chosen, as it should be\footnote{The discrepancy between our result for $a_0(N=3)$ and the value quoted by L\"{u}scher and Weisz is due to a misprint~\cite{luscherPRIV} in ref.~\cite{luscherLW3}.}. Moreover, using \refeq{imprcondmA+one} we see that the values for $a_1^{\rm LW}$ are consistent with the somewhat more accurate values\footnote{L\"{u}scher and Weisz gained two digits by doing a mass calculation in three compact dimensions. Unfortunately this setting is unsuitable for the computation of the other improvement relation by use of scattering theory.} obtained by L\"{u}scher and Weisz~\cite{luscherLW3} (these are copied in \refeq{at1LWresult}). The values for $a_1^{\rm sq}$ agree with our static quark results, \refeq{at1sqresult}.

We would like to report that for the LW~action we have also done a computation of $m_{{\rm A}^{-}}^{(1)}$ (which couples to $\pi_{33}(k_0;0,m,0)$), for $N=3$. As expected, we found a continuum coefficient different from $a_0({\rm A}^{+})$, namely $a_0({\rm A}^{-})=0.0000679470(6)$. A structural check on Symanzik improvement is that $m_{{\rm A}^{-}}^{(1)}$ can be improved simultaneously with $m_{{\rm A}^{+}}^{(1)}$ (and the static quark potential), as we found $a_1^{\rm LW}({\rm A}^{-}) = -0.0208011(3)$. For $N=2$ we did a run for the Wilson action, and found $a_0({\rm A}^{-})=0.00006029(14)$. Our results for $a_0({\rm A}^{\pm})$ agree with those obtained from dimensional regularization by van~Baal~\cite{PvBspin}, which imply the analytic spin-splitting formulas
\be
a_0({\rm A}^{+})-a_0({\rm A}^{-}) = \left\{\begin{array}{ll}
-\mbox{\large$\frac{1}{6\pi^2}$} & {\rm for\ }N=2\\
-\mbox{\large$\frac{3}{8\pi^2}$}\rule[-1mm]{0mm}{8mm} & {\rm for\ } N=3. \end{array} \right.
\ee

\subsection{Effective coupling constant}
\label{subsec-twisted-coupling}

In ref.~\cite{luscherLW1} L\"{u}scher and Weisz define an effective coupling constant $\lambda$ through
\be
\sqrt{Z(\myvec{k}) Z(\myvec{p}) Z(\myvec{q})} \sum_{j=1}^2 e_j \Gamma_3(k,1;p,2;q,j) = i\lambda f(k,p,q).
\label{eq:deflambda}
\ee
A number of new symbols show up. The external lines $(k,1)$ and $(p,2)$ correspond to on-shell A particles with positive spin, while $(q,e)$ corresponds to an on-shell B particle with positive spin:
\bea
k=(i E(\myvec{k}), \myvec{k});&&\ \myvec{k} = (0,m,ir),
\label{eq:momk}\\
p=(-i E(\myvec{p}), \myvec{p});&&\ \myvec{p} = (m,0,ir),
\label{eq:momp}\\
q=(0, \myvec{q});&&\ \myvec{q} = (-m,-m,-2ir),
\label{eq:momq}\\
e = (0;1,-1,0).&&
\label{eq:pospolB}
\eea
Note that these particles and polarizations are completely physical, as they can be created by Polyakov lines, see section~\ref{sec-backgroundtwist}. From now on $k$, $p$ and $q$ always have the above special meaning. The energy $E$ and wave function renormalization $Z$ were defined in the previous subsection\footnote{Momenta implicitly label the particle type, and hence no additional label for the $Z$~factors is necessary.}. The value of $r$ is defined by $E(\myvec{q})=0$ (we choose $r>0$). Note that at tree level, \refeq{simpleprop} applies to all external propagators, with $\varepsilon_{\mu}=\delta_{\mu,1}$, $\delta_{\mu,2}$ or $e_{\mu}$. In particular $E(\myvec{k}) = r  = \half\sqrt{2} m$ to leading order in $m$ and $g_0$. Furthermore the equalities
\be
E(\myvec{p}) = E(\myvec{k}),\ \ \ Z(\myvec{p}) = Z(\myvec{k})
\label{ZpeqZk}
\ee
hold to all orders in perturbation theory, because $\myvec{k}$, $\myvec{p}$ are related by symmetries of the discretized twisted tube~\cite{luscherLW3}. $\Gamma_3(k,1;p,2;q,j)$ is the three-point function $\langle \tilde{A}_1(k) \tilde{A}_2(p) \tilde{A}_j(q)\rangle$, but dropping the trivial factor $\delta(k+p+q)$ and amputating the external lines in the usual way. Finally, the color factor $f(k,p,q)$ is defined in appendix~\ref{app-twisted}. The coupling $\lambda$ is a suitable parameter for on-shell improvement, because its square is proportional to the residue of the pole in the scattering amplitude for $({\rm A}^{+},{\rm A}^{+})\rightarrow({\rm A}^{+},{\rm A}^{+})$, appearing due to ${\rm B}^{+}$~exchange~\cite{luscherLW1}.

We expand
\be
\lambda = g_0 \left\{ \lambda^{(0)} + g_0^2 \lambda^{(1)} + \cO(g_0^4) \right\}.
\label{eq:explambda}
\ee
The computation of the tree-level value $\lambda^{(0)}$ simply amounts to substituting the momenta $k$, $p$ and $q$ in the relevant expressions in appendices~\ref{app-prop}, \ref{app-vert} and \ref{app-twisted} (only $V^{(i=2)}_{\mu\nu\rho}$ is not listed, but it can be looked up in ref.~\cite{weiszII}). The result reads
\bea
&&Z_0(\myvec{k}) = 1-\left( \tilde{c}_1 - c_2 + \twelfth \right) m^2 + \cO(m^4),
\label{eq:Z0k}\\
&&Z_0(\myvec{q}) = 1+\left( \tilde{c}_1 - c_2 \right) m^2 + \cO(m^4),
\label{eq:Z0q}\\
&&\sum_j e_j \Gamma_3^{\rm tree}(k,1;p,2;q,j) = -2 f(k,p,q) g_0 \sum_j  e_j V_{12j}(k,p,q)\nonumber\\
&&  \hspace{1.5cm}= i f(k,p,q)  \times (-8 g_0 m)\!\left[ 1\! - \left(\! 4 \tilde{c}_1 \! - 3 c_2 +\breuk{7}{24} \right) m^2 \! + \cO(m^4) \right]\!.
\label{eq:Gam30}
\eea
It immediately follows that~\cite{luscherLW1}
\be
\lambda^{(0)} = -8 m \left\{ 1- \half m^2 \left[ 9\left( \tilde{c}_1 - c_2 + \twelfth \right) + 2c_2 \right] + \cO(m^4) \right\},
\label{eq:lambda0res}
\ee
so that $c_2 = 0$ is indeed the second on-shell improvement condition at tree-level (\refeq{onshelltree}). From now on we assume $c_2=0$ and $\tilde{c}_1 -c_2 = -\twelfth$.

At one-loop level the following formula can be derived from \refeq{deflambda} and the expressions given in the previous subsection (which can easily be extended to the ${\rm B}^{+}$~particle):
\bea
\frac{\lambda^{(1)}}{m} &=& \left(1-\breuk{1}{24} m^2 \right) \frac{\Gamma^{(1)}}{m} - \left. \frac{4}{k_0} \frac{d}{d k_0} \pi_{11}(k) \right|_{k_0 = i E(\myvec{k})}\nonumber\\
&& -2 \left(1-\twelfth m^2 \right) \left. \frac{d^2}{d q_0^2} \left(\half\sum_{i,j}e_i e_j \pi_{ij}(q) \right)\right|_{q_0=0} + \cO(m^4),
\label{eq:lambda1}
\eea
where $\Gamma^{(1)}$ is defined through
\be
\sum_{j=1}^2 e_j \Gamma_3(k,1;p,2;q,j) = i g_0 f(k,p,q) \left\{ \Gamma^{(0)} + g_0^2 \Gamma^{(1)} + \cO(g_0^4) \right\}.
\label{eq:defGam1}
\ee
In the above equations we may use tree-level expressions for the external momenta, i.e.\ we are free to redefine
\bea
k=(i E_0(\myvec{k}), \myvec{k});&&\ \myvec{k} = (0,m,ir_0),
\label{eq:momk0}\\
p=(-i E_0(\myvec{p}), \myvec{p});&&\ \myvec{p} = (m,0,ir_0),
\label{eq:momp0}\\
q=(0, \myvec{q});&&\ \myvec{q} = (-m,-m,-2ir_0),
\label{eq:momq0}
\eea
where $r_0>0$ is the solution to $E_0(\myvec{q})=0$. This redefinition brings about only $g_0^4$ corrections to \refeq{explambda}. One should however be aware of the fact that in the derivation of \refeq{lambda0res} we used eqs.~\refnn{momk0}--\refnn{momq0} rather than eqs.~\refnn{momk}--\refnn{momq}. A priori this can cause corrections to $\lambda^{(1)}$. However, we have checked that all such corrections are at least of order $m^4$. For example in the second term of \refeq{Zoneloopgen}, $E_0$ and (at least for a physical polarization) $E_1$ are of order $m$, while $((2k_0)^{-1}d/d k_0)^2 d_{(0)}^{-1}$ is of order $m^2$ due to tree-level improvement.

In terms of Feynman diagrams, $\Gamma^{(1)}$ is represented by
\bea
\threegltot &=&\! \threeglone +\! \threegltwoA +\! \threegltwoB +\! \threegltwoC \nonumber\\
&&\nonumber\\
&&\nonumber\\
&& +\! \threeglthreeB +\! \threeglthreeA +\! \threeglfourVA +\! \threeglfourVB \nonumber\\
&&\nonumber\\
&&\nonumber\\
&& +\! \threeglfourVC +\! \threeglfourWA +\! \threeglfourWB +\! \threeglfourWC \nonumber\\
&&\nonumber\\
&&\nonumber\\
&& +\! \threeglfive +\! \threeglsix.
\label{eq:threegltot}\\
&&\nonumber\\
&&\nonumber
\eea
Correspondingly we write
\bea
\Gamma^{(1)} &=& \Gamma' + \Gamma^{\rm gh1} + \Gamma^{\rm gh2} + \Gamma^{\rm gh3} + \Gamma^{\rm gh4} + \Gamma^{\rm gh5}\nonumber\\
&& + \Gamma^{V_4 1} + \Gamma^{V_4 2} + \Gamma^{V_4 3} + \Gamma^{W1} + \Gamma^{W2} + \Gamma^{W3} + \Gamma^{V_3} + \Gamma^{V_5},
\label{eq:Gamcomps}
\eea
where
\bea
\Gamma' & = & 2 i \sum_j e_j \sum_i c'_i V^{(i)}_{12j}(k,p,q),
\label{eq:Gamins}\\
\Gamma^{\rm gh1} & = & 0,
\label{eq:Gamgh1}\\
\Gamma^{\rm gh2} & = & \breuk{1}{12} \dollar{k',k''}\, \delta(k'+k''-k) G_{k,p,q}^{k',k'',k'} \frac{\hat{k}'_2\hat{k}''_2\stackrel{\myhat}{(k'-k'')}_1}{\hat{k}^{'2}\hat{k}^{''2}},
\label{eq:Gamgh2}\\
\Gamma^{\rm gh3} & = & \Gamma^{\rm gh2},
\label{eq:Gamgh3}\\
\Gamma^{\rm gh4} & = & -\!\dollar{k',k'',k'''} \delta(k+k'\!-k''')\delta(p+k''\!-k') G_{k,p,q}^{k',k'',k'''} \frac{\hat{k}'_1 \hat{k}''_2 c'''_1 c'_2 (\hat{k}'''_1 c''_1 \!-\! \hat{k}'''_2 c''_2)}{\hat{k}^{'2}\hat{k}^{''2}\hat{k}^{'''2}}, \label{eq:Gamgh4}\\
\Gamma^{\rm gh5} & = & \Gamma^{\rm gh4},
\label{eq:Gamgh5}\\
\Gamma^{V_4 1} & = & \sixth i \sum_j e_j \dollar{k',k''}\, \delta(k'+k''-q) D_{\lambda\lambda'}(k') D_{\rho\rho'}(k'') V_{\lambda\rho j}(-k',-k'',q) \nonumber\\
&&\times \left\{ g_{-}^2(k',k'') \tilde{V}_{12\lambda'\rho'}(k,p,k',k'') - 2 G_{k,p,q}^{k',k'',k'} \tilde{V}_{1\lambda'2\rho'}(k,k',p,k'') \right\}\!,
\label{eq:GamV41}\\
\Gamma^{V_4 2} & = & -\sixth i \sum_j e_j \dollar{k',k''}\, \delta(k'+k''-k) D_{\lambda\lambda'}(k') D_{\rho\rho'}(k'') V_{\lambda\rho 1}(-k',-k'',k) \nonumber\\
&&\times \left\{ g_{-}^2(k',k'') \tilde{V}_{j2\lambda'\rho'}(q,p,k',k'') + 2 G_{k,p,q}^{k',k'',k'} \tilde{V}_{j\lambda'2\rho'}(q,k',p,k'') \right\}\!,
\label{eq:GamV42}\\
\Gamma^{V_4 3} & = & \Gamma^{V_4 2},
\label{eq:GamV43}\\
\Gamma^{W1} & = & \breuk{1}{48} i \sum_j e_j \dollar{k',k''} \delta(k'\!+k''\!-q) D_{\lambda\lambda'}(k') D_{\rho\rho'}(k'') V_{\lambda\rho j}(-k',-k'',q) \frac{g_{-}(k',k'')}{g_{-}(k,p)} \nonumber\\
&&\times\! \left\{ g_{+}(k',k'')g_{+}(k,p) + (k'\!\!\leftrightarrow\! k) + (k'\!\!\leftrightarrow\! p) \right\} W_{12\lambda'\rho'}(k,p,k',k''),
\label{eq:GamW1}\\
\Gamma^{W2} & = & \breuk{1}{48} i \sum_j e_j \dollar{k',k''} \delta(k'\!+k''\!-k) D_{\lambda\lambda'}(k') D_{\rho\rho'}(k'') V_{\lambda\rho 1}(-k',-k'',k) \frac{g_{-}(k',k'')}{g_{-}(k,p)} \nonumber\\
&&\times\! \left\{ g_{+}(k',k'')g_{+}(q,p) + (k'\!\!\leftrightarrow\! q) + (k'\!\!\leftrightarrow\! p) \right\} W_{j2\lambda'\rho'}(q,p,k',k''),
\label{eq:GamW2}\\
\Gamma^{W3} & = & \Gamma^{W2},
\label{eq:GamW3}\\
\Gamma^{V_3} & = & -i \sum_j e_j\dollar{k',k'',k'''} \delta(k+k'\!-k''')\delta(p+k''\!-k') D_{\lambda\lambda'}(k') D_{\rho\rho'}(k'') D_{\tau\tau'}(k''') \nonumber\\
&& \times G_{k,p,q}^{k',k'',k'''} V_{\lambda'\tau 1}(k',-k''',k) V_{\rho'\lambda 2}(k'',-k',p) V_{\tau'\rho j}(k''',-k'',q),
\label{eq:GamV3}\\
\Gamma^{V_5} & = & \breuk{1}{24} i \sum_j e_j \,\dollar{k'}\, D_{\lambda\rho}(k')\chi_{k'} \left\{ \rule{0mm}{4mm} \left[ V_{12j\lambda\rho}(k,p,q,k',-k') + \mbox{2 cyclic\ perms} \right] \right.\nonumber\\
&& \left. +\mbox{$\frac{1}{g_{-}(k,p)}$} \left[ g_{-}(k,p+2k')V_{1\lambda2j\rho}(k,k',p,q,-k')  + \mbox{2 cyclic p.}\right]\right\}.
\label{eq:GamV5}
\eea
In the latter equation, the cyclic permutations act on $(k,1)$, $(p,2)$ and $(q,j)$. Furthermore, $g_{-}$ is defined in \refeq{defg-} and
\bea
g_{+}(k_1,k_2) &\equiv& 2 \cos\left(\frac{\pi}{N}\bra{k_1}{k_2}\right),
\label{eq:defg+}\\
G_{k_1,k_2,k_3}^{k_4,k_5,k_6} &\equiv& \frac{g_{-}(k_1,k_4)g_{-}(k_2,k_5) g_{-}(k_3,k_6)}{g_{-}(k_1,k_2)},
\label{eq:defG}\\
\tilde{V}_{\mu_1\mu_2\mu_3\mu_4}(k_1,k_2,k_3,k_4) &\equiv& V_{\mu_1\mu_2\mu_3\mu_4}(k_1,k_2,k_3,k_4) - V_{\mu_2\mu_1\mu_3\mu_4}(k_2,k_1,k_3,k_4).\label{eq:defVtilde}
\eea
The equality of various Feynman diagrams is due to lattice symmetries and properties of $k,p$ and $q$. We checked that in leading order in $m$ the expressions reduce to their infinite volume counterparts, cf.\ the discussion below \refeq{defg-}. Numerically we found that the $\Gamma^{W}$ contributions are at least of order $m^4$.

The analytic expressions for the second and third terms in \refeq{lambda1} can be found by differentiating eqs.~\refnn{piprimetwist}--\refnn{piWtwist} (the differentiation can be brought over to the integrands without problem). We checked the corresponding computer programs against our old programs for $\pi_{\mu\nu}(\tilde{k})$ by running the latter for near values of $\tilde{k}_0$.

The diagrams proportional to $c'_i$ can be calculated analytically. We separate their total contribution (which can be read off from \refeq{lambda0res}):
\be
\frac{\lambda^{(1)}}{m} = \frac{\tilde{\lambda}^{(1)}}{m} + 4 m^2\left[9(\tilde{c}'_1 - c'_2) + 2 c'_2 \right] + \cO(m^4).
\label{eq:deflamt1}
\ee
Like $\tilde{m}^{(1)}_{{\rm A}^{+}}$, $\tilde{\lambda}^{(1)}$ is to be evaluated numerically for a number of lattice sizes. The remarks in the previous subsection concerning the integration routine are also valid in the present case. Thus the translation of the analytic expressions to computer programs is straightforward. The resulting data is listed in table~\ref{tab:lamdata}.
\begin{table}
\begin{center}
{
\small
\begin{tabular}{|r|c|c||c|c|}
\hline
&\multicolumn{2}{c||}{\LW}&\multicolumn{2}{c|}{square}\\
\hline
\multicolumn{1}{|c|}{L}&N=2&N=3&N=2&N=3\\
\hline
4& -0.78341711803619& -1.47968692446631& -0.79859872384707& -1.51787874584542\\
\hline
6& -0.99849061664701& -1.78876943596789& -1.01681082402891& -1.81613837864479\\
\hline
8& -1.13963010017710& -1.98424388671692& -1.15382728105870& -2.00463419279200\\
\hline
10& -1.24091978060661& -2.12599811096867& -1.25207004886790& -2.14258204874697\\
\hline
12& -1.31941153232092& -2.23737552735037& -1.32855710935494& -2.25172969936027\\
\hline
14& -1.38345219535201& -2.32928849922146& -1.39125843561358& -2.34224049339430\\
\hline
16& -1.43757858326173& -2.40765025540042& -1.44445943234635& -2.41966813031578\\
\hline
18& -1.48449250619699& -2.47601715882150& -1.49071210672691& -2.48738335434952\\
\hline
20& -1.52592324035698& -2.53669577194007& -1.53165594879818& -2.54759007845263\\
\hline
\end{tabular}
}
\caption{Raw data for $\tilde{\lambda}^{(1)}/m$. For each entry the absolute accuracy is estimated to be $10^{-13}$. Note that the inexact formula \refnn{lambda1} was used, so that the data has systematic deviations of order $m^4$.}
\label{tab:lamdata}
\end{center}
\end{table}

The data is expected to be of the form~\cite{luscherLW3} (with new coefficients $a_i$ and $b_i$)
\be
\frac{\tilde{\lambda}^{(1)}}{m} = a_0 + b_0 \ln m + a_1 m^2 + a_2 m^4 + b_2 m^4 \ln m+ a_3 m^6 + b_3 m^6 \ln m + \cO(m^8).
\label{eq:lamt1exp}
\ee
Indeed it can be checked that our data is consistent with the absence of odd powers in $m$. Also we checked the absence of an $m^2 \ln m$ term, expected due to tree-level improvement. Furthermore, since  $\lim_{m\rightarrow0}(\lambda/m)_{\rm tree~level} = -8 g_0$ and $\lambda$ is a renormalizable parameter, $b_0$ should equal
\be
b_0 = 8\beta_0 = \frac{11 N}{6 \pi^2}.
\label{eq:valueb0}
\ee
The fit \refnn{lamt1exp} (dropping $a_3$ or $b_3$) to our data for $L=10\cdots20$ reproduces this value to six  digits. This a non-trivial check against programming errors (and other errors)\footnote{Useful intermediate checks can be obtained by extracting the coefficients of the $m^0 \ln m$ terms for individual diagrams, and comparing them to the $1/(4-d)$ poles of their dimensionally regularized continuum counterparts.}. Moreover, by fixing $b_0$ to have the exact value \refnn{valueb0} the other coefficients can be fitted to a higher accuracy. In this way we obtain
\bea
a_0^{\rm LW} &=& \left\{\begin{array}{ll}
-0.84832346(3) & {\rm for\ }N=2\\
-1.28773532(5) & {\rm for\ } N=3, \end{array} \right.
\label{eq:a0gLW}\\
a_1^{\rm LW} &=& \left\{\begin{array}{ll}
0.419861(6) & {\rm for\ }N=2\\
0.78417(3) & {\rm for\ } N=3, \end{array} \right.
\label{eq:a1gLW}\\
a_0^{\rm sq} &=& \left\{\begin{array}{ll}
-0.85183887(3) & {\rm for\ }N=2\\
-1.29656105(4) & {\rm for\ } N=3, \end{array} \right.
\label{eq:a0gsq}\\
a_1^{\rm sq} &=& \left\{\begin{array}{ll}
0.324745(5) & {\rm for\ }N=2\\
0.59095(2) & {\rm for\ } N=3, \end{array} \right.
\label{eq:a1gsq}
\eea
where the error-estimate procedure described in the previous subsection was used.

The LW coefficients agree with those obtained by L\"{u}scher and Weisz~\cite{luscherLW3}: $a_0^{\rm LW}(N=2) = -0.8483231(3)$, $a_0^{\rm LW}(N=3) = -1.2877352(1)$, $a_1^{\rm LW}(N=2) = 0.41988(3)$, $a_1^{\rm LW}(N=3) = 0.78412(5)$. The $a_0$ coefficient for the square and LW actions differ. This is a renormalization effect. Like in subsection~\ref{subsec-statquark-gen} one can see that the $a_0$ coefficient is related to the Lambda parameter:
\be
\frac{\Lambda^{*}}{\Lambda} = e^{\frac{1}{b_0}\left(a_0(\{c_i^{*}\}) - a_0(\{c_i\})\right)}.
\label{eq:Lamratiolam}
\ee
We thus find
\be
\frac{\Lambda_{\rm LW}}{\Lambda_{\rm sq}} = \left\{\begin{array}{ll}
1.0095074(2) & {\rm for\ }N=2\\
1.0159636(2) & {\rm for\ } N=3, \end{array} \right.
\label{eq:LamLWsqlamres}
\ee
in complete agreement with \refeq{Lamresults}. Incidentally, for $N=2$ we also did a run for the Wilson action and found
\be
a_0^{\rm W}(N=2) = -1.37530949(10),
\label{eq:a0gW}
\ee
so that
\bea
\frac{\Lambda_{\rm LW}}{\Lambda_{\rm W}} = \begin{array}{ll}
4.1308934(14) & {\rm for\ }N=2, \end{array}\\
\frac{\Lambda_{\rm sq}}{\Lambda_{\rm W}} = \begin{array}{ll}
4.0919894(14) & {\rm for\ }N=2. \end{array}
\label{eq:Lamlamres}
\eea

After renormalization one reads off the improvement condition:
\be
4 \left[9(\tilde{c}'_1 - c'_2) + 2 c'_2 \right] = -a_1.
\label{eq:impr1lam}
\ee
Hence together with the results from the previous subsection we have completely determined the one-loop Symanzik coefficients for the square (and \LW) action.

We are convinced of the correctness of our new result \refnn{a1gsq} because of the many internal checks in our computation. In particular, it is extremely implausible that an expansion giving the correct values for the leading coefficients $b_0$ and $a_0$ (as verified from the Lambda ratios) as well as for the subleading coefficient $b_1$, would produce an incorrect value for the other subleading coefficient $a_1$. The fact that for the LW action all results by L\"{u}scher and Weisz have been reproduced, confirms our confidence to a large extent.

\section{Summary}
\label{sec-summary}

In this paper we computed Lambda ratios, one-loop Symanzik coefficients and the tadpole parameter for the \LW\ and square actions. The results are summarized in table~\ref{tab:coeff}. We remind the reader that coefficient combinations other than $\tilde{c}_1(g_0^2)\equiv c_1(g_0^2)+4c_4(g_0^2)$ and $c_2(g_0^2)$ are unimportant because only $\tilde{c}_1(g_0^2)$ and $c_2(g_0^2)$ couple to the $a^2$ corrections of any on-shell quantities (see section~\ref{sec-onshell}). 
\begin{table}[htb]
\begin{center}
\begin{tabular}{|c|r@{.}ll||r@{.}ll|}
\hline
&\multicolumn{3}{c||}{\LW}&\multicolumn{3}{c|}{square}\\
\hline
$\tilde{c}_0'$ & 0&135160(13) & $(N=2)$ & 0&113417(11) & $(N=2)$\\
               & 0&23709(6)   & $(N=3)$ & 0&19320(4) &   $(N=3)$\\
\hline
$\tilde{c}_1'$ & -0&0139519(8) & $(N=2)$ & -0&0112766(7) & $(N=2)$\\
               & -0&025218(4)  & $(N=3)$ & -0&019799(2)  & $(N=3)$\\
\hline
$c_2'$ & -0&0029431(8) & $(N=2)$ & -0&0029005(7) & $(N=2)$\\
               & -0&004418(4)  & $(N=3)$ & -0&004351(2) &  $(N=3)$\\
\hline
$\Lambda/\Lambda_{\mbox{\scriptsize Wilson}}$ & 4&1308935(3) & $(N=2)$ & 4&0919901(2) & $(N=2)$\\
               & 5&2921038(3) & $(N=3)$ & 5&2089503(2) & $(N=3)$\\
\hline
$\WWsym(1,1)$     & \multicolumn{3}{c||}{0.366262680(2)} & \multicolumn{3}{c|}{0.3587838551(1)}\\
\hline
$\WWsym(1,2)$     & \multicolumn{3}{c||}{0.662626785(2)} & \multicolumn{3}{c|}{0.6542934512(1)}\\
\hline
$\WWsym(2,2)$     & \multicolumn{3}{c||}{1.098143594(2)} & \multicolumn{3}{c|}{1.0887235337(1)}\\
\hline
\end{tabular}
\caption{One-loop improvement coefficients, Lambda parameter ratios and one-loop expectation values of small Wilson loops for the \LW\ and square Symanzik actions. For quantities that were extracted both from the static quark potential and from the twisted spectroscopy, we used the most accurate result. In particular we made use of ref.~\protect\cite{luscherLW3} for the combination $\tilde{c}_1-c_2$, cf.\ subsection~\protect\ref{subsec-statquark-part}, which we verified to somewhat lower accuracy in subsection~\protect\ref{subsec-twisted-mass}. For the definition of $\protect\WWsym(L,T)$ and its relation to the tadpole parameter, see eqs.~\protect\refnn{defwn}, \refnn{u0def} and \refnn{w1extractCf}.}
\label{tab:coeff}
\end{center}
\end{table}

For the \LW\ action we found complete agreement with previous calculations~\cite{weiszII,luscherLW3}. In particular we reproduced the values for the Lambda parameter and the one-loop coefficients, as well as for $\WWsym(1,1)$ and $\WWsym(1,2)$ (which are related to the tadpole parameter $u_0$). Especially the agreement with ref.~\cite{luscherLW3} is non-trivial because we used covariant instead of Coulomb gauge fixing.  
For the square action we computed the combination $\tilde{c}'_1 - c'_2$ and the Lambda parameter in two independent ways, namely using the static quark potential and finite volume spectroscopy. The Lambda parameter was also computed in ref.~\cite{PvBLamsq} from a background field method. The agreement between all results leaves us with no doubt that our values for $(\tilde{c}'_1 - c'_2)_{\rm square}$ and $\Lambda_{\rm square}$ are correct. We are also convinced of the correctness of our result for the combination $(9(\tilde{c}'_1 - c'_2) + 2c'_2)_{\rm square}$, because it was found from the same numerical data that gave correct results for three verifiable coefficients.

It is interesting that tadpole improvement of the tree-level square action captures 79\% (\SU{3}) to 80\% (\SU{2}) of the one-loop correction to the appropriate ratio of the coefficients $c_i(g_0^2)$~\cite{squarelett}, especially because this is similar to the prediction found for the \LW\ action (76\% resp. 80\%).

\section*{Acknowledgments}

The author wishes to thank Pierre van Baal, Margarita Garc\'{\i}a P\'erez, Martin L\"{u}scher and Peter Weisz for discussions and correspondence.

\appendix

\section{Structure of the Feynman rules}
\label{app-struct}

In this appendix we expand the total action, \refeq{Stotal}, in powers of $g_0$, largely adopting the notation of refs.~\cite{weiszI,weiszII,luscherLW3}. We use Fourier space:
\be
A_{\mu}(x) = g_0 \sum_{b=1}^{N^2 - 1} \int_{k} e^{i k (x+\half a \hat{\mu})} \tA_{\mu}^b(k) T^b,
\label{eq:Fourierperiodic}
\ee
where
\be
\int_{k} \equiv \prod_{\mu = 0}^3 \left(\int_{-\pi/a}^{\pi/a} \frac{dk_{\mu}}{2\pi}\right).
\label{eq:defdollarperiodic}
\ee
It should be noted that the above definitions hold for the infinite-volume case. The modifications for the finite twisted volume are discussed in appendix~\ref{app-twisted}. From now on, summations over \SU{N} and Lorentz indices will be implicit.

The expansion of the measure and ghost actions can be found in refs.~\cite{weiszII,luscherLW3}. For completeness we copy the results\footnote{There is a slight overall difference with ref.~\protect\cite{weiszII} because we explicitly keep a factor $(-1/g_0^2)$ in the path integral, eq.~(\protect\ref{eq:pathint}), instead of absorbing it into the action.}:
\be
\action{measure} = \frac{N}{24} g_0^2 \delta_{\mu_1 \mu_2}\delta_{a_1 a_2} \frac{1}{a^2}\int_{k_1,k_2}\momconsper{k_1+k_2} \tA_{\mu_1}^{a_1}(k_1)\tA_{\mu_2}^{a_2}(k_2) + \cO(g_0^4),
\label{eq:expSmeas}
\ee
\bea
\action{ghost} &=& \int_{k_1,k_2}\bar{\tilde{c}}^{a_1}(k_1) \tilde{c}^{a_2}(k_2) \left[\momconsper{k_1 + k_2} \delta_{a_1 a_2} \hat{k}_1^2 + i g_0 f_{a_1 a_2 a_3} \right. \nonumber\\
&& \times\int_{k_3}\momconsper{k_1 + k_2 + k_3}\tA_{\mu}^{a_3}(k_3) \hat{k}_{1\mu}c_{2\mu} + \frac{1}{12} g_0^2 \delta_{\mu_3 \mu_4} f_{a_1 a_3 e} f_{a_2 a_4 e}\nonumber\\
&& \left. \times\ a^2\!\! \int_{k_3,k_4}\! \momconsper{k_1\! + k_2\! + k_3\! + k_4}\tA_{\mu_3}^{a_3}(k_3) \tA_{\mu_4}^{a_4}(k_4)\hat{k}_{1\mu_3}\hat{k}_{2\mu_3}  + \cO(g_0^4) \right]. \nonumber\\
&& \label{eq:expSghost}
\eea
Here $\tilde{c}^a$, $\bar{\tilde{c}}^a$ are the Fourier-transformed ghost fields. Also we adopted the conventional notation $\hat{k}^2 = \sum_{\mu=0}^3 \hat{k}_{\mu}^2$ and
\bea
\hat{k}_{\mu} &=& \mbox{\small $\frac{2}{a}$}\sin(\mbox{\small $\frac{a}{2}$} k_{\mu}),\nonumber\\
c_{\mu} &=& \cos(\mbox{\small $\frac{a}{2}$} k_{\mu}).
\label{eq:khatandc}
\eea
The difference between $c_{\mu}$ and the action coefficients $c_i$ should always be clear from the context.

The remaining part of $\action{total}/g_0^2$ is expanded as follows:
\be
\frac{1}{g_0^2}(S(\{c_i(g_0^2)\}) + \action{gf}) = \sum_{n=2}^{\infty} \frac{g_0^{n-2}}{n!} S_n(\{c_i(g_0^2)\}).
\label{eq:expSrest}
\ee
$S_n(\{c_i(g_0^2)\})$ is linear in $c_i(g_0^2)$ and therefore we have
\be
S_n(\{c_i(g_0^2)\}) = S_n(\{c_i\}) + g_0^2 (S_n(\{c'_i\})-S_n(\{0\})) + \cO(g_0^4).
\label{eq:insertions}
\ee
In sections~\ref{sec-statquark} and~\ref{sec-twisted} the terms $g_0^2 (S_n(\{c'_i\})-S_n(\{0\}))$ are treated as insertions.

Only $S_2$, $S_3$, $S_4$ and $S_5$ are needed in our calculations. These can be written in the following way:
\bea
S_2(\{c_i\}) &=& \delta_{a_1 a_2} \int_{k_1,k_2}\momconsper{k_1+k_2} \tA_{\mu_1}^{a_1}(k_1)\tA_{\mu_2}^{a_2}(k_2) (\prop^{-1})_{\mu1\mu2}(k_1),
\label{eq:expS2}\\
S_3(\{c_i\}) &=& f_{a_1 a_2 a_3} \int_{k_1,k_2,k_3}\momconsper{k_1+k_2+k_3} \nonumber\\ && \times\tA_{\mu_1}^{a_1}(k_1)\tA_{\mu_2}^{a_2}(k_2)\tA_{\mu_3}^{a_3}(k_3) V_{\mu_1\mu_2\mu_3}(k_1,k_2,k_3),
\label{eq:expS3}\\
S_4(\{c_i\}) &=& \int_{k_1,k_2,k_3,k_4}\momconsper{k_1+k_2+k_3+k_4} \tA_{\mu_1}^{a_1}(k_1)\tA_{\mu_2}^{a_2}(k_2) \tA_{\mu_3}^{a_3}(k_3)\tA_{\mu_4}^{a_4}(k_4)\nonumber\\ 
&&\times\left[ f_{a_1 a_2 e}f_{a_3 a_4 e}(V_{\mu_1\mu_2\mu_3\mu_4}(k_1,k_2,k_3,k_4) - V_{\mu_2\mu_1\mu_3\mu_4}(k_2,k_1,k_3,k_4)) \right. \nonumber\\ 
&& - \left. S_{a_1 a_2 a_3 a_4} W_{\mu_1\mu_2\mu_3\mu_4}(k_1,k_2,k_3,k_4) \right],
\label{eq:expS4}\\
S_5(\{c_i\}) &=& C_{a_1 a_2 a_3 a_4 a_5} \int_{k_1,k_2,k_3,k_4,k_5}\momconsper{k_1+k_2+k_3+k_4+k_5} \nonumber\\ && \times\tA_{\mu_1}^{a_1}(k_1)\cdots\tA_{\mu_5}^{a_5}(k_5) V_{\mu_1\mu_2\mu_3\mu_4\mu_5}(k_1,k_2,k_3,k_4,k_5),
\label{eq:expS5}
\eea
where we defined
\bea
V_{\mu_1\mu_2\mu_3} = \sum_i c_i V^{(i)}_{\mu_1\mu_2\mu_3}, &&
V_{\mu_1\mu_2\mu_3\mu_4} = \sum_i c_i V^{(i)}_{\mu_1\mu_2\mu_3\mu_4}, \nonumber\\
W_{\mu_1\mu_2\mu_3\mu_4} = \sum_i c_i W^{(i)}_{\mu_1\mu_2\mu_3\mu_4}, && V_{\mu_1\mu_2\mu_3\mu_4\mu_5} = \sum_i c_i V^{(i)}_{\mu_1\mu_2\mu_3\mu_4\mu_5}.
\label{eq:Vcexplicit}
\eea
The definitions of the color factors appearing in \refeqs{expS4}{expS5} are:
\bea
S_{a b c d} &=& \frac{1}{24} \Tr\left(T^a T^b T^c T^d + \mbox{\ 23 permutations}\right),
\label{eq:Glebsch4sym}\\
C_{a b c d e} &=& \Tr\left( T^a T^b T^c T^d T^e - T^e T^d T^c T^b T^a \right).
\label{eq:Glebsch5}
\eea
For our purposes it is not necessary to work out these factors for general values of the indices. Note that $W_{\mu_1\mu_2\mu_3\mu_4}(k_1,k_2,k_3,k_4)$ is completely symmetric under permutations of $1\cdots4$ and vanishes in the continuum limit. 

The Feynman propagator $D_{\mu\nu}$ and its inverse are discussed in appendix~\ref{app-prop}, while in appendix~\ref{app-vert} the vertices are given. The following parametrization, valid as long as only planar Wilson loops are included in the action, is convenient:
\pagebreak
\bea
&&V_{\mu_1\mu_2\mu_3}(k_1,k_2,k_3) = f^{(3)}_{\mu_1}(k_1,k_2,k_3)\delta_{\mu_1\mu_2\mu_3} \nonumber\\
&&\hspace{2cm} + \left[ g^{(3)}_{\mu_1\mu_3}(k_1,k_2,k_3)\delta_{\mu_1\mu_2} + \mbox{\ 2 cyclic permutations} \right],
\label{eq:dissectV3}\\
&&V_{\mu_1\mu_2\mu_3\mu_4}(k_1,k_2,k_3,k_4) = f^{(4)}_{\mu_1}(k_1,k_2,k_3,k_4)\delta_{\mu_1\mu_2\mu_3\mu_4} \nonumber\\
&&\hspace{2cm} + \left[ g^{(4)}_{\mu_1\mu_4}(k_1,k_2,k_3,k_4)\delta_{\mu_1\mu_2\mu_3} + \mbox{\ 3 cyclic perms} \right] \nonumber\\
&&\hspace{2cm} + h^{(4)}_{\mu_1\mu_2}(k_1,k_2,k_3,k_4)\delta_{\mu_1\mu_3} \delta_{\mu_2\mu_4}  \nonumber\\
&&\hspace{2cm} + \left[ h'^{(4)}_{\mu_1\mu_3}(k_1,k_2,k_3,k_4)\delta_{\mu_1\mu_2} \delta_{\mu_3\mu_4}  + \mbox{\ 1 cyclic perm} \right],
\label{eq:dissectV4}\\
&&W_{\mu_1\mu_2\mu_3\mu_4}(k_1,k_2,k_3,k_4) = f^{(W)}_{\mu_1}(k_1,k_2,k_3,k_4)\delta_{\mu_1\mu_2\mu_3\mu_4} \nonumber\\
&&\hspace{2cm} + \left[ g^{(W)}_{\mu_1\mu_4}(k_1,k_2,k_3,k_4)\delta_{\mu_1\mu_2\mu_3} + \mbox{\ 3 cyclic perms} \right] \nonumber\\
&&\hspace{2cm} + \left[ h^{(W)}_{\mu_1\mu_3}(k_1,k_2,k_3,k_4)\delta_{\mu_1\mu_2} \delta_{\mu_3\mu_4} + (2 \leftrightarrow 3) + (2 \leftrightarrow 4) \right],
\label{eq:dissectW}\\
&&V_{\mu_1\mu_2\mu_3\mu_4\mu_5}(k_1,k_2,k_3,k_4,k_5) = f^{(5)}_{\mu_1}(k_1,k_2,k_3,k_4,k_5)\delta_{\mu_1\mu_2\mu_3\mu_4\mu_5} \nonumber\\
&&\hspace{2cm} + \left[ g^{(5)}_{\mu_1\mu_5}(k_1,k_2,k_3,k_4,k_5)\delta_{\mu_1\mu_2\mu_3\mu_4} + \mbox{\ 4 cyclic perms} \right] \nonumber\\
&&\hspace{2cm} + \left[ h^{(5)}_{\mu_1\mu_2}(k_1,k_2,k_3,k_4,k_5)\delta_{\mu_1\mu_3\mu_5} \delta_{\mu_2\mu_4}  + \mbox{\ 4 cyclic perms} \right] \nonumber\\
&&\hspace{2cm} + \left[ h'^{(5)}_{\mu_1\mu_4}(k_1,k_2,k_3,k_4,k_5)\delta_{\mu_1\mu_2\mu_3} \delta_{\mu_4\mu_5}  + \mbox{\ 4 cyclic perms} \right],
\label{eq:dissectV5}
\eea
where $\delta_{\mu_1\cdots\mu_n} \equiv \delta_{\mu_1\mu_2}\cdots \delta_{\mu_1\mu_n}$. Permutations act simultaneously on Lorentz indices and momenta.
From considerations given in ref.~\cite{luscherLW3} it follows that vertices $V_{\mu_1\cdots\mu_n}$, $n$ even, are real and invariant under inversion of all momenta. For $n$ odd they are imaginary and odd under inversion. Moreover, the components have the following properties with respect to permutations:
\bea
f^{(3)}_{\mu}(k_1,k_2,k_3) &=& -f^{(3)}_{\mu}(k_3,k_2,k_1) = f^{(3)}_{\mu}(k_2,k_3,k_1), \nonumber\\
g^{(3)}_{\mu\nu}(k_1,k_2,k_3) &=& -g^{(3)}_{\mu\nu}(k_2,k_1,k_3), \nonumber\\
f^{(4)}_{\mu}(k_1,k_2,k_3,k_4) &=& f^{(4)}_{\mu}(k_4,k_3,k_2,k_1) = f^{(4)}_{\mu}(k_2,k_3,k_4,k_1), \nonumber\\
g^{(4)}_{\mu\nu}(k_1,k_2,k_3,k_4) &=& g^{(4)}_{\mu\nu}(k_3,k_2,k_1,k_4), \nonumber\\
h^{(4)}_{\mu\nu}(k_1,k_2,k_3,k_4) &=& h^{(4)}_{\nu\mu}(k_4,k_3,k_2,k_1) = h^{(4)}_{\nu\mu}(k_2,k_3,k_4,k_1), \nonumber\\
h'^{(4)}_{\mu\nu}(k_1,k_2,k_3,k_4) &=& h'^{(4)}_{\nu\mu}(k_4,k_3,k_2,k_1) = h'^{(4)}_{\nu\mu}(k_3,k_4,k_1,k_2), \nonumber\\
f^{(W)}_{\mu}(k_1,k_2,k_3,k_4) &=& f^{(W)}_{\mu}(k_{P(1)},k_{P(1)},k_{P(1)},k_{P(1)})\ \mbox{for all perms }P, \nonumber\\
g^{(W)}_{\mu\nu}(k_1,k_2,k_3,k_4) &=& g^{(W)}_{\mu\nu}(k_3,k_2,k_1,k_4) = g^{(W)}_{\mu\nu}(k_2,k_3,k_1,k_4), \nonumber\\
h^{(W)}_{\mu\nu}(k_1,k_2,k_3,k_4) &=& h^{(W)}_{\mu\nu}(k_2,k_1,k_3,k_4) = h^{(W)}_{\mu\nu}(k_1,k_2,k_4,k_3) = \nonumber\\ 
&& h^{(W)}_{\nu\mu}(k_3,k_4,k_1,k_2), \nonumber\\
f^{(5)}_{\mu}(k_1,k_2,k_3,k_4,k_5) &=& -f^{(5)}_{\mu}(k_5,k_4,k_3,k_2,k_1)= f^{(5)}_{\mu}(k_2,k_3,k_4,k_5,k_1), \nonumber\\
g^{(5)}_{\mu\nu}(k_1,k_2,k_3,k_4,k_5) &=& -g^{(5)}_{\mu\nu}(k_4,k_3,k_2,k_1,k_5), \nonumber\\
h^{(5)}_{\mu\nu}(k_1,k_2,k_3,k_4,k_5) &=& -h^{(5)}_{\mu\nu}(k_5,k_4,k_3,k_2,k_1), \nonumber\\
h'^{(5)}_{\mu\nu}(k_1,k_2,k_3,k_4,k_5) &=& -h'^{(5)}_{\mu\nu}(k_3,k_2,k_1,k_5,k_4).
\label{eq:dissectsym}
\eea

\section{Propagator}
\label{app-prop}\vspace{-1mm}

Following the notation of ref.~\cite{weiszI}, the inverse propagator as defined in \refeq{expS2} reads
\be
(\prop^{-1})_{\mu\nu}(k) = \left[\sum_{\rho} q_{\mu\rho}(k)\hat{k}_{\rho}^2 \right] \delta_{\mu\nu} - \left[ q_{\mu\nu}(k) - \frac{1}{\alpha} \right] \hat{k}_{\mu}\hat{k}_{\nu}.\vspace{-1mm}
\label{eq:propinv}
\ee
For the action in \refeq{squareaction} the tensor $q_{\mu\nu}$ equals\vspace{-1mm}
\bea
q_{\mu\nu}(k) &=& (1-\delta_{\mu\nu})s_{\mu\nu}(k), \nonumber\\
s_{\mu\nu}(k) &=& (c_0 + 8 c_1 + 8 c_2 + 16 c_4) - a^2 (c_1 -c_2 + 4c_4) (\hat{k}_{\mu}^2 + \hat{k}_{\nu}^2) \nonumber\\
&& -  a^2 c_2\hat{k}^2 + a^4 c_4 \hat{k}_{\mu}^2 \hat{k}_{\nu}^2).
\label{eq:qgeneral}
\eea
(In view of the discussion in section~\ref{sec-onshell} we do not include $c_3$). Note that, by definition, this tensor factorizes for the square action ($c_2 = 0$, $c_0\, c_4 = c_1^2$).

The inverse of \refeq{propinv} reads
\vspace{-1mm}
\be
\prop_{\mu\nu}(k) = \frac{1}{(\hat{k}^2)^2}\left\{ \left[\sum_{\rho} A_{\mu\rho}(k)\hat{k}_{\rho}^2 \right] \delta_{\mu\nu} - \left[ A_{\mu\nu}(k) - \alpha \right] \hat{k}_{\mu}\hat{k}_{\nu} \right\},\vspace{-3mm}
\label{eq:propgen}
\ee
with $A_{\mu\nu}$ satisfying
\vspace{-2mm}
\be
A_{\mu\mu}(k) = 0\ (\forall \mu),\ \ \ A_{\mu\nu}(k) = A_{\nu\mu}(k) = A_{\mu\nu}(-k),\ \ \ \lim_{a\rightarrow0} A_{\mu\nu}(k) \!\stackrel{\mu\not=\nu}{=}\!\! \frac{1}{s_{\mu\nu}(0)}.\vspace{-2mm}
\label{eq:Aproperties}
\ee
In ref.~\cite{weiszI} the general form of $A_{\mu\nu}$ in terms of $q_{\mu\nu}$ can be found. Using the notation
\vspace{-2mm}
\bea
\hat{k}^{(n)} &=& \sum_{\rho=0}^3 \hat{k}_{\rho}^n, 
\label{eq:hatkn}\\
\tilde{\mu},\ \tilde{\nu} &:& \mbox{such that $\mu\not=\nu\not=\tilde{\mu}\not=\tilde{\nu}$}, 
\label{eq:mutnut}\\
P_{\mu\nu} &=& \hat{k}_{\tilde{\mu}}^2 \hat{k}_{\tilde{\nu}}^2, 
\label{eq:Ptens}\\
Q_{\mu\nu} &=& \hat{k}_{\tilde{\mu}}^2 + \hat{k}_{\tilde{\nu}}^2,
\label{eq:Qtens}
\eea
we here give the explicit formulas for the Wilson (W), \LW\ (LW)~\cite{weiszI,weiszII} and square (sq) cases:
\bea
A^{\mbox{\scriptsize(W)}}_{\mu\nu}(k) &=& 1 - \delta_{\mu\nu},
\label{eq:Awil}
\label{eq:AWil}\\
A^{\mbox{\scriptsize(LW)}}_{\mu\nu}(k) &=&
\frac{1-\delta_{\mu\nu}}{\Delta^{\mbox{\scriptsize(LW)}}} \left(\hat{k}^2 + \twelfth (\hat{k}^{(4)} + \hat{k}^2 \hat{k}_{\tilde{\mu}}^2)\right)
\left(\hat{k}^2 + \twelfth (\hat{k}^{(4)} + \hat{k}^2 \hat{k}_{\tilde{\nu}}^2)\right),
\label{eq:ALW}\\
\Delta^{\mbox{\scriptsize(LW)}} &=& \left( \hk^2 + \twelfth \hk^{(4)} \right) \left[ \hk^2 + \twelfth \left( (\hk^2)^2 + \hk^{(4)} \right) \right. \nonumber\\
&& \left. + \twoeighteighth \left( (\hk^2)^3 - \hk^2 \hk^{(4)} + 2 \hk^{(6)} \right) \right] + \forthreetwoth \hk^2 \prod_{\rho=0}^3 \hk_{\rho}^2,
\label{eq:DelLW}\\
A^{\mbox{\scriptsize(sq)}}_{\mu\nu}(k) &=&
\frac{1-\delta_{\mu\nu}}{\Delta^{\mbox{\scriptsize(sq)}}}
\left\{ \left( \hk^2 + \twelfth \hk^{(4)} \right)
\left[ \hk^2 + \twelfth \hk^{(4)} \right. \right. \nonumber\\
&& \left. + \twelfth \left( \hk^2 - \twentyfourth (\hk^2)^2 + \eighth \hk^{(4)} \right) Q_{\mu\nu} + \seventysecond \hk^2 Q_{\mu\nu}^2 - \fourtyeighth Q_{\mu\nu}^3 \right]  \nonumber\\
&& + P_{\mu\nu} \left[ \onefourfourth \left( -(\hk^2)^2 - \sixth \hk^2 \hk^{(4)} - \onefourfourth (\hk^2)^2 \hk^{(4)} + \seventysecond (\hk^{(4)})^2 + \sixth \hk^{(6)} \right) \right. \nonumber\\
&& \left. + \twentyfourth\! \left( \hk^2 + \eighth \hk^{(4)} + \twoeighteighth \hk^2 \hk^{(4)} \right)\! Q_{\mu\nu} + \eightsixfourth \left( \hk^2 - \sixth \hk^{(4)} \right)\! Q_{\mu\nu}^2 - \twoeighteighth Q_{\mu\nu}^3 \right] \nonumber\\
&& \left. + P_{\mu\nu}^2 \left( \eightsixfourth(-\hk^2 + \sixth \hk^{(4)}) + \onefourfourth Q_{\mu\nu} \right) \right\},
\label{eq:Asq}\\
\Delta^{\mbox{\scriptsize(sq)}} &=& \left( \hk^2 + \twelfth \hk^{(4)} \right)^2 \prod_{\rho=0}^3 \left( 1 + \twelfth \hk_{\rho}^2 \right).
\label{eq:Delsq}
\eea
In these formulas we used lattice units. The $a$ dependence can be reinstated by substituting $\hk \rightarrow a\hk$.

For special momenta and polarizations the propagator greatly  simplifies~\cite{weiszII,luscherLW3}. From \refeqs{propinv}{qgeneral} one deduces that vectors $\varepsilon$ satisfying the following two conditions are eigenvectors of $(\prop^{-1})_{\mu\nu}(k)$.
\begin{enumerate}
\item The gauge condition $\sum_{\mu}\hat{k}_{\mu}\varepsilon_{\mu}=0$ is satisfied;
\item $\exists p\in\real$ (or $\complex$) such that $\forall \mu\in\{\nu|\varepsilon_{\nu}\not=0\}\,\, k_{\mu}\in\{p,-p\}$.
\end{enumerate}
(Solutions to these conditions are for example $\{\varepsilon=(1,0,0,0);\ k=(0,k_1,k_2,k_3)\}$ and $\{\varepsilon=(0,1,-1,0);\ k=(k_0,k_1,k_1,k_3)\}$). Moreover, the eigenvalue equals simply $\sum_{\rho}s_{\bar{\mu}\rho}(k)\hat{k}_{\rho}^2$, where $\bar{\mu}$ is an arbitrary element of $\{\mu|\varepsilon_{\mu}\not=0\}$. Therefore,
whenever the two conditions are satisfied, $\varepsilon$ is an eigenvector of the propagator with eigenvalue
\be
\frac{\sum_{\mu\nu}\varepsilon_{\mu}\prop_{\mu\nu}(k)\varepsilon_{\nu}}{\sum_{\mu} \varepsilon_{\mu}^2} = d_{\bar{\mu}}(k),\ \ \ d_{\bar{\mu}}(k)\equiv\frac{1}{\sum_{\rho}s_{\bar{\mu}\rho}(k)\hat{k}_{\rho}^2}.
\label{eq:simpleprop}
\ee

In section~\ref{sec-twisted} we need the (tree-level) energy $E_0(\myvec{k})$ for $d_1(k)$, as well as the associated wave function renormalization $Z_0(\myvec{k})$. They are defined through (cf.\ subsection~\ref{subsec-twisted-mass})
\be
d_1(k) = \frac{Z_0(\myvec{k})}{k_0^2 + E_0^2(\myvec{k})} + ({\rm regular\ in\ }k_0),
\label{eq:d1aroundpole}
\ee
for $k$ close to the physical mass shell. An equivalent definition of $Z_0(\myvec{k})$ is
\be
Z_0^{-1}(\myvec{k}) = \left.\frac{1}{2 k_0} \frac{\partial}{\partial k_0} d_1^{-1}(k)\right|_{k_0 = \pm i E_0(\myvec{k})}.
\label{eq:Zpract}
\ee
Analytic expressions for $E_0(\myvec{k})$ and $Z_0(\myvec{k})$ can be easily obtained because $d^{-1}_1(k)$ is quadratic in $\hat{k}_0^2$ for any values of $c_0$, $c_1$, $c_2$ and $c_4$. When $c_0$ is fixed by \refeq{contcond} the results read
\bea
E_0(\myvec{k}) &=& 2{\rm asinh}\medbreuk{1}{2}\sqrt{-\frac{B}{2A}\left( 1-\sqrt{1+4 A C/B^2} \right)},\nonumber\\
Z_0(\myvec{k}) &=& \left(\frac{{\rm sinh} \, E_0(\myvec{k})}{E_0(\myvec{k})} B \sqrt{1+4 A C/B^2}\right)^{-1},
\label{eq:ZEtreeexact}
\eea
where ($\tilde{c}_1\equiv c_1+4c_4$)
\bea
A &=& \tilde{c}_1 - c_4 \hat{k}_1^2,\nonumber\\
B &=& 1 - (\tilde{c}_1 - c_2) \hat{k}_1^2 - 2c_2 \hat{\myvec{k}}^2,\nonumber\\
C &=& \left( 1-c_2 \hat{\myvec{k}}^2 \right) \hat{\myvec{k}}^2 - (\tilde{c}_1 - c_2) \left( \hat{k}_1^2 \hat{\myvec{k}}^2 + \hat{\myvec{k}}^{(4)} \right) + c_4 \hat{k}_1^2  \hat{\myvec{k}}^{(4)},
\label{eq:defABC}
\eea
with $\hat{\myvec{k}}^{(4)} = \sum_{j=1}^{3} k_j^4
$. For $c_4 = 0$, $E_0(\myvec{k})$ reduces to the expression found in ref.~\cite{luscherLW3}, where however $Z_0(\myvec{k})$ is incorrect as it has the wrong limit for $k \rightarrow0$ (which corresponds to the continuum limit). The small-$k$ behavior of $E_0(\myvec{k})$ and $Z_0(\myvec{k})$ is
\bea
E_0(\myvec{k}) &=& |\myvec{k}|\left[ 1-\half(\tilde{c_1}-c_2 + \twelfth) \left(\myvec{k}^2 + \frac{\myvec{k}^{(4)}}{\myvec{k}^2} \right) + \cO(k^4) \right],\nonumber\\
Z_0(\myvec{k}) &=& 1 - 2(\tilde{c_1}-c_2 + \twelfth) \myvec{k}^2 + (\tilde{c_1}-c_2) k_1^2 + \cO(k^4).
\label{eq:ZEsmalla}
\eea
Note that unlike in the continuum, $Z_0(\myvec{k})$ need not be 1.

\section{Vertex components}
\label{app-vert}

In this appendix we list the vertex components defined in eqs.~(\ref{eq:dissectV3})--(\ref{eq:dissectV5}). We restrict ourselves to the contributions of Wilson loops with non-zero tree-level coefficients in the square action, i.e.\ the $1\times1$ ($i=0)$, $1\times2$ ($i=1)$ and $2\times2$ ($i=4)$ loops. The results for $i=0$ and $i=1$, as well as for the non-planar cases $i=2$ and $i=3$ were obtained before by Weisz and Wohlert~\cite{weiszII}. We have greatly benefited from comparing to their expressions. For our calculation we used Mathematica\cite{wolfram}. We automized the translation to \LaTeX, only doing by hand line-breaking and some minor changes to improve readability. We therefore expect the expressions below to be free of typesetting errors. We are also confident that no other mistakes were made because 1) the agreement with Weisz and Wohlert's expressions\footnote{Apart from some more or less obvious typographic errors in ref.~\protect\cite{weiszII}, we only disagree on the overall sign of the 3-vertex.}; 2) the correctness of our expressions for small lattice spacings $a$. An example of the second point is given at the end of this appendix.

For shortness we suppress the momentum arguments $(k_1,\cdots,k_n)$ on the left hand sides of the formulas below, and set $a=1$. The lattice spacing can be reinstated by dimensional analysis: the mass dimension of $k_{i}$ equals $+1$, that of any $n$-vertex component equals $4-n$.

Due to the absence of Lorentz symmetry, explicit expressions for the vertices tend to be long, especially for $i\not=0$ and $n\geq4$. We have put some effort in the search for relatively short forms. Nevertheless, we apologize for the lengthy formulas below.
\bea
f_{\mu}^{(3,i=0)} &=& 0,
\label{eq:f30expl}\\
f_{\mu}^{(3,i=1)} &=& -i\,\left\{ {{\hat{k}_1}^2}\, c_{1\mu}
      \stackrel{\myhat}{(k_{2} - k_{3})}_{\mu} +\mbox{ 2 cyclic perms of the momenta} \right\}\! ,
\label{eq:f31expl}\\
f_{\mu}^{(3,i=4)} &=& -i\left\{ {{\widehat{(2k_1)}}^2} c_{1\mu}\!
      \stackrel{\myhat}{(k_{2} - k_{3})}_{\mu} +\mbox{ 2 cyclic perms of the momenta} \right\}\!,
\label{eq:f34expl}\\
g_{\mu\nu}^{(3,i=0)} &=& -i\,{c_{3\mu}}\,{\stackrel{\myhat}{(k_{1} - k_{2})}_{\nu}},
\label{eq:g30expl}\\
g_{\mu\nu}^{(3,i=1)} &=& -i\,\left\{ 4\,\mbox{\Large$($} {\cos(k_{3\mu})}\,{c_{1\mu}}\,{c_{2\mu}} + 
        {\cos\half(k_{1} - k_{2})_{\nu}}\,{c_{3\mu}}\,{c_{3\nu}} \mbox{\Large$)$} \,
      {\stackrel{\myhat}{(k_{1} - k_{2})}_{\nu}} \right. \nonumber\\
&& \left. - \half{{{\widehat{(2k)}_{3\mu}}\,
         {\hat{k}_{3\nu}}\,{\stackrel{\myhat}{(k_{1} - k_{2})}_{\mu}}}} \right\},
\label{eq:g31expl}\\
g_{\mu\nu}^{(3,i=4)} &=& -i\,{c_{3\nu}}\,\left\{ 8\,{\cos(k_{3\mu})}\,{c_{1\mu}}\,
      {c_{2\mu}}\,{\stackrel{\myhatA}{(2(k_{1} - k_{2}))}_{\nu}}  - {\widehat{(2k)}_{3\mu}}\,{\widehat{(2k)}_{3\nu}} \,{\stackrel{\myhat}{(k_{1} - k_{2})}_{\mu}} \right\}
\label{eq:g34expl},\\
f_{\mu}^{(4,i=0)} &=& \sum_{\rho=0}^3
\sixth\! \left\{\! {\stackrel{\myhat}{(k_{1} + k_{3})}_{\rho}^2}\! - \half {\stackrel{\myhat}{(k_{1} + k_{2})}_{\rho}^2}\! - \half {\stackrel{\myhat}{(k_{1} + k_{4})}_{\rho}^2}\! + {\hat{k}_{1\rho}}\,{\hat{k}_{2\rho}}\,
       {\hat{k}_{3\rho}}\,{\hat{k}_{4\rho}}\!\right\},
\label{eq:f40expl}\\
f_{\mu}^{(4,i=1)} &=& \sum_{\rho=0}^3 \left\{\left[
\twelfth {{{\hat{k}_{1\rho}}}^2}\,\left( 40 - 2\,{{{\hat{k}_{1\mu}}}^2}  - 
         3\,{{{\hat{k}_{2\mu}^2}}}  - 
         3\,{{{\hat{k}_{4\mu}^2}}} - 3\,{{{\stackrel{\myhat}{(k_{1} + k_{2})}_{\mu}^2}}}  - 3\,{{{\stackrel{\myhat}{(k_{1} + k_{4})}_{\mu}^2}}}-  2\,{{{\hat{k}_{1\rho}}}^2}  \right) \right.\right. \nonumber\\
&&       \left.  + \eighth
   {{{\stackrel{\myhat}{(k_{1} + k_{2})}_{\rho}^2}}}
       \left( \rule{0cm}{6mm}\!\! -20 + 2\,{{{\hat{k}_{1\mu}}}^2} + 2\,{{{\hat{k}_{2\mu}}}^2} + {{{\stackrel{\myhat}{(k_{1} + k_{2})}_{\mu}^{\!2}}}}  + 2{{{\stackrel{\myhat}{(k_{1} + k_{4})}_{\mu}^{2}}}} + 
         {{{\stackrel{\myhat}{(k_{1} + k_{2})}_{\rho}^{2}}}} \right)\!\right]  \nonumber\\
&&  \left. \rule{0cm}{6mm} +\mbox{3 cyclic perms of the momenta} \right\},
\label{eq:f41expl}\\
f_{\mu}^{(4,i=4)} &=& \sum_{\rho=0}^3 \left\{\left[
\twelfth {{{\widehat{(2k)}_{1\rho}}}^2}\,
       \left( 32 - 2\,{{{\hat{k}_{1\mu}}}^2}  - 
         3\,{{{\hat{k}_{2\mu}^2}}}  - 
         3\,{{{\hat{k}_{4\mu}^2}}} - 3\,{{{\stackrel{\myhat}{(k_{1} + k_{2})}_{\mu}^2}}}- 3\,{{{\stackrel{\myhat}{(k_{1} + k_{4})}_{\mu}^2}}} \right) \right.\right. \nonumber\\
&&   \left.+ \eighth {{{\stackrel{\myhatA}{(2(k_{1} + k_{2}))}_{\rho}^2}}}\,
       \left( \rule{0cm}{6mm}\! -16 + 2\,{{{\hat{k}_{1\mu}}}^2} + 
         2\,{{{\hat{k}_{2\mu}}}^2} + {{{\stackrel{\myhat}{(k_{1} + k_{2})}_{\mu}^2}}} + 2 {{{\stackrel{\myhat}{(k_{1} + k_{4})}_{\mu}^2}}} \right)\right] \nonumber\\
&& \left. \rule{0cm}{6mm} + \mbox{ 3 cyclic perms of the momenta} \right\}\! ,
\label{eq:f44expl}\\
g_{\mu\nu}^{(4,i=0)} &=& \sixth{{\left( {c_{3\nu}}\,{\stackrel{\myhat}{(k_{1} - k_{2})}_{\nu}} - 
        {c_{1\nu}}\,{\stackrel{\myhat}{(k_{2} - k_{3})}_{\nu}} - 
        {\hat{k}_{1\nu}}\,{\hat{k}_{2\nu}}\,{\hat{k}_{3\nu}} \right) \,
      {\hat{k}_{4\mu}}}},
\label{eq:g40expl}\\
g_{\mu\nu}^{(4,i=1)} &=& - {{{c_{4\nu}}\,\left( {\cos(k_{1} - k_{3})_{\nu}}\,{\widehat{(2k)}_{2\nu}} + \third {\widehat{(2k)}_{4\nu}}
          \right) \,{\hat{k}_{4\mu}}}} + 
   {\cos\half(k_{1} - k_{3})_{\nu}} \,{\hat{k}_{2\nu}} \nonumber\\
&& \times \,\left({\cos(k_{4\mu})}\,\mbox{\Large$($}2\,{c_{1\mu}}\,{c_{3\mu}} \,{\hat{k}_{2\mu}} -{c_{2\mu}}\,{\stackrel{\myhat}{(k_{1} + k_{3})}_{\mu}}\mbox{\Large$)$} -4\,{c_{1\mu}}\,{c_{2\mu}}\,
       {c_{3\mu}}\,{\widehat{(2k)}_{4\mu}} 
       \right) \nonumber\\
&& - \twelfth {{{\widehat{(2k)}_{4\mu}} \,
       {\hat{k}_{4\nu}}\,\left( 16\,{c_{1\mu}}\,{c_{2\mu}}\,
          {c_{3\mu}} - {\stackrel{\myhat}{(k_{1} + k_{3})}_{\mu}}\,{\hat{k}_{2\mu}} + 
         2\,{c_{2\mu}}\,{\hat{k}_{1\mu}}\,{\hat{k}_{3\mu}} \right)}} \nonumber\\
&& - {\cos(k_{4\mu})}\,{c_{2\mu}}\,{c_{2\nu}}\,
      {\stackrel{\myhat}{(k_{1} - k_{3})}_{\mu}}\,{\stackrel{\myhat}{(k_{1} - k_{3})}_{\nu}},
\label{eq:g41expl}\\
g_{\mu\nu}^{(4,i=4)} &=& 2\,{\cos(k_{1} - k_{3})_{\nu}}\,{c_{4\nu}}\,{\widehat{(2k)}_{2\nu}} \nonumber\\
&& \times \,\left({\cos(k_{4\mu})}\,\mbox{\Large$($}2\,{c_{1\mu}}\,{c_{3\mu}} \,{\hat{k}_{2\mu}} -{c_{2\mu}}\,{\stackrel{\myhat}{(k_{1} + k_{3})}_{\mu}}\mbox{\Large$)$} -4\,{c_{1\mu}}\,{c_{2\mu}}\,
       {c_{3\mu}}\,{\widehat{(2k)}_{4\mu}} 
       \right) \nonumber\\
&& - \sixth {{{c_{4\nu}}\,{\widehat{(2k)}_{4\mu}}\,{\widehat{(2k)}_{4\nu}}\,
       \left( 16\,{c_{1\mu}}\,{c_{2\mu}}\,{c_{3\mu}} - 
         {\stackrel{\myhat}{(k_{1} + k_{3})}_{\mu}}\,{\hat{k}_{2\mu}} + 
         2\,{c_{2\mu}}\,{\hat{k}_{1\mu}}\,{\hat{k}_{3\mu}} \right) }} \nonumber\\
&& -2\,{\cos(k_{2\nu})}\,{\cos(k_{4\mu})}\,{c_{2\mu}}\,{c_{4\nu}} \,{\stackrel{\myhat}{(k_{1} - k_{3})}_{\mu}}\,
    {\stackrel{\myhatA}{(2(k_{1} - k_{3}))}_{\nu}},
\label{eq:g44expl}\\
h_{\mu\nu}^{(4,i=0)} &=& 2\,{\cos\half(k_{2} - k_{4})_{\mu}}\,{\cos\half(k_{1} - k_{3})_{\nu}},
\label{eq:h40expl}\\
h_{\mu\nu}^{(4,i=1)} &=& 8\,{c_{1\mu}}\,{c_{3\mu}}\,{\cos(k_{2} - k_{4})_{\mu}}\,{\cos\half(k_{1} - k_{3})_{\nu}}  + 8\,{c_{2\nu}}\,{c_{4\nu}}\,{\cos\half(k_{2} - k_{4})_{\mu}}\,{\cos(k_{1} - k_{3})_{\nu}},\nonumber\\
&&\label{eq:h41expl}\\
h_{\mu\nu}^{(4,i=4)} &=& 32\,{c_{1\mu}}\,{c_{3\mu}}\, {c_{2\nu}}\,{c_{4\nu}}\,{\cos(k_{2} - k_{4})_{\mu}}\,{\cos(k_{1} - k_{3})_{\nu}},
\label{eq:h44expl}\\
{h'}_{\mu\nu}^{(4,i=0)} &=& -{\cos\half(k_{3} - k_{4})_{\mu}}\,{\cos\half(k_{1} - k_{2})_{\nu}}  + 
   \fourth {\hat{k}_{3\mu}}\,
       {\hat{k}_{4\mu}}\,{{{\hat{k}_{1\nu}}\,{\hat{k}_{2\nu}}}},
\label{eq:h4p0expl}\\
{h'}_{\mu\nu}^{(4,i=1)} &=& \mbox{\Large$[$} -4\,{c_{1\mu}}\,
    {c_{2\mu}}\,{\cos(k_{3} - k_{4})_{\mu}}\,{\cos\half(k_{1} - k_{2})_{\nu}} - \half {{{\cos\half(k_{1} + k_{2})_{\nu}}}}\,{\stackrel{\myhat}{(k_{1} - k_{2})}_{\mu}} \nonumber\\
&& \times{\stackrel{\myhatA}{(2(k_{3} - k_{4}))}_{\mu}} \!+ {c_{1\mu}}\,{c_{2\mu}}\,{\widehat{(2k)}_{3\mu}}\,
    {\widehat{(2k)}_{4\mu}}\,{\hat{k}_{1\nu}}\,{\hat{k}_{2\nu}} \mbox{\Large$]$} \!+\! (k_1\!\leftrightarrow\! k_3,k_2\!\leftrightarrow\! k_4,\mu\!\leftrightarrow\!\nu), \label{eq:h4p1expl}\\
{h'}_{\mu\nu}^{(4,i=4)} &=& 4\,\left( -4\,{\cos(k_{3} - k_{4})_{\mu}}\,{\cos(k_{1} - k_{2})_{\nu}} + 
      {\widehat{(2k)}_{3\mu}}\,
       {\widehat{(2k)}_{4\mu}}\,{\widehat{(2k)}_{1\nu}}\,{\widehat{(2k)}_{2\nu}} \right) {c_{1\mu}}\,{c_{2\mu}}\,{c_{3\nu}}\,{c_{4\nu}}  \nonumber\\
&&  - 2\,\left( {c_{3\nu}}\,{c_{4\nu}}\,{\cos(k_{1} + k_{2})_{\nu}}\,{\stackrel{\myhat}{(k_{1} - k_{2})}_{\mu}}\, 
    {\stackrel{\myhatA}{(2(k_{3} - k_{4}))}_{\mu}} + {c_{1\mu}}\,{c_{2\mu}}\,{\cos(k_{3} + k_{4})_{\mu}} \right. \nonumber\\
&&  \left. \times {\stackrel{\myhat}{(k_{3} - k_{4})}_{\nu}}\, {\stackrel{\myhatA}{(2(k_{1} - k_{2}))}_{\nu}}\! \right)\! -\! \fourth {\stackrel{\myhat}{(k_{1} - k_{2})}_{\mu}}{\stackrel{\myhatA}{(2(k_{3} + k_{4}))}_{\mu}}
       {\stackrel{\myhat}{(k_{3} - k_{4})}_{\nu}}{{{\stackrel{\myhatA}{(2(k_{1} + k_{2}))}_{\nu}}}}, \nonumber\\
&&\label{eq:h4p4expl}\\
f_{\mu}^{(W,i)} &=& \sum_{\rho=0}^3 {\hat{k}_{1\rho}} {\hat{k}_{2\rho}} {\hat{k}_{3\rho}} {\hat{k}_{4\rho}} K^{(i)}_{\rho\mu}(k_1,k_2,k_3,k_4),
\label{eq:fWsimplel}\\
g_{\mu\nu}^{(W,i)} &=& - {\hat{k}_{4\mu}} {\hat{k}_{1\nu}} {\hat{k}_{2\nu}} {\hat{k}_{3\nu}} K^{(i)}_{\mu\nu}(k_1,k_2,k_3,k_4),
\label{eq:gWsimplel}\\
h_{\mu\nu}^{(W,i)} &=& {\hat{k}_{3\mu}} {\hat{k}_{4\mu}} {\hat{k}_{1\nu}} {\hat{k}_{2\nu}} K^{(i)}_{\mu\nu}(k_1,k_2,k_3,k_4).
\label{eq:hWsimplel}
\eea
In the last three equations we introduced
\bea
K^{(i=0)}_{\mu\nu}(k_1,k_2,k_3,k_4) &\equiv& 2,
\label{eq:Kmunu0}\\
K^{(i=1)}_{\mu\nu}(k_1,k_2,k_3,k_4) &\equiv& 32\left( c_{1\mu} c_{2\mu} c_{3\mu} c_{4\mu}  + c_{1\nu} c_{2\nu} c_{3\nu} c_{4\nu} \right),
\label{eq:Kmunu1}\\
K^{(i=4)}_{\mu\nu}(k_1,k_2,k_3,k_4) &\equiv& 512\left( c_{1\mu} c_{2\mu} c_{3\mu} c_{4\mu}\,\,  c_{1\nu} c_{2\nu} c_{3\nu} c_{4\nu} \right).
\label{eq:Kmunu4}
\eea

The 5-vertex component $f^{(5,i)}$ is not needed in the calculation of the one-loop Symanzik coefficients. The other components are expressed below as polynomials in a small number of functions, namely $\hat{k}_{j\lambda}$, $c_{j\lambda}$ and
\bea
& c^{\pm}_{ij\lambda} \equiv \cos\half(k_i \pm k_j)_{\lambda}; & s^{\pm}_{ij\lambda} \equiv \half\,{\stackrel{\myhat}{(k_{i} \pm k_{j})}_{\lambda}},\nonumber\\
& c^{2\pm}_{ij\lambda} \equiv \cos(k_i \pm k_j)_{\lambda}; & s^{2\pm}_{ij\lambda} \equiv \half\,{\stackrel{\myhatA}{(2(k_{i} \pm k_{j}))}_{\lambda}}.
\label{eq:sijmuetc}
\eea
For each component only those functions are used that are even or odd with respect to the symmetry operation belonging to that component (see \refeq{dissectsym}).
\bea
g_{\mu\nu}^{(5,i=0)} &=& 4 i {c_{5\mu}} \left\{ ( -3 {c^{-}_{23\nu}} + 
        4 {c^{+}_{23\nu}} )  {s^{-}_{14\nu}} - 
     3 {c^{-}_{14\nu}} {s^{-}_{23\nu}} \right\},
\label{eq:g50expl}\\
g_{\mu\nu}^{(5,i=1)} &=& 8 i \left\{ \rule{0mm}{5mm}{c_{5\mu}} {c_{5\nu}} 
      \left( ( -3 {c^{2-}_{23\nu}} + 4 {c^{2+}_{23\nu}} )  
         {s^{2-}_{14\nu}} - 3 {c^{2-}_{14\nu}} {s^{2-}_{23\nu}}
         \right) \right. \nonumber\\
&&  + {\cos(k_{5\mu})} \left( \rule{0mm}{5mm} 3 \left( {c^{-}_{23\nu}} {s^{-}_{14\nu}} + 
           {c^{-}_{14\nu}} {s^{-}_{23\nu}} \right)  
         \left( - {c^{-}_{23\mu}} 
              ( {c^{-}_{14\mu}} + 2 {c^{+}_{14\mu}} ) - ( 2 {c^{-}_{14\mu}} + 3 {c^{+}_{14\mu}} )  
            {c^{+}_{23\mu}} \right. \right. \nonumber\\
&& \left. 
              + {s^{-}_{14\mu}} {s^{-}_{23\mu}} - 
           {s^{+}_{14\mu}} {s^{+}_{23\mu}} \right) + 
        2 {c^{+}_{23\nu}} {s^{-}_{14\nu}} 
         \left( ( {c^{-}_{14\mu}} + {c^{+}_{14\mu}} )  
            ( 3 {c^{-}_{23\mu}} + 5 {c^{+}_{23\mu}} )  \right. \nonumber\\
&& \left.\left. - 
           3 {s^{-}_{14\mu}} {s^{-}_{23\mu}} + 
           {s^{+}_{14\mu}} {s^{+}_{23\mu}} \right) + 
        2 {c^{-}_{14\nu}} \left( -3 {s^{-}_{23\mu}} 
            {s^{+}_{14\mu}} + {s^{-}_{14\mu}} {s^{+}_{23\mu}} \right) 
           {s^{+}_{23\nu}} \rule{0mm}{5mm} \right)   \nonumber\\
&&  +    \half {(\widehat{2k})_{5\mu}}\left( \rule{0mm}{5mm} 6 
            ( -{c^{-}_{23\nu}} + {c^{+}_{23\nu}} )  
            {s^{-}_{14\nu}} \left( -(  {c^{-}_{23\mu}} + 
                   {c^{+}_{23\mu}} )  {s^{+}_{14\mu}}  + ( {c^{-}_{14\mu}} + {c^{+}_{14\mu}} )  
               {s^{+}_{23\mu}} \right)\right. \nonumber\\
&&  + 
           \left( ( 3 {c^{-}_{23\mu}} + 4 {c^{+}_{23\mu}} ) 
               {s^{-}_{14\mu}} + 3 {c^{-}_{14\mu}} {s^{-}_{23\mu}}
               \right) \left( {c^{+}_{23\nu}} {s^{+}_{14\nu}} + 
              {c^{+}_{14\nu}} {s^{+}_{23\nu}} \right) \nonumber\\
&&   + 
           6 {c^{-}_{14\nu}} 
            \left[ {s^{-}_{23\nu}} 
               \left( ( {c^{-}_{23\mu}} + {c^{+}_{23\mu}} )  
                  {s^{+}_{14\mu}} -  ( {c^{-}_{14\mu}} + {c^{+}_{14\mu}} )  
                  {s^{+}_{23\mu}} \right) \right. \nonumber\\
&& \left. \left. \left.  - 
              \left( ( {c^{-}_{23\mu}} + {c^{+}_{23\mu}} )  
                  {s^{-}_{14\mu}} + 
                 ( {c^{-}_{14\mu}} + {c^{+}_{14\mu}} )  
                  {s^{-}_{23\mu}} \right)  {s^{+}_{23\nu}} \right] 
           \rule{0mm}{5mm} \right)   \right\}, 
\label{eq:g51expl}\\
g_{\mu\nu}^{(5,i=4)} &=& 16 i {c_{5\nu}} \left\{ {\cos(k_{5\mu})} 
      \left( \rule{0mm}{5mm} 2 {c^{2-}_{14\nu}} {s^{2+}_{23\nu}} 
         ( -3 {s^{-}_{23\mu}} {s^{+}_{14\mu}} + 
           {s^{-}_{14\mu}} {s^{+}_{23\mu}} ) + 
        3 \left( {c^{2-}_{23\nu}} {s^{2-}_{14\nu}} + 
           {c^{2-}_{14\nu}} {s^{2-}_{23\nu}} \right) \right. \right. \nonumber\\
&&  \times  \left( - {c^{-}_{23\mu}} 
              ( {c^{-}_{14\mu}} + 2 {c^{+}_{14\mu}} ) - ( 2 {c^{-}_{14\mu}} + 3 {c^{+}_{14\mu}} )  
            {c^{+}_{23\mu}} + {s^{-}_{14\mu}} {s^{-}_{23\mu}} - 
           {s^{+}_{14\mu}} {s^{+}_{23\mu}} \right) \nonumber\\
&& \left. + 
        2 {c^{2+}_{23\nu}} {s^{2-}_{14\nu}} 
         \left( ( {c^{-}_{14\mu}} + {c^{+}_{14\mu}} )  
            ( 3 {c^{-}_{23\mu}} + 5 {c^{+}_{23\mu}} )  - 
           3 {s^{-}_{14\mu}} {s^{-}_{23\mu}} + 
           {s^{+}_{14\mu}} {s^{+}_{23\mu}} \right) \rule{0mm}{5mm} \right) \nonumber\\
&&  + {(\widehat{2k})_{5\mu}} \left( \rule{0mm}{5mm} 6 {c^{-}_{14\nu}} 
         ( -{c^{2-}_{23\nu}} + {c^{2+}_{23\nu}} )  
         {s^{-}_{14\nu}} \left( - ( {c^{-}_{23\mu}} + 
                {c^{+}_{23\mu}} )  {s^{+}_{14\mu}} + 
           ( {c^{-}_{14\mu}} + {c^{+}_{14\mu}} )  
            {s^{+}_{23\mu}} \right)  \right. \nonumber\\
&&  + 
        \left( ( 3 {c^{-}_{23\mu}} + 4 {c^{+}_{23\mu}} )  
            {s^{-}_{14\mu}} + 3 {c^{-}_{14\mu}} {s^{-}_{23\mu}}
            \right) \left( {c^{2+}_{23\nu}} {c^{+}_{14\nu}} 
            {s^{+}_{14\nu}} + 
           {c^{2+}_{14\nu}} {c^{+}_{23\nu}} {s^{+}_{23\nu}} \right) \nonumber\\
&&  +
          6 {c^{2-}_{14\nu}} 
         \left[ {c^{-}_{23\nu}} {s^{-}_{23\nu}} 
            \left( ( {c^{-}_{23\mu}} + {c^{+}_{23\mu}} )  
               {s^{+}_{14\mu}} - ( {c^{-}_{14\mu}} + {c^{+}_{14\mu}} )  
               {s^{+}_{23\mu}} \right) \right. \nonumber\\
&& \left. \left. \left.   - 
           {c^{+}_{23\nu}} \left( ( {c^{-}_{23\mu}} + 
                 {c^{+}_{23\mu}} )  {s^{-}_{14\mu}}  + ( {c^{-}_{14\mu}} + {c^{+}_{14\mu}} )  
               {s^{-}_{23\mu}} \right)  {s^{+}_{23\nu}} \right] \rule{0mm}{5mm} \right) 
      \right\},
\label{eq:g54expl}\\
h_{\mu\nu}^{(5,i=0)} &=& -24 i {s^{-}_{24\mu}} \left( {c_{3\nu}} {c^{+}_{15\nu}} + 
     {{\half{\hat{k}_{3\nu}} {s^{+}_{15\nu}}}} \right),
\label{eq:h50expl} \\
h_{\mu\nu}^{(5,i=1)} &=& -48 i \left\{ ( {c^{-}_{24\nu}} + {c^{+}_{24\nu}} )  
      {s^{-}_{24\mu}} \left( {\cos(k_{3\nu})} {c^{2+}_{15\nu}} + 
        \half{(\widehat{2k})_{3\nu}} {s^{2+}_{15\nu}} \right) +  2 {c_{3\mu}} ( {c^{-}_{15\mu}} + {c^{+}_{15\mu}} )
        {s^{2-}_{24\mu}} \right. \nonumber\\
&& \left. \times \left( {c_{3\nu}} {c^{+}_{15\nu}} + 
        \half{\hat{k}_{3\nu}} {s^{+}_{15\nu}} \right) + 2 {c^{2-}_{24\mu}} {c_{3\mu}} {s^{-}_{15\mu}} 
      \left( {c_{3\nu}} {c^{+}_{15\nu}} + 
        \half{\hat{k}_{3\nu}} {s^{+}_{15\nu}} \right)  \right\},
\label{eq:h51expl} \\
h_{\mu\nu}^{(5,i=4)} &=& -192 i\, {c_{3\mu}} ( {c^{-}_{24\nu}} \!+\! {c^{+}_{24\nu}}
      )\!  \left( {s^{2-}_{24\mu}} ( {c^{-}_{15\mu}} \!+\! 
     {c^{+}_{15\mu}} ) + 
     {c^{2-}_{24\mu}} {s^{-}_{15\mu}} \right) \!\!\left( {\cos(k_{3\nu})} {c^{2+}_{15\nu}} \!+ 
     \half{(\widehat{2k})_{3\nu}} {s^{2+}_{15\nu}} \right),\nonumber\\
&&\label{eq:h54expl}\\
{h'}_{\mu\nu}^{(5,i=0)} &=& 8 i \left( {c^{+}_{45\nu}} {s^{-}_{45\mu}} + 
     \threehalf {\hat{k}_{2\nu}} {s^{-}_{13\nu}} {s^{+}_{45\mu}}
      \right),
\label{eq:hp50expl}\\
{h'}_{\mu\nu}^{(5,i=1)} &=& 16 i \left\{ \left( 3 {c^{-}_{13\mu}} {c_{2\mu}} + 
        {c^{+}_{45\mu}} \right)  {c^{+}_{45\nu}} {s^{2-}_{45\mu}} + 
     3 {c_{2\mu}} \left( - {c^{-}_{13\nu}} {c_{2\nu}} 
           {c^{2+}_{45\mu}} + ( -{c^{2-}_{45\mu}} + {c^{2+}_{45\mu}} )  
         {c^{+}_{45\nu}} \right)  {s^{-}_{13\mu}} \right. \nonumber\\
&&  + 
     {c^{2+}_{45\nu}} ( {c^{-}_{45\nu}} + {c^{+}_{45\nu}} )
        {s^{-}_{45\mu}} + \threehalf {\hat{k}_{2\nu}} {s^{-}_{13\nu}} \left[ ( {c^{-}_{13\mu}} + {c^{+}_{13\mu}} )  
            \left( 2 {c_{2\mu}} {s^{2+}_{45\mu}} - \half {c^{2+}_{45\mu}} {\hat{k}_{2\mu}} \right) + {c_{2\mu}} {c^{2+}_{45\mu}} \right. \nonumber\\
&& \left.\left.  
           \times{s^{+}_{13\mu}} \right]+ \threehalf ( {c^{-}_{45\nu}} + {c^{+}_{45\nu}} )  {(\widehat{2k})_{2\nu}} {s^{2-}_{13\nu}} {s^{+}_{45\mu}} - 
     {s^{-}_{45\nu}} \left( \threehalf {c^{2-}_{13\nu}} {(\widehat{2k})_{2\nu}} + {s^{2+}_{45\nu}} \right)  {s^{+}_{45\mu}} \right\},
\label{eq:hp51expl}\\
{h'}_{\mu\nu}^{(5,i=4)} &=& 32 i \left\{ \rule{0mm}{5mm} {c^{2+}_{45\nu}} 
      \left( 3 {c^{-}_{13\mu}} {c_{2\mu}} + {c^{+}_{45\mu}} \right)
        ( {c^{-}_{45\nu}} + {c^{+}_{45\nu}} )  
      {s^{2-}_{45\mu}} \right. \nonumber\\
&& - 3 {c_{2\mu}} 
      \left( {c^{2-}_{13\nu}} {\cos(k_{2\nu})} {c^{2+}_{45\mu}} + 
        ( {c^{2-}_{45\mu}} - {c^{2+}_{45\mu}} )  
         {c^{2+}_{45\nu}} \right)  
      ( {c^{-}_{45\nu}} + {c^{+}_{45\nu}} )  {s^{-}_{13\mu}}  \nonumber\\
&& + \threehalf ( {c^{-}_{45\nu}} + {c^{+}_{45\nu}} )  
         {(\widehat{2k})_{2\nu}} {s^{2-}_{13\nu}} \left[ ( {c^{-}_{13\mu}} + {c^{+}_{13\mu}} )  
            \left( 2 {c_{2\mu}} {s^{2+}_{45\mu}} - \half {c^{2+}_{45\mu}} {\hat{k}_{2\mu}} \right) + {c_{2\mu}} {c^{2+}_{45\mu}} {s^{+}_{13\mu}} \right] \nonumber\\
&&  - {s^{-}_{45\nu}} 
      \left( \rule{0mm}{5mm} 3 {\cos(k_{2\nu})} {c_{2\mu}} {c^{2+}_{45\mu}} 
         {s^{2-}_{13\nu}} {s^{-}_{13\mu}} + \left( 3 {c^{-}_{13\mu}} {c_{2\mu}} + {c^{+}_{45\mu}}
            \right)  {s^{2+}_{45\mu}} {s^{2+}_{45\nu}} \right. \nonumber\\
&& \left. \left. + \threehalf {c^{2-}_{13\nu}} {(\widehat{2k})_{2\nu}} \left[ ( {c^{-}_{13\mu}} + {c^{+}_{13\mu}} )  
               \left( 2 {c_{2\mu}} {s^{2+}_{45\mu}} - \half {c^{2+}_{45\mu}} {\hat{k}_{2\mu}} \right) 
                + {c_{2\mu}} {c^{2+}_{45\mu}} {s^{+}_{13\mu}}
               \right] \rule{0mm}{5mm} \right)  \right\}.
\label{eq:hp54expl}
\eea

A powerful tool for tracking errors is the check whether the $a\rightarrow0$ behavior of the vertices is consistent with \refeq{expandac}. As a non-trivial example we perform this analysis for the 5-vertex. 

Reinstating the lattice spacing $a$, the $c_i$-weighted sums of eqs~(\ref{eq:g50expl})--(\ref{eq:hp54expl}) equal, up to $\cO(a^4)$ corrections,
\bea
g^{(5)}_{\mu\nu} &=& 2i a^2 (c_0+20c_1+64c_4) \left\{ (k_1-k_4)_{\nu} - 3 (k_2-k_3)_{\nu} \right\}, \nonumber\\
h^{(5)}_{\mu\nu} &=& -12i a^2\left\{(c_0+20c_1+64c_4) (k_2-k_4)_{\mu} + 4 (c_1+4c_4) (k_1-k_5)_{\mu} \right\}, \nonumber\\
h^{'(5)}_{\mu\nu} &=& 4i a^2\left\{(c_0+20c_1+64c_4) (k_4-k_5)_{\mu} - 6 (c_1+4c_4) (k_1-k_3)_{\mu} \right\}.
\label{eq:smalla5}
\eea
Eq.~(\ref{eq:expandac}) seems to be violated because $h$ and $h'$ are not proportional to $(c_0+20c_1+64c_4)$. However, we should take into account the remark below \refeq{Lieconv}: eq.~(\ref{eq:expandac}) only holds in terms of the field $\bar{A}_{\mu}$, defined by
\bea
&& U_{\mu}(x) = \Pexp\left(\int_0^a\!\! ds\, \bar{A}_{\mu}(x+s \hat{\mu}) \right)\nonumber\\
&&= 1+\sum_{m=1}^{\infty} \int_{0<s_1<s_2<\cdots<s_m<a}d^m\! s\, \bar{A}_{\mu}(x+s_1 \hat{\mu}) \cdots \bar{A}_{\mu}(x+ s_m \hat{\mu}).
\label{eq:Upath}
\eea
By expanding \refeqs{Upath}{Upert} with respect to $a$, the relation between $\bar{A}_{\mu}$ and $A_{\mu}$ is found to be
\be
A_{\mu}(x) = \bar{A}_{\mu}(x') + \twelfth a^2 \left\{ \half \partial_{\mu}^2 \bar{A}_{\mu}(x') + [\bar{A}_{\mu}(x'),\partial_{\mu} \bar{A}_{\mu}(x')]\right\} + \cO(a^3),
\label{eq:AinAbar}
\ee
where $x'=x+\half a\hat{\mu}$. Below we neglect the shift over $\half a\hat{\mu}$, because it only contributes total derivatives to the Lagrangian density that hence drop out of the action. 

The commutator term in \refeq{AinAbar} is interesting because it mixes $n$ and $n+1$ vertices. In particular if we expand \refeq{expSrest} with respect to $\bar{A}$ instead of $A$, denoting terms by $\bar{S}_n$, we find the relation
\be
\frac{g_0^3}{5!}  \bar{S}_5 = \frac{g_0^3}{5!}   S_5\, \rule[-2.5mm]{.15mm}{6mm}_{A_{\mu} \rightarrow \bar{A}_{\mu}} +  \frac{g_0^2}{4!}  \left( S_4\, \rule[-2.5mm]{.15mm}{6mm}_{A_{\mu} \rightarrow \bar{A}_{\mu}+ \twelfth a^2 [\bar{A}_{\mu},\partial_{\mu} \bar{A}_{\mu}]}\right)_{\!{\rm terms}\sim\bar{A}^5} + \cO(a^4).
\label{eq:S5tobarS5}
\ee
It is clear that only the $S_4$ components with a non-vanishing continuum limit, i.e.\ $h^{(4)}$ and $h^{'(4)}$, contribute to $\bar{S}_5$ at order $a^2$. By comparing the Lorentz-index structures of eqs.~\refnn{dissectV4} and~\refnn{dissectV5} one easily deduces that only $\bar{h}^{(5)}$ and $\bar{h}^{'(5)}$ are affected by this contribution, where of course $\bar{h}^{(5)}$, $\bar{h}^{'(5)}$, $\cdots$ have the same meaning with respect to $\bar{A}$ as $h^{(5)}$, $h^{'(5)}$, $\cdots$ have with respect to $A$. After some algebra, including the disentanglement of the \SU{N} structure, one finds the explicit result
\bea
\bar{g}^{(5)}_{\mu\nu} &=& 2i a^2 (c_0+20c_1+64c_4) \left\{ (k_1-k_4)_{\nu} - 3 (k_2-k_3)_{\nu} \right\} + \cO(a^4), \nonumber\\
\bar{h}^{(5)}_{\mu\nu} &=& -4i a^2(c_0+20c_1+64c_4) \left\{ 3(k_2-k_4)_{\mu} + (k_1-k_5)_{\mu} \right\} + \cO(a^4), \nonumber\\
\bar{h}^{'(5)}_{\mu\nu} &=& 2i a^2(c_0+20c_1+64c_4) \left\{ 2 (k_4-k_5)_{\mu} - (k_1-k_3)_{\mu} \right\} + \cO(a^4),
\label{eq:smalla5bar}
\eea
which is consistent with \refeq{expandac} (as it should be). We stress that this is a rather stringent test on the structure of $g^{(5,i)}$, $h^{(5,i)}$ and $h^{'(5,i)}$, and in particular on the signs and numerical prefactors of $h^{(5,i)}$ and $h^{'(5,i)}$.

\section{Feynman rules with a twist}
\label{app-twisted}
In a periodic finite volume the Fourier representations given in appendix~\ref{app-struct} are still valid, as long as one replaces the integral \refnn{defdollarperiodic} by a sum. In a twisted finite volume the situation is different due to color-momentum mixing. The Fourier representation appropriate to the `twisted tube' (see section~\ref{sec-twisted}) is given below.

We use lattice units in this appendix. For shortness we adopt the summation convention, both for Lorentz indices and for `color momentum' $e$ in the twisted directions $1$ and $2$ ($e_{\nu}/m$ ($\nu=1,2$) runs over $\{0,1,\cdots,N-1\}$, where $m\equiv 2\pi/(NL)$).
\bea
\action{measure} &=& \frac{N}{24} g_0^2 \delta_{\mu_1 \mu_2}\dollar{k_1,k_2}\delta(k_1+k_2) \tA_{\mu_1}(k_1)\tA_{\mu_2}(k_2) \!\left(-2z^{\half\brs{k}{k}}\right)\! + \!\cO(g_0^4),\label{eq:expSmeastwist}\\
\action{ghost} &=& 
\dollar{k_1,k_2}\bar{\tilde{c}}(k_1) \tilde{c}(k_2) \left[\rule{0mm}{5mm}\right.\delta(k_1 + k_2) \hat{k}_1^2 + i g_0 \dollar{k_3}\,\,\delta(k_1 + k_2 + k_3) \tA_{\mu}(k_3) f(k_1,k_2,k_3) \nonumber\\
&& \times z^{-\half\brs{k_1}{k_1}}
\hat{k}_{1\mu}c_{2\mu}  + \frac{1}{12} g_0^2 \delta_{\mu_3 \mu_4} \dollar{k_3,k_4} \delta(k_1 + k_2 + k_3 + k_4)  \tA_{\mu_3}(k_3) \tA_{\mu_4}(k_4) \nonumber\\
&& \left. \times 
f(k_1,k_3,e)f(k_2,k_4,-e)z^{-\half\brs{e}{e}}z^{-\half\brs{k_1}{k_1}}
\hat{k}_{1\mu_3}\hat{k}_{2\mu_3}  + \cO(g_0^4) \rule{0mm}{5mm}\right]\!,
\label{eq:expSghosttwist}\\
S_2(\{c_i\}) &=& \dollar{k_1,k_2}\delta(k_1+k_2) \tA_{\mu_1}(k_1)\tA_{\mu_2}(k_2) \left(-2z^{\half\brs{k}{k}}\right)
(\prop^{-1})_{\mu1\mu2}(k_1),
\label{eq:expS2twist}\\
S_3(\{c_i\}) &=&  \dollar{k_1,k_2,k_3}\delta(k_1+k_2+k_3) \tA_{\mu_1}(k_1)\tA_{\mu_2}(k_2)\tA_{\mu_3}(k_3) \nonumber\\
&& \times \left(-2 f(k_1,k_2,k_3)\right) V_{\mu_1\mu_2\mu_3}(k_1,k_2,k_3),
\label{eq:expS3twist}\\
S_4(\{c_i\}) &=& \dollar{k_1,k_2,k_3,k_4}\delta(k_1+k_2+k_3+k_4) \tA_{\mu_1}(k_1)\tA_{\mu_2}(k_2) \tA_{\mu_3}(k_3)\tA_{\mu_4}(k_4)\nonumber\\ 
&& \times\left[ -2f(k_1,k_2,e)f(k_3,k_4,-e) z^{-\half\brs{e}{e}} \right.\nonumber\\
&& \times (V_{\mu_1\mu_2\mu_3\mu_4}(k_1,k_2,k_3,k_4) - V_{\mu_2\mu_1\mu_3\mu_4}(k_2,k_1,k_3,k_4)) \nonumber\\ 
&& - \left. S(k_1,k_2,k_3,k_4) W_{\mu_1\mu_2\mu_3\mu_4}(k_1,k_2,k_3,k_4) \right],
\label{eq:expS4twist}\\
S_5(\{c_i\}) &=& 
\dollar{k_1,k_2,k_3,k_4,k_5}\delta(k_1+k_2+k_3+k_4+k_5) \tA_{\mu_1}(k_1)\cdots\tA_{\mu_5}(k_5) \nonumber\\
&& \times C(k_1 ,k_2 ,k_3 ,k_4 ,k_5) V_{\mu_1\mu_2\mu_3\mu_4\mu_5}(k_1,k_2,k_3,k_4,k_5).
\label{eq:expS5twist}
\eea
In these equations, $\prop_{\mu\nu}$ and the vertices $V$, $W$ are precisely the same as in infinite volume. However, the color factors are now functions of the momenta (in the twisted directions only),
\bea
f(k_1,k_2,k_3) &\equiv& \frac{1}{N}\Tr\left(\left[\Gamma_{k_1},\Gamma_{k_2}\right] \Gamma_{k_3}\right) \nonumber\\
&=& 2i \left(1-\chi_{k_1+k_2+k_3}\right) \nonumber\\
&& \times z^{\half \brs{k_1}{k_1} + \half \brs{k_2}{k_2} + \half \brs{k_1}{k_2}} \sin\left( \frac{\pi}{N}\bra{k_1}{k_2} \right),
\label{eq:Glebsch3wist}\\
S(k_1,k_2,k_3,k_4) &\equiv& \frac{1}{24N}\Tr\left(\Gamma_{k_1}\Gamma_{k_2} \Gamma_{k_3}\Gamma_{k_4} + \mbox{\ 23 permutations} \right) \nonumber\\
&=& \third \left(1-\chi_{k_1+k_2+k_3+k_4}\right) z^{-\half \sum_{1\leq i<j\leq4} \brs{k_i}{k_j} } \nonumber\\
&& \times \left\{ z^{\half\bra{k_1+k_2}{k_3+k_4}} \cos\left( \frac{\pi}{N}\bra{k_1}{k_2} \right) \cos\left( \frac{\pi}{N}\bra{k_3}{k_4} \right) \right.\nonumber\\
&& \left. + \left(k_2\leftrightarrow k_3 \right) + \left(k_2\leftrightarrow k_4 \right) \rule{0mm}{5mm} \right\},
\label{eq:Glebsch4symtwist}\\
C(k_1,k_2,k_3,k_4,k_5) &\equiv& \frac{1}{N}\Tr\left(\Gamma_{k_1}\Gamma_{k_2} \Gamma_{k_3}\Gamma_{k_4}\Gamma_{k_5} - \Gamma_{k_5}\Gamma_{k_4} \Gamma_{k_3}\Gamma_{k_3}\Gamma_{k_1} \right) \nonumber\\
&=& 2i \left(1-\chi_{k_1+k_2+k_3+k_4+k_5}\right) z^{-\half \sum_{1\leq i<j\leq5} \brs{k_i}{k_j} }  \nonumber\\
&& \times \sin\!\left( \frac{\pi}{N} \sum_{1\leq i<j\leq5} \bra{k_i}{k_j} \right).
\label{eq:Glebsch5twist}
\eea
The notation is explained in subsection~\ref{subsec-twisted-form}. Note that the factor $(-2z^{\half\brs{k}{k}})$ appearing in $S_2\{c_i\}$ is a color factor, brought about by $N^{-1}\Tr(\Gamma_k\Gamma_{-k}) = z^{\half\brs{k}{k}}$ as opposed to $\Tr(T^a T^b) = -\half\delta_{ab}$.

\section{Positivity of the square action}
\label{app-positive}

For many values of the coefficients $c_i(g_0^2)$, the action~\refnn{squareaction} is not positive~\cite{luscherLW2}. Such a choice of coefficients is unacceptable because the `vacuum' $A_{\mu} = 0$ (i.e.\ $U_{\mu}=1$), which has zero action, would not be a minimal-action configuration, and hence not a correct field configuration to expand about in perturbation theory. It is therefore important to prove that the square action is positive.

The basic ingredient~\cite{luscherLW2} for the proof is the statement that for any two unitary matrices $U$ and $V$
\be
\RE\Tr(1-UV)\leq 2\RE\Tr(1-U) + 2\RE\Tr(1-V).
\label{eq:CScorol}
\ee
This can be directly applied to \refeq{squareaction} by expressing the $1\times2$, $2\times2$ and corner Wilson loops as products of smaller loops (we choose $c_3=0$). For example, one may write
\be
\twoplaqbare = \twoplaqsplitA = \twoplaqup \times \twoplaqdown.
\label{eq:loopprod}
\ee
Noting that any unitary matrix $U$ satisfies
\be
\RE\Tr(1-U)\geq 0,
\label{eq:oneminUpos}
\ee
one quickly derives a rough lower bound:
\be
S(\{c_i(g_0^2)\}) \geq \left( c_0(g_0^2) + 8 \overline{(c_1(g_0^2) + 2\overline{c_4(g_0^2)})} + {\scriptstyle \medbreuk{80}{3}} \overline{c_2(g_0^2)} \right) S_{\rm Wilson},
\label{eq:Sgeq0cond}
\ee
where $\overline{x} \equiv \min(x,0)$. The Wilson action $S_{\rm Wilson}$ is positive due to \refeq{oneminUpos}. Therefore the action~(\ref{eq:squareaction}) is positive if the prefactor on the right hand side of \refeq{Sgeq0cond} is.

At tree level, the \LW\cite{luscherLW2} as well as the square action satisfy \refeq{Sgeq0cond} with a positive prefactor. If one includes the one-loop corrections that are computed in this paper, table~\ref{tab:coeff}, the prefactor is still positive for any reasonable value of $g_0^2$, as long as the free parameter $c_4'$ is chosen not too big. For example, the one-loop square action is positive for $c_4'=0$; $0<g_0^2<10.9$, and also for $g_0^2 = 1$; $144\,c'_4 < 7.2$ ($N=3$).

\end{document}